%% file: WittenDiagramBootstrap.tex
\documentclass[12pt]{article}

\usepackage[utf8]{inputenc}
\usepackage{amssymb,amsmath,amsfonts}
\usepackage{amsthm}
\usepackage[
  pdfauthor={Aleix Gimenez-Grau},
  pdftitle={The Witten Diagram Bootstrap for Holographic Defects},
  bookmarksnumbered,
  hidelinks,
  colorlinks,
  citecolor=blue,
  urlcolor=blue,
  linkcolor=blue,
  linktoc=page
]{hyperref}
\usepackage{dsfont}
\usepackage{graphicx}
\usepackage{caption}
\usepackage{subcaption}
\usepackage{pgfplots}
\usepackage{slashed}
\usepackage{cite}
\pgfplotsset{compat=1.15}

\graphicspath{{figures/}}

\usepackage{tikz}
\usetikzlibrary{shapes}

\usepackage[titles]{tocloft}
\usepackage[font=small,labelfont=bf,width=0.95\textwidth]{caption}

\def\Cm{{\mathcal{C}}}
\def\Dm{{\mathcal{D}}}

\def\Fm{{\mathcal{F}}}

\def\Lm{{\mathcal{L}}}
\def\Mm{{\mathcal{M}}}
\def\Nm{{\mathcal{N}}}
\def\Om{{\mathcal{O}}}

\def\Sm{{\mathcal{S}}}

\def\a{{\alpha}}
\def\b{{\beta}}

\newcommand\wh\widehat

\newcommand{\vev}[1]{\langle #1 \rangle}

\newcommand\veps{{\varepsilon}}
\newcommand\zb{{\bar z}}

\newcommand\Dh{{\widehat{\Delta}}}

\newcommand\pmin{{p_\mathrm{min}}}
\newcommand\AdS{\mathrm{AdS}}

\def\mathematica{{\texttt{Mathematica}}}

\def \ph{\phantom}

\DeclareMathOperator{\tr}{tr}

\DeclareMathOperator{\dDisc}{dDisc}
\DeclareMathOperator{\Disc}{Disc}

\DeclareMathOperator{\re}{Re}
\DeclareMathOperator{\Pexp}{Pexp}
\DeclareMathOperator{\sgn}{sgn}

\usepackage{tikz}
\usetikzlibrary{shapes}
\usetikzlibrary{shapes.misc}

\tikzset{
  vtx/.style={
    circle,
    draw=blue,
    fill=blue,
    inner sep=1pt
  },
  wcirc/.style={
    circle,
    draw=white,
    fill=white,
    inner sep=2pt
  },
  bcirc/.style={
    circle,
    draw=black,
    fill=black,
    inner sep=1pt
  },
  dcirc/.style={
    circle,
    draw=blue,
    fill=blue,
    inner sep=1pt
  },
  rcirc/.style={
    circle,
    draw=red,
    fill=red,
    inner sep=1pt
  },
  phi/.style={
    thick
  },
  sigma/.style={
    thick,
    dashed
  },
  vl1/.style={
    thick,
    blue
  },
  vl2/.style={
    thick,
    dashed,
    blue
  },
  valign/.style={
    baseline={([yshift=-.55ex]current bounding box.center)}
  }
}

\title{\LARGE \bf The Witten Diagram Bootstrap \\ for Holographic Defects \\[0.5em]}
\author{
{\large Aleix Gimenez-Grau\thanks{gimenez@ihes.fr}} \\ ~ \\[0em]
{\normalsize \emph{Institut des Hautes Études Scientifiques, 91440 Bures-sur-Yvette, France}} \\[1em] }
\date{}

\begin{document}

\clearpage\maketitle
\thispagestyle{empty}

\begin{abstract}
~\\
\noindent{
We study the AdS/CFT correspondence with a brane extending in AdS, a setup which is dual to CFT in the presence of a defect.
We focus on the correlation function of two local operators and the defect, which is the simplest observable with non-trivial dependence on kinematical invariants.
We propose a method to bootstrap this observable which relies on supersymmetry, but does not require detailed knowledge of the supergravity and brane effective actions.
After developing the method in full generality, we turn to the case of two chiral-primary operators and a half-BPS Wilson loop in $\Nm=4$ SYM.
Working in the leading supergravity approximation, we determine the correlator in closed form for chiral-primary operators of arbitrary length.
The result has elegant expressions in position and Mellin space, and it agrees with localization and an explicit calculation up to contact terms.
More generally, we expect our method to be suitable in other holographic setups in the presence of supersymmetric defects.
}

\end{abstract}

\newpage

\tableofcontents

\newpage

\section{Introduction}

One of the most exciting discoveries in high energy theory is that certain quantum gravity theories in Anti de-Sitter (AdS) are dual to conformal field theories \cite{Maldacena:1997re,Gubser:1998bc,Witten:1998qj}.
This goes under the name AdS/CFT correspondence, and it has been a fruitful line of research for the last 25 years.
A natural observable, which is studied since the early days of AdS/CFT, is the correlation function of local operators \cite{Freedman:1998tz,DHoker:1999kzh,Arutyunov:2000py}.
These correlators are the AdS analog of scattering amplitudes, an analogy that can be made precise in Mellin space \cite{Penedones:2010ue}.
More specifically, a simple formula relates the flat-space limit of Mellin amplitudes to flat-space string amplitudes \cite{Okuda:2010ym,Fitzpatrick:2011hu,Fitzpatrick:2011jn}.
In AdS/CFT, correlators of local operators can be calculated as sums of Witten diagrams, although this is notoriously hard due to the proliferation of diagrams and the complicated form of the supergravity effective actions \cite{Kim:1985ez,Lee:1998bxa,Arutyunov:1998hf,Arutyunov:1999en,Arutyunov:1999fb}.
Several years ago, the work of Rastelli and Zhou \cite{Rastelli:2016nze,Rastelli:2017udc} bypassed these complications by making an ansatz for four-point correlators, and fixing it with basic consistency conditions, such as crossing symmetry and supersymmetry.
Since then, the study of correlators in the AdS/CFT correspondence has received renewed attention.
To mention only some of the many developments, there are now results going beyond the supergravity approximation \cite{Alday:2019nin,Alday:2020tgi,Alday:2021ajh,Alday:2022rly,Binder:2019jwn,Aprile:2017bgs,Aprile:2017qoy,Chester:2020dja}, results for five-point functions \cite{Goncalves:2019znr,Alday:2022lkk,Goncalves:2023oyx}, or results in setups without maximal supersymmetry \cite{Alday:2021odx,Alday:2021ajh,Behan:2023fqq}.

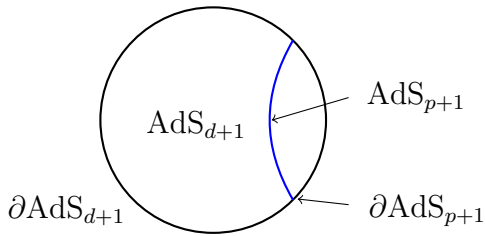
\begin{figure}[h]
   \centering
   \begin{tikzpicture}[valign,scale=1.5]
    \pgfmathsetmacro{\x}{sqrt(2)/2}
    \pgfmathsetmacro{\y}{sqrt(2)/2}
    \draw [thick] (0,0) circle [radius=1];
    \draw [thick, blue] (+\x,+\y) to[out=240,in=120] (+\x,-\y);
    \draw [->] (1.2, 0.2) -- (0.52,0);
    \draw [->] (1.2,-0.75) -- (0.75,-0.7);
    \node at (-0.15, -0.05) {$\AdS_{d+1}$};
    \node at (-1.3, -0.8) {$\partial\AdS_{d+1}$};
    \node at ( 1.8,  0.2) {$\AdS_{p+1}$};
    \node at ( 1.9, -0.8) {$\partial\AdS_{p+1}$};
  \end{tikzpicture}
  \caption{The bulk $\AdS_{d+1}$ space is enclosed by the black circle.
  The brane is represented in blue and it extends in $\AdS_{p+1} \subset \AdS_{d+1}$.
  The boundary theory is a CFT in the presence of a $p$-dimensional conformal defect.}
  \label{fig:setup}
\end{figure}

In the present work, we generalize the standard setup in yet another way: we scatter local operators off an extended object.
We refer to the extended object as a brane, which could be for example a long fundamental string, a probe D$p$-brane, etc.
More precisely, we consider a $(p+1)$-dimensional brane that extends in $\AdS_{p+1} \subset \AdS_{d+1}$.
From the point of view of the dual CFT, there is an extended operator where the brane intersects with the boundary of $\AdS_{d+1}$.
Because extended operators are often called defects, we refer to this setup as a holographic defect, see figure \ref{fig:setup}.

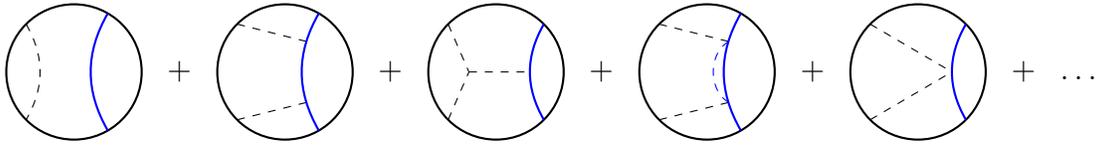
\begin{figure}
 \begin{align*}
 \begin{tikzpicture}[valign,scale=0.9]
    \tikzstyle{every node}=[font=\scriptsize]
    \pgfmathsetmacro{\x}{sqrt(1)/2}
    \pgfmathsetmacro{\y}{sqrt(3)/2}
    \pgfmathsetmacro{\xx}{sqrt(2)/2}
    \pgfmathsetmacro{\yy}{sqrt(2)/2}
    \pgfmathsetmacro{\z}{0.5}
    \pgfmathsetmacro{\rx}{0.32}
    \pgfmathsetmacro{\ry}{0.45}
    \draw [thick] (0,0) circle [radius=1];
    \draw [dashed]      (-\xx,+\yy) to[out=-60,in=60] (-\xx,-\yy);
    \draw [thick, blue] (\x,+\y) to[out=240,in=120] (\x,-\y);
 \end{tikzpicture}
 \;\;+\;\;
 \begin{tikzpicture}[valign,scale=0.9]
    \tikzstyle{every node}=[font=\scriptsize]
    \pgfmathsetmacro{\x}{sqrt(1)/2}
    \pgfmathsetmacro{\y}{sqrt(3)/2}
    \pgfmathsetmacro{\xx}{sqrt(2)/2}
    \pgfmathsetmacro{\yy}{sqrt(2)/2}
    \pgfmathsetmacro{\z}{0.5}
    \pgfmathsetmacro{\rx}{0.32}
    \pgfmathsetmacro{\ry}{0.45}
    \draw [thick] (0,0) circle [radius=1];
    \draw [dashed]      (-\xx,+\yy) -- (\rx,\ry);
    \draw [dashed]      (-\xx,-\yy) -- (\rx,-\ry);
    \draw [thick, blue] (\x,+\y) to[out=240,in=120] (\x,-\y);
 \end{tikzpicture}
 \;\;+\;\;
   \begin{tikzpicture}[valign,scale=0.9]
    \tikzstyle{every node}=[font=\scriptsize]
    \pgfmathsetmacro{\x}{-sqrt(2)/2}
    \pgfmathsetmacro{\y}{sqrt(2)/2}
    \pgfmathsetmacro{\z}{-0.5}
    \pgfmathsetmacro{\r}{0.4}
    \draw [thick] (0,0) circle [radius=1];
    \draw [thick, blue] (-\x,+\y) to[out=240,in=120] (-\x,-\y);
    \draw [dashed] (+\x,+\y) -- (-\r,0);
    \draw [dashed] (+\x,-\y) -- (-\r,0);
    \draw [dashed] (-\r,  0) -- (-\z,0);
 \end{tikzpicture}
 \;\;+\;\;
 \begin{tikzpicture}[valign,scale=0.9]
    \tikzstyle{every node}=[font=\scriptsize]
    \pgfmathsetmacro{\x}{sqrt(1)/2}
    \pgfmathsetmacro{\y}{sqrt(3)/2}
    \pgfmathsetmacro{\xx}{sqrt(2)/2}
    \pgfmathsetmacro{\yy}{sqrt(2)/2}
    \pgfmathsetmacro{\z}{0.5}
    \pgfmathsetmacro{\rx}{0.32}
    \pgfmathsetmacro{\ry}{0.45}
    \draw [thick] (0,0) circle [radius=1];
    \draw [dashed]      (-\xx,+\yy) -- (\rx,\ry);
    \draw [dashed]      (-\xx,-\yy) -- (\rx,-\ry);
    \draw [dashed,blue] (\rx,\ry) to[out=210,in=150] (\rx,-\ry);
    \draw [thick, blue] (\x,+\y) to[out=240,in=120] (\x,-\y);
 \end{tikzpicture}
 \;\;+\;\;
 \begin{tikzpicture}[valign,scale=0.9]
    \tikzstyle{every node}=[font=\scriptsize]
    \pgfmathsetmacro{\x}{sqrt(2)/2}
    \pgfmathsetmacro{\y}{sqrt(2)/2}
    \pgfmathsetmacro{\z}{0.5}
    \pgfmathsetmacro{\r}{-0.75}
    \draw [thick] (0,0) circle [radius=1];
    \draw [thick, blue] (+\x,+\y) to[out=240,in=120] (+\x,-\y);
    \draw [dashed]      (-\x,+\y) -- (\z,0);
    \draw [dashed]      (-\x,-\y) -- (\z,0);
  \end{tikzpicture}
  \;\;+\;\;
  \ldots
\end{align*}
\caption{Low-lying Witten diagrams for a scattering process of two local operators and a brane. Black dashed lines are fields propagating in the bulk, and blue dashed lines are fields propagating on the brane. }
\label{fig:witten}
\end{figure}

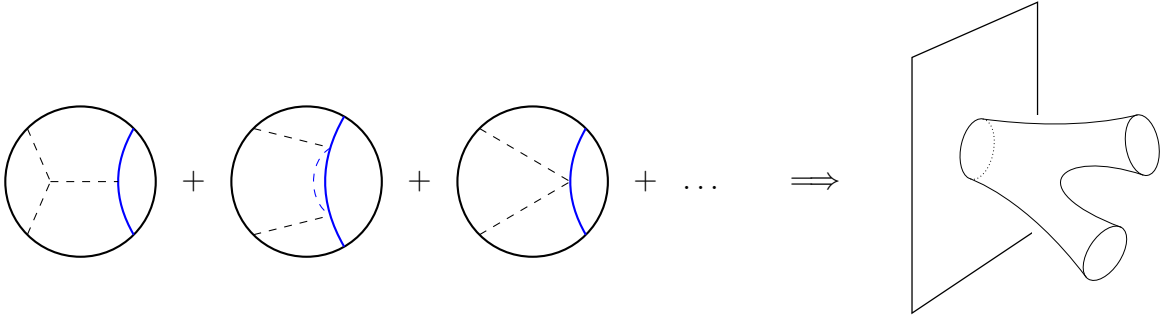
\begin{figure}
 \begin{align*}
 \begin{tikzpicture}[valign,scale=1]
    \tikzstyle{every node}=[font=\scriptsize]
    \pgfmathsetmacro{\x}{-sqrt(2)/2}
    \pgfmathsetmacro{\y}{sqrt(2)/2}
    \pgfmathsetmacro{\z}{-0.5}
    \pgfmathsetmacro{\r}{0.4}
    \draw [thick] (0,0) circle [radius=1];
    \draw [thick, blue] (-\x,+\y) to[out=240,in=120] (-\x,-\y);
    \draw [dashed] (+\x,+\y) -- (-\r,0);
    \draw [dashed] (+\x,-\y) -- (-\r,0);
    \draw [dashed] (-\r,  0) -- (-\z,0);
 \end{tikzpicture}
 \;\;+\;\;
 \begin{tikzpicture}[valign,scale=1]
    \tikzstyle{every node}=[font=\scriptsize]
    \pgfmathsetmacro{\x}{sqrt(1)/2}
    \pgfmathsetmacro{\y}{sqrt(3)/2}
    \pgfmathsetmacro{\xx}{sqrt(2)/2}
    \pgfmathsetmacro{\yy}{sqrt(2)/2}
    \pgfmathsetmacro{\z}{0.5}
    \pgfmathsetmacro{\rx}{0.32}
    \pgfmathsetmacro{\ry}{0.45}
    \draw [thick] (0,0) circle [radius=1];
    \draw [dashed]      (-\xx,+\yy) -- (\rx,\ry);
    \draw [dashed]      (-\xx,-\yy) -- (\rx,-\ry);
    \draw [dashed,blue] (\rx,\ry) to[out=210,in=150] (\rx,-\ry);
    \draw [thick, blue] (\x,+\y) to[out=240,in=120] (\x,-\y);
 \end{tikzpicture}
 \;\;+\;\;
 \begin{tikzpicture}[valign,scale=1]
    \tikzstyle{every node}=[font=\scriptsize]
    \pgfmathsetmacro{\x}{sqrt(2)/2}
    \pgfmathsetmacro{\y}{sqrt(2)/2}
    \pgfmathsetmacro{\z}{0.5}
    \pgfmathsetmacro{\r}{-0.75}
    \draw [thick] (0,0) circle [radius=1];
    \draw [thick, blue] (+\x,+\y) to[out=240,in=120] (+\x,-\y);
    \draw [dashed]      (-\x,+\y) -- (\z,0);
    \draw [dashed]      (-\x,-\y) -- (\z,0);
  \end{tikzpicture}
  \;\;+\;\;
  \ldots
  \qquad \Longrightarrow \qquad
  \raisebox{-0.4\height}{\includegraphics[width=0.2\linewidth]{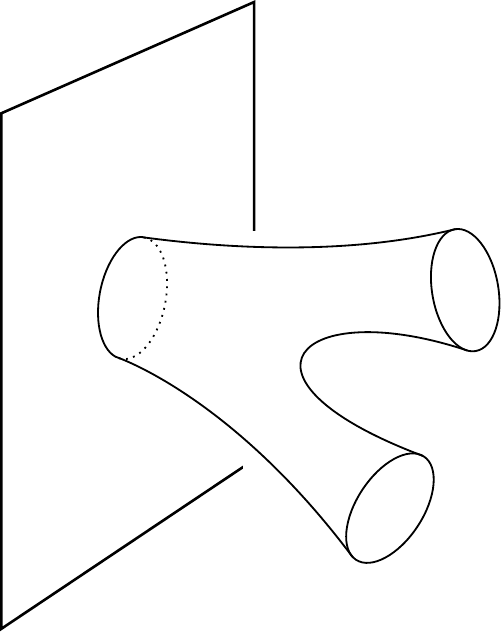}}
\end{align*}
\caption{The connected correlator, to be defined below, should map in the flat-space limit to a scattering amplitude of closed strings off an extended brane.}
\label{fig:flat-space}
\end{figure}

\noindent
To probe the system, we prepare local operators at the boundary of $\AdS_{d+1}$, and scatter them off the brane in the bulk.
These correlation functions admit an expansion in Witten diagrams, with extra vertices due to interactions between bulk and brane fields.
This work focuses on the case of two local operators, for which some low-lying Witten diagrams are shown in figure \ref{fig:witten}.
This setup is the AdS/CFT analog of a $1 \to 1$ scattering experiment off a fixed extended target.
This analogy is particularly compelling in Mellin space, see \cite{Goncalves:2018fwx} and section \ref{sec:mellin} below.
We also expect a connection between Mellin amplitudes and flat-space string amplitudes, see figure \ref{fig:flat-space}, although this connection has not been worked out in the literature.

To make the discussion more concrete, let us consider a prototypical example of a holographic defect, namely a fundamental string in $\AdS_5 \times S^5$.
This string intersects the boundary of $\AdS_5$ at a contour $\gamma$, and has Dirichlet boundary conditions on the $S^5$.
As shown in \cite{JJ1,sjrey}, this configuration is dual to $\Nm=4$ SYM with a half-BPS Wilson loop in the fundamental representation.
The Wilson operator $W$ is supported on the contour $\gamma$, and when this contour is a straight line, the fundamental string extends in the bulk in an $\AdS_2 \subset \AdS_5$.
This setup has been studied extensively in the literature, since it provides a rich interplay of techniques such as holography \cite{Berenstein:1998ij,Giombi:2006de,Giombi:2009ds,Giombi:2017cqn,Giombi:2022pas}, localization \cite{Pestun:2007rz,Pufu:2023vwo}, integrability \cite{Giombi:2018qox,Giombi:2018hsx,Cavaglia:2022yvv,Cavaglia:2021bnz}, perturbation theory \cite{Kiryu:2018phb,Barrat:2021tpn,Barrat:2022eim} or bootstrap \cite{Liendo:2018ukf,Ferrero:2021bsb,Cavaglia:2021bnz,Cavaglia:2022qpg}.\footnote{This list covers only a small fraction of the literature, and we apologize for omissions.}
Many of these works focus on observables like the expectation value of the loop $\vev{W}$, one-point functions of local operators $\vev{\Om W}$, or correlators of fields inserted on the defect $\vev{W[\wh \Om_1 \wh \Om_2 \wh \Om_3 \wh \Om_4]}$.
However, the observable that concerns us is the two-point function of local operators $\vev{\Om_1 \Om_2 W}$, which has received comparatively little attention, but see e.g. \cite{Buchbinder:2012vr,Barrat:2020vch,Barrat:2021yvp,Barrat:2022psm}.

Motivated by this lack of results, here we present an efficient bootstrap method to compute two-point functions of local operators in the presence of a holographic defect.
The idea is that of the position-space method of Rastelli and Zhou \cite{Rastelli:2016nze,Rastelli:2017udc}: one makes an ansatz in terms of Witten diagrams as in figure \ref{fig:witten}, and fixes the relative coefficients demanding that superconformal Ward identities are satisfied.
The use of an ansatz allows to bypass detailed knowledge of the bulk and brane effective actions, which can be quite complicated.
Because the method relies on supersymmetry, we expect it to work for half-BPS branes in maximally supersymmetric holography, but it is unclear whether it will work with less supersymmetry.

Two ingredients are necessary to implement the bootstrap.
The first ingredient are closed form expressions for Witten diagrams like the ones in figure \ref{fig:witten}.
Witten diagrams were studied for codimension-one branes, namely $p=d-1$, in \cite{RastZhouMell,Mazac:2018biw,Kaviraj:2018tfd}.
That case is simpler, because internal lines contain only scalar fields, and the result depends on only one cross-ratio.
Some of these results were generalized to $p<d-1$ in \cite{Goncalves:2018fwx}.
However, for arbitrary $p$, one needs diagrams with spinning fields in internal lines, which have not been worked out yet.
To fill this gap in the literature, sections \ref{sec:pos} and \ref{sec:mellin} present a comprehensive computation of Witten diagrams in position and Mellin space respectively.
The second ingredient of the bootstrap method are the superconformal Ward identities.
Superconformal Ward identities are known in several interesting cases, such as half-BPS interfaces and lines in $\Nm=4$ SYM \cite{Liendo:2016ymz}, or surface defects in $6d$ $(2,0)$ theory \cite{Meneghelli:2022gps}.
Although no general results are available, we expect that the need to derive Ward identities on a case-by-case basis will not be an obstacle to bootstrap other setups.

Section \ref{sec:MWL} combines the results of sections \ref{sec:pos} and \ref{sec:mellin} to bootstrap $\Nm=4$ SYM correlators in the presence of a half-BPS Wilson line.
More precisely, we bootstrap $\vev{S_{p_1} S_{p_2} W}$, the correlator of two local operators and a half-BPS Wilson line.
The local operators $S_k$ are superprimaries of half-BPS multiplets, and transform in rank-$k$ symmetric-traceless representations of $SO(6)_R$.
The half-BPS Wilson line $W$ was described above, and is dual to a fundamental string.
Because we work in the supergravity approximation, the correlator is computed in the planar limit $N \to \infty$ at fixed but large 't Hooft coupling $\lambda = g^2 N \to \infty$.
The perturbative expansion is a double power series in $\frac{1}{N^2}$ and $\frac{1}{\sqrt{\lambda}}$
\begin{align}
 \vev{S_{p_1} S_{p_2} W}_c
 = \frac{1}{N^2} \left(
    \sqrt{\lambda} \, \Fm_{-\frac12,2}
 +  \Fm_{0,2}
 + \ldots \right)
 + \frac{1}{N^4} \left(
    \lambda \, \Fm_{-1,4}
 +  \sqrt{\lambda} \, \Fm_{-\frac12,4}
 + \ldots \right)
 + \ldots \, .
 \label{eq:series}
\end{align}
Loosely speaking, powers of $\frac{1}{N^2}$ count loops in the bulk, and powers of $\frac{1}{\sqrt{\lambda}}$ count loops on the string worldsheet.
In this work, we determine the leading term $\Fm_{-\frac12,2}$, hoping this will be the first step towards determining higher-order corrections, see \cite{Pufu:2023vwo}.

In equation \eqref{eq:series} we are considering the connected part of the correlator $\vev{S_{p_1} S_{p_2} W}_c$, which is the sum of connected Witten diagrams, compare figures \ref{fig:witten} and \ref{fig:flat-space}.
The disconnected correlator is somewhat trivial, and will be ignored in most of this work.
Our main focus is $\Fm_{-\frac12,2}$, which consists of a non-trivial function of three kinematical invariants.
For chiral-primaries of length $p_1,p_2 \le 4$, the correlators $\Fm_{-\frac12,2}$ were obtained recently in \cite{Barrat:2021yvp,Barrat:2022psm} using analytic bootstrap.\footnote{The correlator $\vev{S_2 S_2 W}$ was also studied in \cite{Barrat:2020vch}, but in the planar limit at weak 't Hooft coupling $\lambda \ll 1$. Other works \cite{Giombi:2009ds,Beccaria:2020ykg} have computed $\vev{S_{p_1} S_{p_2} W}$ with localization, but their results apply to very special kinematics, where the cross-ratio dependence drops out, and the correlator reduces to a constant.
Unlike their results, our correlator contains non-trivial information of long operators through OPE expansions.}
Here we improve their results in several ways.
On one hand, thanks to the more powerful methods in the present work, we succeed in determining the correlator in closed form for arbitrary $p_1,p_2$.
The result is particularly elegant in Mellin space, see equation \eqref{eq:mell-res}.
On the other hand, the analytic bootstrap method \cite{Barrat:2021yvp,Barrat:2022psm} suffered from low-spin ambiguities that were  fixed in a somewhat ad-hoc way.
Here that calculation is put on solid ground by showing that low-spin ambiguities are due to Witten diagrams that exchange fields on the brane, as in the fourth diagram of figure \ref{fig:witten}.
Finally, we also do a first-principles computation of the correlators, which requires us to determine the interaction vertices of the effective action.
The result agrees with the bootstrap, up to contact terms that we do not know how to obtain in the first-principles calculation.

The main sections of our work are \ref{sec:pos}--\ref{sec:MWL}, the contents of which we just outlined.
Besides them, section \ref{sec:prelim} provides a more thorough description of the setup and its kinematics, reviews standard material and sets the notation.
We also outline possible future work in section \ref{sec:conclusion}, including a list of setups where we expect our methods to be applicable.
Finally, we relegate a number of technical details to appendix \ref{app:spintwo}--\ref{sec:wi-mell}.

\section{Preliminaries}
\label{sec:prelim}

\subsection{Setup}

As explained in the introduction, this work studies $(d+1)$-dimensional anti-de Sitter space containing a $(p+1)$-dimensional brane for $p<d$, see figure \ref{fig:setup}.
We restrict our attention to a brane that extends in an $\AdS_{p+1}$ submanifold, so in particular the brane breaks the isometries of $\AdS_{d+1}$ as follows
\begin{align}
 SO(d+1,1) \; \to \; SO(p+1,1) \oplus SO(d-p) \, .
 \label{eq:symbreak}
\end{align}
In the standard AdS/CFT dictionary, the isometry group $SO(d+1,1)$ maps to the conformal symmetry that acts on the boundary theory.
In the present case, the symmetry of the boundary theory is \eqref{eq:symbreak}, which is precisely the group preserved by CFT in the presence of a flat conformal defect \cite{Billo:2016cpy}.
Borrowing common terminology of defect CFT, we use the words brane and defect interchangeably for the $\AdS_{p+1}$.
The connection to defect CFT is important for our purposes, and we discuss it in more detail in section \ref{sec:dcft}.

To be more precise, we work in Euclidean signature and parametrize $\AdS_{d+1}$ in Poincare coordinates $z^\mu \in \AdS_{d+1}$, with $\mu = 0, 1, \ldots, d$.
The conformal boundary of $\AdS_{d+1}$ can be identified with $\mathbb R^d$, and is approached in the limit $z^0 \to 0$.
In particular, we denote boundary points as $x \in \mathbb R^d$.
It is convenient to split the coordinates of $\AdS_{d+1}$ in directions parallel and orthogonal to the brane
\begin{align}
 z^\mu = (z^0, z^a, z^i)
 \quad\; \text{with} \quad\;
 a = 1, \ldots, p \, , \quad
 i = p+1, \ldots, d \, .
\end{align}
For points in $\AdS_{p+1}$ we use notation $\wh z^{\,\a} = (\wh z^{\,0}, \wh z^{\,a})$, and points at the boundary are denoted $x^a \in \mathbb R^p$.
In Poincare coordinates, the $(d+1)$-- and $(p+1)$--dimensional metrics read respectively
\begin{align}
 g_{\mu\nu} = \frac{\delta_{\mu \nu}}{(z^0)^2} \, , \qquad
 \wh g_{\a\b} = \frac{\delta_{\a\b}}{(\wh z^0)^2} \, .
 \label{eq:metric}
\end{align}
These metrics define covariant derivatives and an integration measure in the standard way.

\subsection{Effective action and Witten diagrams}

The focus of this paper are correlation functions of local operators in the setup just described, which are computed perturbatively with Witten diagrams.
Here we present a simple example of a bulk scalar $\phi$ and a brane scalar $\wh \phi$, which illustrates the main ingredients necessary in the rest of this work.

Let's start with the bulk scalar $\phi$, which has mass $m^2 = \Delta(\Delta-d)$ and a cubic self-coupling
\begin{align}
 S_{\text{bulk}}
 =
 \int_{\AdS_{d+1}} \! \frac{d^{d+1} z}{(z_0)^{d+1}} \left(
  \frac{1}{2} \nabla^\mu \phi \nabla_\mu \phi
  + \frac{1}{2} m^2 \phi^2
  + \frac{1}{3!} g \phi^3
 \right) \, .
 \label{eq:ex-bulk}
\end{align}
Besides the bulk field, we also consider a brane scalar $\wh \phi$ of mass $\wh m^2 = \Dh(\Dh-p)$.
The bulk field can couple directly to the brane or it can couple to $\wh \phi$.
In this simplified example, we take the brane action to be
\begin{align}
 S_{\text{brane}}
 =
 \int_{\AdS_{p+1}} \! \frac{d^{p+1}\wh z}{(\wh z_0)^{p+1}} \left(
    \frac{1}{2} \nabla^\a \wh \phi \, \nabla_\a \wh \phi
  + \frac{1}{2} \wh m^2 \wh \phi^{\,2}
  + \lambda \phi
  + \theta \phi^2
  + \mu \phi \wh \phi
 \right) \, .
 \label{eq:ex-brane}
\end{align}
Although a toy model, this action resembles realistic examples such as the half-BPS Wilson line of section \ref{sec:MWL}.
The major difference is that the action for the half-BPS Wilson line contains many more terms, some of which are bulk spin-two fields or brane fields charged under $SO(d-p)$, see \eqref{eq:symbreak}.
Here we ignore these complications.

In what follows, we treat the couplings $g$, $\lambda$, $\theta$ and $\mu$ as small perturbations.
We can then compute correlation functions of the fields $\phi$ and $\wh \phi$ using position space Feynman rules.
In AdS/CFT, position-space diagrams are conventionally called Witten diagrams, and are built with free propagators in the standard way.
For example, the propagator for $\phi$ is
\begin{align}
 \left( -\nabla^2 + m^2 \right) G_{\Delta}(z_1,z_2)
 = (z_0)^{d+1} \, \delta^{(d+1)}(z_1-z_2)  \, .
 \label{eq:eom}
\end{align}
The solution $G_{\Delta}$ is called bulk-to-bulk propagator, and it can be obtained in closed form \cite{DHoker:1999mqo,Penedones:2016voo}
\begin{align}
  G_{\Delta}(z_1,z_2)
& = \frac{\Cm_\Delta \left(z^0_1 z^0_2\right)^{\Delta}}
         {(z_1 - z_2)^{2\Delta}} \,
    {}_2F_1 \left(
    \Delta, \Delta-\frac d2+\frac12, 2\Delta - d + 1,
    -\frac{4 z^0_1 z^0_2}{(z_1 - z_2)^2}
    \right) \, , \label{eq:blk-blk-G} \\
  \Cm_\Delta
& = \frac{\Gamma(\Delta)}{2 \pi^{d/2} \Gamma(\Delta-d/2+1)} \, .
\label{eq:norm-G}
\end{align}
Similarly, the propagator for $\wh \phi$ is called brane-to-brane propagator.
It is defined as in \eqref{eq:blk-blk-G} with $d \to p$, and we denote it $\wh G_{\wh\Delta}(\wh z_1, \wh z_2)$.

In the computation of Witten diagrams, one often encounters fields that propagate from the boundary of $\AdS$ into the bulk.
One then uses the bulk-to-boundary propagator, which follows from \eqref{eq:blk-blk-G} sending one point to the boundary $z \to \partial $AdS$_{d+1}$ and rescaling the result
\begin{align}
 K_{\Delta}(x,z)
 = \left( \frac{z_0}{(z-x)^2} \right)^{\Delta} \, .
 \label{eq:K-blk-bdy}
\end{align}
In the literature, the bulk-to-boundary propagator sometimes includes a normalization factor, but we find more convenient to use it unnormalized.
There is also a brane-to-boundary propagator, but it is given exactly by the same expression \eqref{eq:K-blk-bdy}.

It is well known that processes that start with $\phi$ at point $x$ in the boundary of $\AdS_{d+1}$ can be interpreted as correlation functions of a dual CFT operator $\Om(x)$.
When the mass of the scalar is $m^2 = \Delta(\Delta-d)$, then the CFT operator has scaling dimension $\Delta$.
In the present context, we compute processes with a brane in the bulk, which in CFT corresponds to inserting a non-local defect operator $\Dm$ in the expectation value.
In the example of a fundamental string in $\AdS_5\times S^5$, the defect operator is the half-BPS Wilson line operator $\Dm = W \sim \tr \mathrm{P} e^{\int \! A_\tau + i \phi_6}$.
More generally, depending on the type of brane in the bulk, $\Dm$ is instead a 't Hooft line operator, surface operator, etc.

The simplest process that involves both local and non-local operators is the one-point function $\vev{\Om\Dm}$.
The first few Witten diagrams are
\begin{align}
 \langle \Om \Dm \rangle
 \;\;=\;\;
 \begin{tikzpicture}[valign,scale=0.8]
    \tikzstyle{every node}=[font=\scriptsize]
    \pgfmathsetmacro{\x}{sqrt(2)/2}
    \pgfmathsetmacro{\y}{sqrt(2)/2}
    \pgfmathsetmacro{\z}{0.5}
    \pgfmathsetmacro{\r}{-0.75}
    \draw [thick] (0,0) circle [radius=1];
    \draw [dashed] (-1,0) -- (\z,0);
    \node at (-0.2, 0.3) {$\phi$};
    \draw [thick, blue] (+\x,+\y) to[out=240,in=120] (+\x,-\y);
  \end{tikzpicture}
  \;\;+\;\;
  \begin{tikzpicture}[valign,scale=0.8]
    \tikzstyle{every node}=[font=\scriptsize]
    \pgfmathsetmacro{\x}{sqrt(2)/2}
    \pgfmathsetmacro{\y}{sqrt(2)/2}
    \pgfmathsetmacro{\z}{0.5}
    \pgfmathsetmacro{\r}{-0.75}
    \draw [thick] (0,0) circle [radius=1];
    \draw [dashed] (-1,0) -- (-0.1,0);
    \draw [dashed] (-0.1,0) -- (0.57, 0.4);
    \draw [dashed] (-0.1,0) -- (0.57,-0.4);
    \node at (-0.5,  0.3) {$\phi$};
    \node at ( 0.2,  0.5) {$\phi$};
    \node at ( 0.2, -0.55) {$\phi$};
    \draw [thick, blue] (+\x,+\y) to[out=240,in=120] (+\x,-\y);
  \end{tikzpicture}
  \;\;+\;\;
  \begin{tikzpicture}[valign,scale=0.8]
    \tikzstyle{every node}=[font=\scriptsize]
    \pgfmathsetmacro{\x}{sqrt(2)/2}
    \pgfmathsetmacro{\y}{sqrt(2)/2}
    \pgfmathsetmacro{\z}{0.5}
    \pgfmathsetmacro{\r}{-0.75}
    \draw [thick] (0,0) circle [radius=1];
    \draw [dashed] (-1,0) -- (-0.3,0);
    \draw [dashed] (-0.3,0) to[out=60,in=120] (0.5, 0);
    \draw [dashed] (-0.3,0) to[out=-60,in=-120] (0.5, 0);
    \node at (-0.6,  0.3) {$\phi$};
    \node at ( 0.1,  0.5) {$\phi$};
    \node at ( 0.1, -0.55) {$\phi$};
    \draw [thick, blue] (+\x,+\y) to[out=240,in=120] (+\x,-\y);
  \end{tikzpicture}
  \;\;+\;\;
  \begin{tikzpicture}[valign,scale=0.8]
    \tikzstyle{every node}=[font=\scriptsize]
    \pgfmathsetmacro{\x}{sqrt(2)/2}
    \pgfmathsetmacro{\y}{sqrt(2)/2}
    \pgfmathsetmacro{\z}{0.5}
    \pgfmathsetmacro{\r}{-0.75}
    \draw [thick] (0,0) circle [radius=1];
    \draw [dashed] (-1,0) -- (-0.1,0);
    \draw [dashed] (-0.1,0) -- (0.57, 0.4);
    \draw [dashed] (-0.1,0) -- (0.57,-0.4);
    \node at (-0.5,  0.3) {$\phi$};
    \node at ( 0.2,  0.5) {$\phi$};
    \node at ( 0.2, -0.55) {$\phi$};
    \node at ( 0.82,  0) {$\wh\phi$};
    \draw [thick, blue] (+\x,+\y) to[out=240,in=120] (+\x,-\y);
    \draw [dashed, blue] (0.57,0.4) to[out=-70,in=70] (0.57,-0.4);
  \end{tikzpicture}
  \;\;+\;\;
  \ldots \, ,
  \label{eq:example-one-pt}
\end{align}
where the vertices follow from the action \eqref{eq:ex-bulk}-\eqref{eq:ex-brane}.
Each diagram corresponds to a position-space integral with propagators $G_{\Delta}$, $\wh G_{\Dh}$ and $K_\Delta$, which in general can be highly non-trivial to compute.

In this work, we focus mostly on two-point functions of local operators in the presence of the brane, and the first few diagrams read
\begin{align}
 \langle \Om \Om \Dm \rangle
 \;\;=\;\;
 \begin{tikzpicture}[valign,scale=0.8]
    \tikzstyle{every node}=[font=\scriptsize]
    \pgfmathsetmacro{\x}{sqrt(1)/2}
    \pgfmathsetmacro{\y}{sqrt(3)/2}
    \pgfmathsetmacro{\xx}{sqrt(2)/2}
    \pgfmathsetmacro{\yy}{sqrt(2)/2}
    \pgfmathsetmacro{\z}{0.5}
    \pgfmathsetmacro{\rx}{0.32}
    \pgfmathsetmacro{\ry}{0.45}
    \draw [thick] (0,0) circle [radius=1];
    \draw [dashed]      (-\xx,+\yy) to[out=-60,in=60] (-\xx,-\yy);
    \draw [thick, blue] (\x,+\y) to[out=240,in=120] (\x,-\y);
    \node at (-0.2, 0.0) { $\phi$};
 \end{tikzpicture}
 \;\;+\;\;
 \begin{tikzpicture}[valign,scale=0.8]
    \tikzstyle{every node}=[font=\scriptsize]
    \pgfmathsetmacro{\x}{sqrt(1)/2}
    \pgfmathsetmacro{\y}{sqrt(3)/2}
    \pgfmathsetmacro{\xx}{sqrt(2)/2}
    \pgfmathsetmacro{\yy}{sqrt(2)/2}
    \pgfmathsetmacro{\z}{0.5}
    \pgfmathsetmacro{\rx}{0.32}
    \pgfmathsetmacro{\ry}{0.45}
    \draw [thick] (0,0) circle [radius=1];
    \draw [dashed]      (-\xx,+\yy) -- (\rx,\ry);
    \draw [dashed]      (-\xx,-\yy) -- (\rx,-\ry);
    \draw [thick, blue] (\x,+\y) to[out=240,in=120] (\x,-\y);
    \node at (-0.2, 0.3) { $\phi$};
    \node at (-0.2,-0.3) { $\phi$};
 \end{tikzpicture}
 \;\;+\;\;
   \begin{tikzpicture}[valign,scale=0.8]
    \tikzstyle{every node}=[font=\scriptsize]
    \pgfmathsetmacro{\x}{-sqrt(2)/2}
    \pgfmathsetmacro{\y}{sqrt(2)/2}
    \pgfmathsetmacro{\z}{-0.5}
    \pgfmathsetmacro{\r}{0.4}
    \draw [thick] (0,0) circle [radius=1];
    \draw [thick, blue] (-\x,+\y) to[out=240,in=120] (-\x,-\y);
    \draw [dashed] (+\x,+\y) -- (-\r,0);
    \draw [dashed] (+\x,-\y) -- (-\r,0);
    \draw [dashed] (-\r,  0) -- (-\z,0);
    \node at (0.1, 0.2) {$\phi$};
    \node at (-0.35,  0.4) {$\phi$};
    \node at (-0.35, -0.4) {$\phi$};
 \end{tikzpicture}
 \;\;+\;\;
 \begin{tikzpicture}[valign,scale=0.8]
    \tikzstyle{every node}=[font=\scriptsize]
    \pgfmathsetmacro{\x}{sqrt(1)/2}
    \pgfmathsetmacro{\y}{sqrt(3)/2}
    \pgfmathsetmacro{\xx}{sqrt(2)/2}
    \pgfmathsetmacro{\yy}{sqrt(2)/2}
    \pgfmathsetmacro{\z}{0.5}
    \pgfmathsetmacro{\rx}{0.32}
    \pgfmathsetmacro{\ry}{0.45}
    \draw [thick] (0,0) circle [radius=1];
    \draw [dashed]      (-\xx,+\yy) -- (\rx,\ry);
    \draw [dashed]      (-\xx,-\yy) -- (\rx,-\ry);
    \draw [dashed,blue] (\rx,\ry) to[out=210,in=150] (\rx,-\ry);
    \draw [thick, blue] (\x,+\y) to[out=240,in=120] (\x,-\y);
    \node at (-0.15,0.1) { $\wh \phi$};
    \node at (-0.5, 0.4) { $\phi$};
    \node at (-0.5,-0.4) { $\phi$};
 \end{tikzpicture}
 \;\;+\;\;
 \begin{tikzpicture}[valign,scale=0.8]
    \tikzstyle{every node}=[font=\scriptsize]
    \pgfmathsetmacro{\x}{sqrt(2)/2}
    \pgfmathsetmacro{\y}{sqrt(2)/2}
    \pgfmathsetmacro{\z}{0.5}
    \pgfmathsetmacro{\r}{-0.75}
    \draw [thick] (0,0) circle [radius=1];
    \draw [thick, blue] (+\x,+\y) to[out=240,in=120] (+\x,-\y);
    \draw [dashed]      (-\x,+\y) -- (\z,0);
    \draw [dashed]      (-\x,-\y) -- (\z,0);
    \node at (-0.0,  0.55) {$\phi$};
    \node at (-0.0, -0.55) {$\phi$};
  \end{tikzpicture}
  \;\;+\;\;
  \ldots
  \label{eq:example-two-pt}
\end{align}
In sections \ref{sec:pos} and \ref{sec:mellin} we compute these diagrams and their generalizations in position and Mellin space respectively.
As we now explain, these diagrams are constrained by the symmetry group \eqref{eq:symbreak} preserved by the setup, which is that of defect CFT.

\subsection{Conformal symmetry and defect CFT}
\label{sec:dcft}

The isometry group of Euclidean $\AdS_{d+1}$ is the $d$-dimensional conformal group $SO(d+1,1)$.
However, this group is broken by the brane to $SO(p+1,1) \oplus SO(d-p)$, recall \eqref{eq:symbreak}.
This is precisely the symmetry group preserved in defect CFT \cite{Billo:2016cpy}, namely $SO(p+1,1)$ is the conformal group along a $p$-dimensional flat defect, and $SO(d-p)$ are rotations transverse to it.
This means that Witten diagrams such as those in \eqref{eq:example-one-pt} and \eqref{eq:example-two-pt} behave exactly as defect CFT correlators.

The simplest defect CFT observable is the one-point function of a scalar operator.
This is determined up to an overall constant
\begin{align}
 \frac{\vev{\Om(x)\Dm}}{\vev{\Dm}}
 = \frac{a_\Om}{|x^i|^\Delta} \, ,
 \label{eq:onept-def}
\end{align}
where we introduce $|x^i|^2=\sum_{i=p+1}^d (x^i)^2$.
In particular, every diagram in \eqref{eq:example-one-pt} has to be of this form.
Similarly, the two-point function of scalar operators reads
\begin{align}
 \frac{\langle \Om_1(x_1) \Om_2(x_2) \Dm \rangle}{\vev{\Dm}}
 = \frac{F(\xi,\eta)}{|x_1^i|^{\Delta_1} |x_2^i|^{\Delta_2}} \, ,
 \label{eq:twopt-def}
\end{align}
where the correlator depends on two conformal cross ratios:
\begin{align}
 \xi
 = \frac{(x^a_{12})^2 + (x^i_{12})^2}{|x^i_1| |x^i_2|} \, , \qquad
 \eta
 = \frac{x^j_1 x^j_2}{|x^i_1| |x^i_2|} \,.
 \label{eq:cross-ratios-xieta}
\end{align}
Recall that the index $a=1,\ldots,p$ runs parallel to the defect.
There are many conventions for cross-ratios in the literature, which are useful for different purposes.
In this work, we sometimes use cross-ratios $(r,w)$ or $(z,\zb)$ defined as \cite{Lauria:2017wav,Lemos:2017vnx}
\begin{align}
 r + \frac{1}{r}
 = \xi + 2\eta \, , \qquad
 w + \frac{1}{w}
 = 2 \eta \, , \qquad
 z = r w \, , \qquad
 \zb = \frac{r}{w} \, .
 \label{eq:cross-ratios-rw}
\end{align}
Note that we normalize all correlators by the expectation value of the defect $\vev{\Dm}$.
This ensures that in the limit $x_1 \to x_2$, the two-point function \eqref{eq:twopt-def} reduces to a unit-normalized two-point function without a defect.
Equivalently, the correlator of equal operators is normalized as $F(\xi,\eta) \sim \xi^{-\frac{\Delta_1+\Delta_2}{2}}$ in the limit $\xi \to 0$.

An important property of the correlator $F(\xi,\eta)$ is that it admits two different OPE decompositions \cite{Billo:2016cpy}.
Although these OPEs only play a tangential role in the present work, it is still useful to discuss them, because sometimes they help interpret certain results.
The first OPE is the standard one, which replaces the product of two bulk local operators by a sum of local operators, schematically $\Om_1 \Om_2 \sim \sum_\Om \lambda_{\Om_1\Om_2\Om} \Om$.
The second OPE replaces the product of a bulk local operator with the defect $\Dm$ by a sum of local operators on top of the defect, schematically $\Om \Dm \sim \sum_{\wh\Om} b_{\Om\wh\Om} \wh\Om \Dm$.
These two OPEs can be resumed into two different conformal block expansions
\begin{align}
 F(\xi,\eta)
 = \xi^{-\frac{\Delta_1+\Delta_2}{2}}
   \sum_{\Om} 2^{-\ell} \lambda_{\Om_1\Om_2\Om} a_\Om
   f_{\Delta,\ell}(\xi,\eta)
 = \sum_{\wh\Om} 2^{-s} b_{\Om_1\wh\Om} b_{\Om_2\wh\Om}
   \wh f_{\Dh,s}(\xi,\eta) \, ,
\end{align}
where $a_\Om$ are the one-point coefficients \eqref{eq:onept-def}.
See appendix A of \cite{Gimenez-Grau:2022ebb} for a pedagogical introduction, including our normalization conventions and explicit formulas for conformal blocks $f_{\Delta,\ell}$, $\wh f_{\Dh,s}$.

Another important concept is the connected correlator $\vev{ \ldots \Dm}_c$, which for the two-point function is defined by
\begin{align}
 \frac{\langle \Om_1(x_1) \Om_2(x_2) \Dm \rangle}{\vev{\Dm}}
 = \vev{\Om_1(x_1) \Om_2(x_2)}
 + \frac{\vev{\Om_1(x_1) \Dm}}{\vev{\Dm}}
   \frac{\vev{\Om_2(x_2) \Dm}}{\vev{\Dm}}
 + \frac{\langle \Om_1(x_1) \Om_2(x_2) \Dm \rangle_c}{\vev{\Dm}} \, .
 \label{eq:conn-two}
\end{align}
Equivalently, if we use the one-point function \eqref{eq:onept-def} and factor out the prefactor in \eqref{eq:twopt-def}, the connected correlator $F_c$ is
\begin{align}
 F(\xi,\eta)
 = \delta_{\Om_1,\Om_2} \xi^{-\frac{\Delta_1+\Delta_2}{2}}
 + a_{\Om_1} a_{\Om_2}
 + F_c(\xi,\eta) \, .
 \label{eq:conn-F}
\end{align}
The main reason to decompose the correlator in this way is that the first two terms appear in mean-field theory (MFT), while $F_c$ contains the truly non-trivial part of the correlator.
A second reason is that Mellin amplitudes, to be defined in section \ref{sec:mellin}, make sense only for connected correlators.
For example, the connected part of equation \eqref{eq:example-two-pt} is
\begin{align}
 \langle \Om \Om \Dm \rangle_c
 \;\;=\;\;
 \begin{tikzpicture}[valign,scale=0.8]
    \tikzstyle{every node}=[font=\scriptsize]
    \pgfmathsetmacro{\x}{-sqrt(2)/2}
    \pgfmathsetmacro{\y}{sqrt(2)/2}
    \pgfmathsetmacro{\z}{-0.5}
    \pgfmathsetmacro{\r}{0.4}
    \draw [thick] (0,0) circle [radius=1];
    \draw [thick, blue] (-\x,+\y) to[out=240,in=120] (-\x,-\y);
    \draw [dashed] (+\x,+\y) -- (-\r,0);
    \draw [dashed] (+\x,-\y) -- (-\r,0);
    \draw [dashed] (-\r,  0) -- (-\z,0);
    \node at (0.1, 0.2) {$\phi$};
    \node at (-0.35,  0.4) {$\phi$};
    \node at (-0.35, -0.4) {$\phi$};
 \end{tikzpicture}
 \;\;+\;\;
 \begin{tikzpicture}[valign,scale=0.8]
    \tikzstyle{every node}=[font=\scriptsize]
    \pgfmathsetmacro{\x}{sqrt(1)/2}
    \pgfmathsetmacro{\y}{sqrt(3)/2}
    \pgfmathsetmacro{\xx}{sqrt(2)/2}
    \pgfmathsetmacro{\yy}{sqrt(2)/2}
    \pgfmathsetmacro{\z}{0.5}
    \pgfmathsetmacro{\rx}{0.32}
    \pgfmathsetmacro{\ry}{0.45}
    \draw [thick] (0,0) circle [radius=1];
    \draw [dashed]      (-\xx,+\yy) -- (\rx,\ry);
    \draw [dashed]      (-\xx,-\yy) -- (\rx,-\ry);
    \draw [dashed,blue] (\rx,\ry) to[out=210,in=150] (\rx,-\ry);
    \draw [thick, blue] (\x,+\y) to[out=240,in=120] (\x,-\y);
    \node at (-0.15,0.1) { $\wh \phi$};
    \node at (-0.5, 0.4) { $\phi$};
    \node at (-0.5,-0.4) { $\phi$};
 \end{tikzpicture}
 \;\;+\;\;
 \begin{tikzpicture}[valign,scale=0.8]
    \tikzstyle{every node}=[font=\scriptsize]
    \pgfmathsetmacro{\x}{sqrt(2)/2}
    \pgfmathsetmacro{\y}{sqrt(2)/2}
    \pgfmathsetmacro{\z}{0.5}
    \pgfmathsetmacro{\r}{-0.75}
    \draw [thick] (0,0) circle [radius=1];
    \draw [thick, blue] (+\x,+\y) to[out=240,in=120] (+\x,-\y);
    \draw [dashed]      (-\x,+\y) -- (\z,0);
    \draw [dashed]      (-\x,-\y) -- (\z,0);
    \node at (-0.0,  0.55) {$\phi$};
    \node at (-0.0, -0.55) {$\phi$};
  \end{tikzpicture}
  \;\;+\;\;
  \ldots
  \label{eq:example-two-pt-conn}
\end{align}
As its name suggests, the connected correlator is the sum of connected Witten diagrams.
In most of this work, we restrict our attention only to the connected correlator.

\subsection{Bootstrap approach}

To summarize the discussion so far, we have considered a theory in $\AdS_{d+1}$ described by an effective action, that in our example is given in \eqref{eq:ex-bulk}.
Furthermore, we added a brane that extends in an $\AdS_{p+1}$ submanifold.
The brane has degrees of freedom of its own, that interact with bulk fields through another effective action, in our example \eqref{eq:ex-brane}.
With these two actions we can compute Witten diagrams, which define correlation functions in a defect CFT that lives at the boundary of $\AdS_{d+1}$.

In examples arising from string theory, the bulk and brane effective actions are significantly more involved, see for example \cite{Kim:1985ez,Lee:1998bxa,Arutyunov:1998hf,Arutyunov:1999en,Arutyunov:1999fb}.
For this reason, it is extremely valuable to have a method that bypasses detailed knowledge of the effective action.
The idea of \cite{Rastelli:2016nze,Rastelli:2017udc} is that one only needs to know what fields appear in the effective action, but not their precise interactions.
In this case, one can write all Witten diagrams that contribute to a certain correlator, with unspecified coefficients.
The combination of Witten diagrams needs to satisfy certain consistency conditions, such as superconformal Ward identities, or agreement with the flat-space limit or localization.
In many examples, the free coefficients in the ansatz are fully fixed by these requirements.
In this work, we apply the same idea in the presence of a brane in the bulk.
We show in section \ref{sec:MWL} that the method works perfectly for a half-BPS Wilson line in $\Nm=4$ SYM.
More generally, it is plausible that this method applies to many (if not all) half-BPS defects in maximally supersymmetric theories.
We discuss some of these possible setups in the conclusions.

\section{Witten diagrams in position space}
\label{sec:pos}

In this section, we compute Witten diagrams for a two-point functions of local operators in the presence of a brane.
The interested reader can find related work in \cite{RastZhouMell,Goncalves:2018fwx}.
We restrict our attention to diagrams that appear at leading order in the supergravity approximation.
In the future, it would of course be very interesting to extend the analysis to loop diagrams.

\subsection{Simple diagrams}

As a first example, we compute tree-level diagrams for the bulk one-point function $\vev{ \Om \Dm}$ and the bulk-defect two-point function $\vev{ \Om \wh\Om \Dm}$.
Besides the pedagogical value, these results are needed to calculate Mellin amplitudes in section \ref{sec:mellin}.

\subsubsection{Bulk one-point diagram}

We consider a scalar $\phi$ dual to an operator $\Om$ of dimension $\Delta$.
If the scalar couples to a brane as $\int_{\AdS_{p+1}} \phi$, then the leading diagram to the one-point function $\vev{\Om\Dm}$ is
\begin{align}
  I_\Delta(x)
  \;\; = \;\;
  \begin{tikzpicture}[valign]
    \tikzstyle{every node}=[font=\small]
    \pgfmathsetmacro{\x}{sqrt(2)/2}
    \pgfmathsetmacro{\y}{sqrt(2)/2}
    \pgfmathsetmacro{\z}{0.5}
    \pgfmathsetmacro{\r}{-0.75}
    \draw [thick] (0,0) circle [radius=1];
    \draw [thick, blue] (+\x,+\y) to[out=240,in=120] (+\x,-\y);
    \draw [dashed] (-1,0) -- (\z,0);
    \node at (-1.25, 0) {$x$};
    \node at (-\r,0.05) {$\wh z$};
    \node at (-0.2, 0.2) {$\Delta$};
  \end{tikzpicture}
 \;\; = \;\;
 \int_{\wh z} K_{\Delta}(x, \wh z) \, .
 \label{eq:one-pt-diag}
\end{align}
The integral is over $\wh z \in \AdS_{p+1}$ with appropriate measure, and the bulk-boundary propagator is defined in \eqref{eq:K-blk-bdy}, so we have to compute
\begin{align}
 I_\Delta(x)
 =
 \int_0^\infty \frac{d \, \wh z_0}{\wh z_0^{\,p+1}}
 \int_{\mathbb R^p} d^p \wh z^{\,a}
 \left( \frac{\wh z_0}{\wh z_0^{\,2} + (x^i)^2 + (x^a - \wh z^{\,a})^2} \right)^\Delta \, .
\end{align}
The dependence on $x^a$ drops out by translation invariance, so we can use spherical coordinates in the $\wh z^{\,a}$ directions.
Introducing a Schwinger parameter $s$, we find
\begin{align}
 I_\Delta(x)
 = \frac{1}{\Gamma(\Delta)}
   \int_0^\infty \frac{ds}{s} \, s^\Delta e^{-s (x^i)^2}
   \int_0^\infty \frac{d \, \wh z_0}{\wh z_0} \,
   \wh z_0^{\,\Delta-p} e^{-s \wh z_0^2} \,
   \frac{2 \pi^{\frac p2}}{\Gamma(p/2)}
   \int_0^\infty \frac{dr}{r} \, r^p e^{-s r^2} \, .
\end{align}
The integrals over $\wh z_0$ and $r = |\wh z^{\,a}|$ are straightforward, and converge provided $0<p< \re \Delta$.
The final integral over $s$ is elementary, and we find \cite{RastZhouMell,Goncalves:2018fwx}\footnote{There is a minor typo in equation (4.2) of \cite{RastZhouMell}, where $d-2$ should read $d-1$.}
\begin{align}
 I_\Delta(x)
 = \frac{\pi^{\frac p2}
   \Gamma\!\left(\frac{\Delta-p}{2}\right)}{2\Gamma(\Delta)}
   \int_0^\infty \frac{ds}{s} \, s^{\frac\Delta 2} e^{-s (x^i)^2}
 = \frac{a_\Delta}{|x^i|^\Delta} \, , \quad
 a_\Delta
 \equiv \pi ^{p/2} \,
        \frac{\Gamma \! \left(\frac{\Delta }{2}\right)
              \Gamma \! \left(\frac{\Delta -p}{2}\right)}
             {2 \Gamma (\Delta )} \, .
  \label{eq:onept}
\end{align}
This has the correct form dictated by conformal symmetry \cite{Billo:2016cpy}.

\subsubsection{Bulk-defect diagram}
\label{sec:bulk-def}

Now let's imagine a brane scalar $\wh \phi$, dual to a defect operator $\wh \Om$ of dimension $\Dh$.
The scalar $\phi$ can couple to this brane field as $\int_{\AdS_{p+1}} \phi \, \wh \phi$.
In this case, the tree-level correlator $\vev{\Om\wh\Om\Dm}$ consists of the diagram
\begin{align}
  I_{\Delta,\Dh}(x_1,x_2)
  \; =
  \begin{tikzpicture}[valign]
    \tikzstyle{every node}=[font=\small]
    \pgfmathsetmacro{\x}{sqrt(2)/2}
    \pgfmathsetmacro{\y}{sqrt(2)/2}
    \pgfmathsetmacro{\z}{0.5}
    \pgfmathsetmacro{\r}{-0.75}
    \draw [thick] (0,0) circle [radius=1];
    \draw [thick, blue] (+\x,+\y) to[out=240,in=120] (+\x,-\y);
    \draw [dashed, blue] (+\x,+\y) to[out=190,in=140] (+\z,  0);
    \draw [dashed] (-1,0) -- (\z,0);
    \node at (-1.25, 0) {$x_1$};
    \node at (0.95, 0.85) {$x_2$};
    \node at (-\r,0.05) {$\wh z$};
    \node at (-0.3, 0.2) {$\Delta$};
    \node at (-0.2, 0.7)[anchor=north,rotate=70] {$\Dh$};
  \end{tikzpicture}
 = \;
 \int_{\wh z} K_{\Delta}(x_1, \wh z) K_{\Dh}(x_2, \wh z) \, .
 \label{eq:blk-def-diag}
\end{align}
Note that point $x_2$ lives in the boundary of $\AdS_{p+1}$, meaning that $x_2^i = 0$.
As before, we introduce Schwinger parameters and integrate over $\wh z$.
The integrals are elementary and converge provided $0<p<\re\Delta+\re\Dh$:
\begin{align}
 I_{\Delta,\Dh}(x_1,x_2)
 =
 \frac{\pi^{p/2} \Gamma \! \left(\frac{\Delta+\Dh-p}{2}\right) }
      {2 \Gamma (\Delta ) \Gamma (\Dh)}
 \int_0^\infty \frac{ds \, dt}{s \, t}
 \frac{s^{\Delta} t^{\Dh}}{(s+t)^{\frac{\Delta +\Dh}{2}}}
 \exp\left(- \frac{s t}{s+t} (x_{12}^a)^2 - s (x_1^i)^2 \right) \, .
\end{align}
To integrate over $s,t$ we employ a method that we also use repeatedly below.
First multiply by the identity $1 = \int_0^\infty d\lambda \, \delta(\lambda - t)$ and change variables to $s,t \to \lambda s, \lambda t$.
Now use $\delta(\lambda(1-t)) = \lambda^{-1} \delta(1-t)$, so both the $t$ and $\lambda$ integrals become elementary.
Integrating $t$ and $\lambda$ gives
\begin{align}
 I_{\Delta,\Dh}(x_1,x_2)
 =
 \frac{\pi ^{p/2} \Gamma \! \left(\frac{\Delta +\Dh}{2}\right) \Gamma \! \left(\frac{\Delta +\Dh-p}{2}\right)}{2 \Gamma (\Delta ) \Gamma (\Dh)}
 \int_0^\infty \frac{ds}{s} \,
 \frac{s^{\frac{\Delta -\Dh}{2}}}
      {\big((x_{12}^a)^2 + (s+1) (x_1^i)^2 \big)^{(\Delta+\Dh)/2}} \, .
 \label{eq:bd-interm}
\end{align}
The integral over $s$ converges for $\re \Delta > \re \Dh$, giving \cite{RastZhouMell,Goncalves:2018fwx}
\begin{align}
 I_{\Delta,\Dh}(x_1,x_2)
 = \frac{b_{\Delta,\Dh}}{|x_1^i|^{\Delta-\Dh} \big( (x_1^i)^2 + (x_{12}^a)^2 \big)^{\Dh} } \, , \qquad
 b_{\Delta,\Dh}
 = \pi^{p/2} \, \frac{\Gamma\!\left(\frac{\Delta -\Dh}{2}\right) \Gamma\! \left(\frac{\Delta+\Dh-p}{2}\right)}{2 \Gamma (\Delta )} \, .
 \label{eq:bulkbrane}
\end{align}
Once again, the result has the form dictated by conformal symmetry \cite{Billo:2016cpy}.

We can also have brane fields charged under $SO(d-p)$ rotations.
For example, a spin-$s$ field $\wh \phi^{i_1\ldots i_s}$ couples to a bulk scalar as
\begin{align}
 S_{\phi,\wh\phi}
 \, \sim \,
 \int 
 \frac{d^{p+1}\wh z}{\wh z_0^{p+1}} \, \wh z_0^{\,s} \,
 \wh \phi^{\,i_1\ldots i_s} \partial_{i_1} \ldots \partial_{i_s} \phi \, ,
 \label{eq:action-tspin}
\end{align}
where as before $i = p+1, \ldots d$ and the factor $\wh z_0^{\,s}$ is required by dimensional analysis.
As a result, the correlation function $\vev{\Om \wh\Om^{i_1\ldots i_s} \Dm}$ is
\begin{align}
 I^{i_1\ldots i_s}_{\Delta,\Dh}(x_1,x_2)
 = \;
 \int_{\wh z} \wh z_0^{\,s} \, \partial^{\{i_1} \ldots \partial^{i_s\}} K_{\Delta}(x_1, \wh z) K_{\Dh}(x_2, \wh z)
 \propto
 x_1^{\{i_1} \ldots x_1^{i_s\}}
 I_{\Delta+s,\Dh}(x_1,x_2) \, .
\end{align}
The bracket $\{i \ldots j\}$ denotes the symmetric traceless combination.
In the second step we take derivatives of the bulk-to-boundary propagator \eqref{eq:K-blk-bdy}, and observe that the resulting integral is proportional to \eqref{eq:blk-def-diag} with $\Delta \to \Delta+s$.
The upshot is that correlators with defect operators charged under $SO(d-p)$ are simple shifts of the scalar results.
When we consider two-point functions below we will observe the same phenomena.

\subsection{Contact diagram}

Next we consider a coupling between two bulk scalars on the brane $\int_{\AdS_{p+1}} \phi_1 \phi_2$, which contributes to the two-point function $\vev{\Om_1 \Om_2 \Dm}$ of operators dual to $\phi_1$ and $\phi_2$.
We call the resulting Witten diagram a contact diagram:
\begin{align}
  \begin{tikzpicture}[valign]
    \tikzstyle{every node}=[font=\small]
    \pgfmathsetmacro{\x}{sqrt(2)/2}
    \pgfmathsetmacro{\y}{sqrt(2)/2}
    \pgfmathsetmacro{\z}{0.5}
    \pgfmathsetmacro{\r}{-0.75}
    \draw [thick] (0,0) circle [radius=1];
    \draw [thick, blue] (+\x,+\y) to[out=240,in=120] (+\x,-\y);
    \draw [dashed]      (-\x,+\y) -- (\z,0);
    \draw [dashed]      (-\x,-\y) -- (\z,0);
    \node at (-0.95, 0.85) {$x_1$};
    \node at (-0.95,-0.95) {$x_2$};
    \node at (-\r,0.05) {$\wh z$};
    \node at ( 0.3, 0.8) [anchor=north,rotate=-30] {$\Delta_1$};
    \node at (-0.4, 0.15)[anchor=north,rotate= 30] {$\Delta_2$};
  \end{tikzpicture}
 \;\; = \;\;
 \int_{\wh z} 
 K_{\Delta_1}(x_1, \wh z) K_{\Delta_2}(x_2, \wh z)
 \; = \;
 \frac{C_{\Delta_1\Delta_2}(\xi,\eta)}
      {|x^i_1|^{\Delta_1} |x_2^i|^{\Delta_2}} \, .
  \label{eq:def-cont-diag}
\end{align}
As explained in equation \eqref{eq:twopt-def}, the result depends on two conformal cross-ratios $\xi$ and $\eta$.
The calculation proceeds as in section \ref{sec:bulk-def}, but now the result also depends on the parallel distance $x_{12}^a$, and the $s$ integral is not elementary:
\begin{align}
  C_{\Delta_1\Delta_2}(\xi,\eta)
& =
 \frac{\pi^{\frac p2}\Gamma\!\left(\frac{\Delta_1+\Delta_2}{2}\right)
       \Gamma \! \left(\frac{\Delta_1+\Delta_2-p}{2}\right)}
      {2 \Gamma (\Delta_1) \Gamma (\Delta_2)} \notag \\
& \qquad \times
  \int_0^\infty \frac{ds}{s} \,
  \frac{ s^{\Delta_1} |x_1^i|^{\Delta_1} |x_2^i|^{\Delta_2}}
       {\left((\xi + 2\eta - 2) s |x_1^i| |x_2^i| +\left(s |x_1^i|+|x_2^i|\right)^2\right)
      \!{}^{\frac{\Delta_1+\Delta_2}{2}}} \, .
 \label{eq:cont-int}
\end{align}
To simplify the denominator, we used the definition of the cross-ratios $\xi$ and $\eta$ in \eqref{eq:cross-ratios-xieta}.
We shall encounter many similar integrals below.
A convenient way to proceed is to split the denominator with
\begin{align}
 \frac{1}{(A+B)^\Delta}
 = \frac{1}{\Gamma(\Delta)} \int_{-i\infty}^{i\infty}
   \frac{d\tau}{2\pi i}
   \frac{\Gamma(\tau) \Gamma(\Delta-\tau)}{A^\tau B^{\Delta-\tau}} \, ,
 \qquad
 0 < \re \tau < \Delta \, .
 \label{eq:mell-barn-sum}
\end{align}
This identity makes the $s$ integral elementary, giving
\begin{align}
 C_{\Delta_1\Delta_2}(\xi,\eta)
&= \frac{\pi^{\frac{p+1}{2}}
         \Gamma \! \left(\frac{\Delta_1+\Delta_2-p}{2}\right)}
        {2^{\Delta_1+\Delta_2} \Gamma (\Delta_1) \Gamma (\Delta_2)}
 \int \frac{d\tau}{2\pi i}
 \frac{\Gamma (\tau ) \Gamma (\Delta_1-\tau ) \Gamma (\Delta_2-\tau )}
      {\Gamma\!\left(\frac{\Delta_1+\Delta_2+1}{2}-\tau\right)}
 \left(\frac{\xi + 2\eta - 2}{4} \right)^{-\tau} \, .
\end{align}
The remaining $\tau$ integral is the Mellin-Barnes representation of the hypergeometric function, and we find
\begin{align}
 C_{\Delta_1\Delta_2}(\xi,\eta)
&= \frac{\pi ^{\frac{p+1}{2}}}{2^{\Delta_1+\Delta_2}}
   \frac{\Gamma \! \left(\frac{\Delta_1+\Delta_2-p}{2} \right)}
        {\Gamma \! \left(\frac{\Delta_1+\Delta_2+1}{2} \right)} \,
   {}_2F_1 \left( \Delta_1,\Delta_2,
            \frac{\Delta_1+\Delta_2+1}{2};-\frac{\xi + 2\eta - 2}{4} \right) \, .
 \label{eq:cont-form}
\end{align}
An analogous formula was found for boundary CFT in \cite{RastZhouMell}, and we just showed that the result continues to hold with minor modifications for arbitrary codimension.
It is important to note that the contact diagram only depends on $\xi + 2\eta$, or equivalently, it depends on the cross-ratio $r$ defined in \eqref{eq:cross-ratios-rw}.
In fact, for integer $\Delta_1, \Delta_2$ the hypergeometric function reduces to a rational function of $r$ and $\log r$, for example
\begin{subequations}
\label{eq:cont-examp}
 \begin{align}
&C_{11}(r)
 \,\propto\, \frac{r \log r}{r^2-1} \, , \qquad
 C_{22}(r)
 \,\propto\, \frac{r^2 \left(r^2+1\right)}{\left(r^2-1\right)^3} \log r
 - \frac{r^2}{\left(r^2-1\right)^2} \, , \\
&C_{21}(r)
 \,\propto\, \frac{r}{(1+r)^2} \, , \qquad
 C_{31}(r)
 \,\propto\, \frac{r^3+r}{2 \left(r^2-1\right)^2}-\frac{2 r^3}{\left(r^2-1\right)^3} \log r \, .
\end{align}
\end{subequations}
One could compute contact diagrams for interactions with derivatives, but we do not need the results below.

As a final comment, contact diagrams $C_{\Delta_1\Delta_2}$ are analogs of $D_{\Delta_1\Delta_2\Delta_3\Delta_4}$ functions for four-point correlators, see their definition in \cite{DHoker:1999kzh,Dolan:2000ut}.
However, the functions $C_{\Delta_1\Delta_2}$ are simpler than $D_{\Delta_1\Delta_2\Delta_3\Delta_4}$, because they depend on a single cross ratio.
Moreover, the $D$-functions for integer $\Delta_i$ contain dilogarithms, as opposed to just logarithms in \eqref{eq:cont-examp}.

\subsection{Bulk-exchange diagram}
\label{sec:blk-exch-pos}

The simplest interaction in the bulk is a three-point vertex $\int_{\AdS_{d+1}} \phi_1 \phi_2 \phi$, which combined with the one-point coupling to the brane $\int_{\AdS_{p+1}} \phi$, leads to the scalar-exchange Witten diagram
\begin{align}
\label{eq:def-blk-exch}
   \begin{tikzpicture}[valign]
    \tikzstyle{every node}=[font=\small]
    \pgfmathsetmacro{\x}{-sqrt(2)/2}
    \pgfmathsetmacro{\y}{sqrt(2)/2}
    \pgfmathsetmacro{\z}{-0.5}
    \pgfmathsetmacro{\r}{0.4}
    \draw [thick] (0,0) circle [radius=1];
    \draw [thick, blue] (-\x,+\y) to[out=240,in=120] (-\x,-\y);
    \draw [dashed]      (+\x,+\y) -- (-\r,0);
    \draw [dashed]      (+\x,-\y) -- (-\r,0);
    \draw [dashed]      (-\r,  0) -- (-\z,0);
    \node at (0.02, 0.2) {{\footnotesize $(\Delta,\!0)$}};
    \node at (-0.95, 0.85) {$x_1$};
    \node at (-0.95,-0.95) {$x_2$};
    \node at (-0.7,0.05) {$z$};
    \node at (0.75,0.0) {$\wh z$};
    \end{tikzpicture}
 \;\; = \;\;
 \int_{z,\wh z}
    K_{\Delta_1}(x_1,z)
    K_{\Delta_2}(x_2,z)
    G_{\Delta}(z,\wh z)
  \; = \;
  \frac{E^{\Delta,0}_{\Delta_1\Delta_2}(\xi,\eta)}
       {|x^i_1|^{\Delta_1} |x_2^i|^{\Delta_2}} \, .
\end{align}
The calculation of this integral is significantly more involved than above, because the bulk-to-bulk propagator $G_\Delta$ is the hypergeometric function \eqref{eq:blk-blk-G}.
Fortunately, reference \cite{DHoker:1999mqo} developed an efficient method to reduce exchange Witten diagram to sums of contact diagrams, leading to a relation of the form
\begin{align}
   \begin{tikzpicture}[valign, scale=0.75]
    \tikzstyle{every node}=[font=\small]
    \pgfmathsetmacro{\x}{-sqrt(2)/2}
    \pgfmathsetmacro{\y}{sqrt(2)/2}
    \pgfmathsetmacro{\z}{-0.5}
    \pgfmathsetmacro{\r}{0.4}
    \draw [thick] (0,0) circle [radius=1];
    \draw [thick, blue] (-\x,+\y) to[out=240,in=120] (-\x,-\y);
    \draw [dashed]      (+\x,+\y) -- (-\r,0);
    \draw [dashed]      (+\x,-\y) -- (-\r,0);
    \draw [dashed]      (-\r,  0) -- (-\z,0);
    \end{tikzpicture}
 \;\; \sim \;\;
 \sum_n \; a_n \;\;
   \begin{tikzpicture}[valign, scale=0.75]
    \tikzstyle{every node}=[font=\small]
    \pgfmathsetmacro{\x}{sqrt(2)/2}
    \pgfmathsetmacro{\y}{sqrt(2)/2}
    \pgfmathsetmacro{\z}{0.5}
    \pgfmathsetmacro{\r}{-0.75}
    \draw [thick] (0,0) circle [radius=1];
    \draw [thick, blue] (+\x,+\y) to[out=240,in=120] (+\x,-\y);
    \draw [dashed]      (-\x,+\y) -- (\z,0);
    \draw [dashed]      (-\x,-\y) -- (\z,0);
  \end{tikzpicture} \, .
  \label{eq:exch-sum-conts}
\end{align}
Although in general the sum contains infinitely many terms, for many cases of interest it truncates.
More specifically, whenever $\Delta_1+\Delta_2-\Delta \in 2\mathbb{Z}_{>0}$ the exchange diagram reads
\begin{align}
 E_{\Delta_1\Delta_2}^{\Delta,0}(\xi,\eta)
 = \sum_{n=1}^{\frac{\Delta_1+\Delta_2-\Delta}{2}}
   \frac{\left(-\frac{\Delta_1+\Delta_2-\Delta-2}{2}\right)_{n-1}
         \left(-\frac{\Delta_1+\Delta_2+\Delta-d-2}{2}\right)_{n-1}}
        {4 (1-\Delta_1)_n (1-\Delta_2)_n}
   \frac{C_{\Delta_1-n,\Delta_2-n}(\xi,\eta)}{\xi^n} \, .
 \label{eq:blkexch-conts}
\end{align}
We do not repeat the derivation of this formula here, because it follows easily from \cite{DHoker:1999mqo}, and it was reviewed for Witten diagrams with branes in \cite{RastZhouMell}.
For general scaling dimensions, we calculate the bulk-exchange diagram in Mellin space in section \ref{sec:mell-blk-ex}.

Besides scalars, there can also be higher-spin fields in the $\AdS_{d+1}$ effective action.
For example, a spin-two field $\phi_{\mu\nu}$ couples to the brane as $\int_{\AdS_{p+1}} g^{ij} \phi_{ij}$, where we contract orthogonal indices $i,j=p+1,\ldots,d$.
As a result, the CFT operator dual to $\phi_{\mu\nu}$ acquires a one-point function.
Furthermore, if the spin-two field couples to the scalars $\sim\int_{\AdS_{d+1}} \phi_{\mu\nu} \nabla^\mu \phi_1 \nabla^\nu \phi_2$, then it can be exchanged in a two-point function
\begin{align}
   \begin{tikzpicture}[valign]
    \tikzstyle{every node}=[font=\small]
    \pgfmathsetmacro{\x}{-sqrt(2)/2}
    \pgfmathsetmacro{\y}{sqrt(2)/2}
    \pgfmathsetmacro{\z}{-0.5}
    \pgfmathsetmacro{\r}{0.4}
    \draw [thick] (0,0) circle [radius=1];
    \draw [thick, blue] (-\x,+\y) to[out=240,in=120] (-\x,-\y);
    \draw [dashed]      (+\x,+\y) -- (-\r,0);
    \draw [dashed]      (+\x,-\y) -- (-\r,0);
    \draw [dashed]      (-\r,  0) -- (-\z,0);
    \node at (0.02, 0.2) {{\footnotesize $(\Delta,\!2)$}};
    \node at (-0.95, 0.85) {$x_1$};
    \node at (-0.95,-0.95) {$x_2$};
    \node at (-0.7,0.05) {$z$};
    \node at (0.75,0.0) {$\wh z$};
    \end{tikzpicture}
 \;\; = \;\;
  \frac{E^{\Delta,2}_{\Delta_1\Delta_2}(\xi,\eta)}
       {|x^i_1|^{\Delta_1} |x_2^i|^{\Delta_2}} \, .
\end{align}
In many situations of interest, the spin-two diagram reduces to a sum of contact diagrams.
This has been discussed for equal external fields in \cite{Arutyunov:2002fh,Rastelli:2017udc}, and in appendix \ref{app:spintwo} we generalize the discussion to unequal fields.\footnote{Spin-two diagrams for unequal fields were also discussed in \cite{Berdichevsky:2007xd}, but because our formulation is more streamlined than the reference, we believe it is valuable to include a detailed discussion in the appendix.}
Furthermore, we also explain how to apply the truncation in the presence of branes, which is the focus of this work.
Because the appendix is somewhat technical, we attach to this publication a \mathematica~notebook that computes spin-two diagrams $E_{\Delta_1\Delta_2}^{\Delta,2}$.

As a final comment, note that the diagram $E_{\Delta_1\Delta_1}^{d,2}$, which corresponds to the exchange of a bulk graviton, contains spurious divergences.
However, we found that the limit $E_{\Delta_1\Delta_1}^{d,2} \equiv \lim_{\veps \to 0} E_{\Delta_1+\veps,\Delta_1+\veps}^{d+2\veps,2}$ leads to sensible results. To motivate this choice of limit, note that it keeps $\Delta_1+\Delta_2-\Delta \in 2 \mathbb Z_{>0}$, which is required for the formulas in appendix \ref{app:spintwo} to make sense.
As illustration of graviton- and massive-exchange diagrams, for $d=4$ and $p=1$ we have
\begin{align}
 \label{eq:conts-ex}
 E^{42}_{22}
 = \frac{4}{3 \xi } \big( \eta  C_{22} - C_{13} \big) \, , \qquad
 E^{52}_{32}
 = \frac{3}{5 \xi } \big( 2\eta  C_{23} - C_{23} - C_{14} \big) \, .
\end{align}

\subsection{Defect-exchange diagram}

\subsubsection{Scalar diagram}

Lastly, we consider a diagram where a brane field is excited by couplings $\int_{\AdS_{p+1}} \phi_i \, \wh \phi$.
More specifically, we introduce the defect exchange diagram
\begin{align}
\begin{tikzpicture}[valign]
    \tikzstyle{every node}=[font=\small]
    \pgfmathsetmacro{\x}{sqrt(1)/2}
    \pgfmathsetmacro{\y}{sqrt(3)/2}
    \pgfmathsetmacro{\xx}{sqrt(2)/2}
    \pgfmathsetmacro{\yy}{sqrt(2)/2}
    \pgfmathsetmacro{\z}{0.5}
    \pgfmathsetmacro{\rx}{0.32}
    \pgfmathsetmacro{\ry}{0.45}
    \draw [thick] (0,0) circle [radius=1];
    \draw [dashed]      (-\xx,+\yy) -- (\rx,\ry);
    \draw [dashed]      (-\xx,-\yy) -- (\rx,-\ry);
    \draw [dashed,blue] (\rx,\ry) to[out=210,in=150] (\rx,-\ry);
    \draw [thick, blue] (\x,+\y) to[out=240,in=120] (\x,-\y);
    \node at (-0.95, 0.85) {$x_1$};
    \node at (-0.95,-0.95) {$x_2$};
    \node at (0.6, 0.3) {$\wh z_1$};
    \node at (0.6,-0.3) {$\wh z_2$};
    \node at (-0.2,0) { $\Dh$};
    \end{tikzpicture}
 \; & = \;
 \int_{\wh z_1, \wh z_2}
 K_{\Delta_1}(x_1, \wh z_1)
 K_{\Delta_2}(x_2, \wh z_2)
 \wh G_{\wh\Delta}(\wh z_1,\wh z_2)
 \; \equiv \;
  \frac{\wh E^{\wh\Delta,0}_{\Delta_1\Delta_2}(\xi,\eta)}
       {|x^i_1|^{\Delta_1} |x_2^i|^{\Delta_2}} \, ,
  \label{eq:def-exch-def}
\end{align}
where for now we assume the defect field is not charged under transverse $SO(d-p)$ rotations.
The result depends only on $\xi + 2\eta$, or equivalently, it depends only on the cross-ratio $r$ defined in \eqref{eq:cross-ratios-rw}.
This follows by observing that the integrand \eqref{eq:def-exch-def} contains no terms of the form $x_1^i x_2^i$.

Since the exchange diagram depends on a single cross-ratio $r$, it is very efficient to compute it by solving a differential equation.
To obtain the differential equation in question, we employ the method of \cite{Zhou:2018sfz,Mazac:2018biw}.
Start giving a name to the $\wh z_1$ integral in \eqref{eq:def-exch-def}
\begin{align}
 I(x_1, \wh z_2)
 = \int_{\wh z_1}
 \wh G_{\wh\Delta}(\wh z_1,\wh z_2)
 K_{\Delta_1}(x_1, \wh z_1)
 \, .
\end{align}
Because this integral preserves the $p$-dimensional conformal group, we have the relation
\begin{align}
 \big( \wh{\mathbf L}_1
     + \wh \Lm_{\wh z_2} \big)_{AB} \, I(x_1, \wh z_2) = 0 \, ,
\end{align}
where $\wh{\mathbf L}_1$ is a generator of $SO(p+1,1)$ acting on operator 1, and $\wh \Lm_{\wh z}$ is the $\AdS_{p+1}$ symmetry generator.
It then follows that
\begin{align}
 \left( \frac12 \, \wh{\mathbf L}_1^2 + \Dh(\Dh-p)
 \right) I(x_1, \wh z_2)
 = \left(- \nabla^2_{\wh z_2} + \Dh(\Dh-p)
 \right) I(x_1, \wh z_2)
 = K_{\Delta_1}(x_1, \wh z_2) \, ,
 \label{eq:eom-cont}
\end{align}
where the last equality is a consequence of the equations of motion for the brane-to-brane propagator, see \eqref{eq:eom} with $d \to p$.
The final step is to multiply \eqref{eq:eom-cont} by $K_{\Delta_2}(x_2,\wh z_2)$ and integrate over $\wh z_2$.
The left-hand side gives a differential operator acting on $\wh E_{\Delta_1\Delta_2}^{\Dh,0}$, and the right-hand side gives a contact diagram
\begin{align}
 \left[
 - r \partial_r r \partial_r
 + \frac{p r (r^2+1)}{(1-r) (1+r)} \, \partial_r
 + \Dh (\Dh-p)
 \right] \wh E_{\Delta_1,\Delta_2}^{\wh\Delta,0}(r)
 = C_{\Delta_1,\Delta_2}(r) \, .
 \label{eq:diffeq-Eh}
\end{align}
In deriving the differential operator, we used that the diagram depends only on $r$ to simplify expressions.
The most general solution of \eqref{eq:diffeq-Eh} depends on two free parameters, but they can always be fixed demanding the following behavior
\begin{align}
 \wh E_{\Delta_1,\Delta_2}^{\Dh,0}(r)
 = \frac{\pi^{p/2} \Gamma(\Dh)
         \Gamma \! \left(\frac{\Delta_1-\Dh}{2}\right)
         \Gamma \! \left(\frac{\Delta_2-\Dh}{2}\right)
         \Gamma \! \left(\frac{\Delta_1+\Dh-p}{2} \right)
         \Gamma \! \left(\frac{\Delta_2+\Dh-p}{2} \right)}
        {8 \Gamma (\Delta_1) \Gamma (\Delta_2) \Gamma (\Dh+1-\frac p2)} \,
    r^{\Dh}
 + O(r^{\Dh+1}) \, .
 \label{eq:normEh}
\end{align}
To derive this equation we used the split representation, as in section \ref{sec:mell-def-ex}.
First one computes \eqref{eq:Hint} exactly in $r$, and then closes the $\nu$ contour to either left or right. Picking the leading residue gives the contribution \eqref{eq:normEh}.

In practice, we have found that solving the ODE provides a very efficient computational method.
Recall that for integer $\Delta_1$ and $\Delta_2$ the contact diagram is a rational function of $r$ and $\log r$, see \eqref{eq:cont-examp}.
In the examples relevant to section \ref{sec:MWL}, the solution to the ODE also turns out to be a rational function, but now including $\log r$ and $\log (r+1)$.
For example, in $d=4$ and $p=1$ we obtain
\begin{align}
&\wh E_{2,2}^{1,0}
 \,\propto\, \log (r+1)
    - \frac{r^2}{r^2-1} \log r \, , \\
&\wh E_{3,3}^{2,0}
 \,\propto\,
     \frac{-2 r^4+r^3+4 r^2+r-2}{2 \left(r^2-1\right)^2}
   - \frac{\left(r^4-2 r^2+3\right) r^3}{\left(r^2-1\right)^3} \, \log r
   + \frac{1+r^2}{r} \log (r+1)  \, .
\end{align}
Although the ODE works perfectly for the calculations in section \ref{sec:MWL}, let us mention that when $\Delta_1 - \wh\Delta \in 2\mathbb{Z}_{>0}$, we can instead use a truncation method:
\begin{align}
 \wh E_{\Delta_1\Delta_2}^{\wh\Delta,0}(\xi,\eta)
 = \sum_{n=1}^{\frac{\Delta_1 - \wh\Delta}{2}}
  \frac{\left(\!-\frac{\Delta_1-\wh\Delta-2}{2}\right)_{n-1}
        \left(\!-\frac{\Delta_1+\wh\Delta-p-2}{2}\right)_{n-1}}
       {4 (1-\Delta_1)_{2n}}
  C_{\Delta_1-2n,\Delta_2}(\xi,\eta) \, .
 \label{eq:Ehtrunc}
\end{align}
(If $\Delta_2 - \wh\Delta \in 2\mathbb{Z}_{>0}$, one uses the same formula with $\Delta_1 \leftrightarrow \Delta_2$).
Let us stress that the truncation condition on $\Delta_1$, $\Delta_2$ and $\Dh$ does not cover all cases of interest.
As a result, formula \eqref{eq:Ehtrunc} is of little use, and we do not derive it here (in any case, the derivation is identical to \cite{RastZhouMell} with $d-1 \to p$).

\subsubsection{Transverse-spin diagrams}
\label{sec:transpin}

The previous discussion has a simple generalization to operators that are charged under transverse spin, i.e. charged under $SO(d-p)$ rotations.
As a simple example, taking the bulk-brane coupling in \eqref{eq:action-tspin} we find that the spin-one exchange diagram is given by
\begin{align}
 \int_{\wh z_1, \wh z_2} \!
 \wh z_1^{\,0} \wh z_2^{\,0}
 \wh G_{\wh\Delta}(\wh z_1,\wh z_2)
 \partial_{i} K_{\Delta_1}(x_1, \wh z_1)
 \partial_{i} K_{\Delta_2}(x_2, \wh z_2)
 \; \equiv \;
  \frac{\wh E^{\wh\Delta,1}_{\Delta_1\Delta_2}(\xi,\eta)}
       {|x^i_1|^{\Delta_1} |x_2^i|^{\Delta_2}} \, .
\end{align}
A simple calculation shows that the result factorizes as
\begin{align}
 \wh E^{\wh\Delta,1}_{\Delta_1\Delta_2}(\xi,\eta)
 \equiv 4 \Delta_1 \Delta_2 \eta \,
   \wh E^{\wh\Delta,0}_{\Delta_1+1,\Delta_2+1}(\xi,\eta) \, .
 \label{eq:tspinone}
\end{align}
It is important to notice the shift in the arguments $\Delta_1,\Delta_2 \to \Delta_1+1,\Delta_2+1$.
More generally, it is not hard to check that a defect-exchange diagram with transverse spin reduces to a product of a scalar diagram with shifted arguments, and a factor with simple $\eta$ dependence:
\begin{align}
 \wh E^{\wh\Delta,s}_{\Delta_1\Delta_2}(\xi,\eta)
 \propto \eta^s \,
   \wh E^{\wh\Delta,0}_{\Delta_1+s,\Delta_2+s}(\xi,\eta) \, .
\end{align}
However, we only need the $s=0,1$ cases below.

\subsection[A detour: connection to analytic functionals]{A detour: connection to analytic functionals\footnote{This section lies somewhat outside the main scope of this work, and can be safely skipped by readers not interested in analytic functionals.}}
\label{sec:anfuncts}

It is well known that, in CFT without defects, different approaches to analytic bootstrap are closely related to each other.
Indeed, the Lorentzian inversion formula \cite{Caron-Huot:2017vep,Simmons-Duffin:2017nub} was resummed into a dispersion relation \cite{Carmi:2019cub}, which in turn generates the analytic functionals of \cite{Mazac:2019shk}. Furthermore, the dispersion relations in position and Mellin space \cite{Penedones:2019tng} are equivalent \cite{Caron-Huot:2020adz}.
For boundary CFT, some of these connections have been explored in \cite{RastZhouMell,Mazac:2018biw,Kaviraj:2018tfd}.

Motivated by these developments, we show here that in defect CFT there is also a connection between the dispersion relation, a basis of analytic functionals, and exchange Witten diagrams.
Unlike the rest of this work, this section relies heavily on analytic bootstrap methods for conformal defects, and the required background can be found in \cite{Lemos:2017vnx,Liendo:2019jpu,Barrat:2022psm,Bianchi:2022ppi}.
Recall that in defect CFT there are two dispersion relations \cite{Barrat:2022psm,Bianchi:2022ppi} (or equivalently two inversion formulas \cite{Lemos:2017vnx,Liendo:2019jpu}).
Here we focus on the dispersion relation that reconstructs the correlator starting from a single discontinuity $\Disc F(r,w') = F(r, w' + i0) - F(r, w' - i0)$.
We define the Polyakov-Regge block $P_{\Delta_1\Delta_2}^{\Delta,J}$ as the dispersion relation applied to a bulk-channel conformal block $f_{\Delta,J}$.\footnote{In this section, $J$ is the spin of an operator of dimension $\Delta$, and $\ell$ is the spin of a double-twist operator. Similarly, we also distinguish transverse-spin $S$ and $s$ below.}
The expression looks more elegant in the $r$, $w$ cross ratios of equation \eqref{eq:cross-ratios-rw}:
\begin{align}
 P_{\Delta_1\Delta_2}^{\Delta,J}(r,w)
 \equiv \int_0^r \frac{dw'}{2 \pi i}
 \frac{w (1 - w') (1 + w')}{ w' (w'-w) (1 - w w')}
 \Disc\!\Big[
 \xi^{-\frac{\Delta_1+\Delta_2}{2}} f_{\Delta,J}(r,w') \Big] \, .
 \label{eq:PR-def}
\end{align}
The discontinuity vanishes on double-twist operators $\Disc
 \xi^{-\frac{\Delta_1+\Delta_2}{2}} f_{\Delta_1+\Delta_2+\ell+2n,\ell} = 0$, see e.g. \cite{Barrat:2022psm}. As a result, the bulk-channel block expansion of the Polyakov-Regge block is of the form
\begin{align}
 P_{\Delta_1\Delta_2}^{\Delta,J}(r,w)
 = \xi^{-\frac{\Delta_1+\Delta_2}{2}} \bigg[
   f_{\Delta,J}(r,w)
 + \sum_{n,\ell=0}^\infty a_{\ell,n}
   f_{\Delta_1+\Delta_2+\ell+2n,\ell}(r,w) \bigg] \, .
 \label{eq:PR-bulk}
\end{align}
To obtain the defect-channel block expansion, one can apply the inversion formula of \cite{Lemos:2017vnx} to the block $f_{\Delta,J}$.
It is not difficult to see that the expansion has the following structure
\begin{align}
 P_{\Delta_1\Delta_2}^{\Delta,J}(r,w)
 = \sum_{n,\ell=0}^\infty b_{s,n}
   \wh f_{\Delta_1+s+2n,s}(r,w)
 + \sum_{n,\ell=0}^\infty c_{s,n}
   \wh f_{\Delta_2+s+2n,s}(r,w) \, ,
 \label{eq:PR-defect}
\end{align}
although extracting $b_{n,\ell}$ and $c_{n,\ell}$ in closed form is more challenging.
Finally, note that the Polyakov-Regge block decays in the Regge limit
\begin{align}
 \big|P_{\Delta_1\Delta_2}^{\Delta,J}(r,w)\big| < |w|^{-\veps}
 \quad \text{for} \quad w \to \infty \, , \; \veps >0 \, ,
 \label{eq:reggelim}
\end{align}
a property that follows from definition \eqref{eq:PR-def}.
In the terminology of \cite{Mazac:2018biw,Mazac:2019shk}, we say that the Polyakov-Regge block is superbounded.

At this point, we claim that the functions $\{f_{\Delta_1+\Delta_2+\ell+2n,\ell}, \wh f_{\Delta_1+s+2n,s}, \wh f_{\Delta_2+s+2n,s}\}$ form a basis in a suitable space of functions, and restrict our attention to CFT correlators in such a space.\footnote{We expect functions in this space to be Regge superbounded as in \eqref{eq:reggelim}. However, one might need further conditions that ensure that the defect-to-bulk inversion formula \cite{Liendo:2019jpu} converges.}
We do not attempt to prove this claim, although there is evidence in the literature for similar statements being true \cite{Mazac:2018biw,Mazac:2019shk,Giombi:2020xah}.
We can then introduce a dual basis $\{\a_{\ell,n}, \b_{s,n}, \gamma_{s,n} \}$ in the natural way, called the basis of analytic functionals.
The analytic functionals are useful in the  bootstrap program, because when acting on the crossing equation, they generate sum rules obeyed by the CFT data.
For example, we have
\begin{align}
 \sum_{\Om} \lambda_{\Om_1\Om_2\Om} a_\Om \,
 \a_{\ell,n}[f_{\Delta,J}]
 - \sum_{\wh\Om} b_{\Om_1\wh\Om} b_{\Om_2\wh\Om} \,
 \a_{\ell,n}[\wh f_{\Dh,s}]
 = 0 \, ,
\end{align}
and two more expressions with $\b_{s,n}$ and $\gamma_{s,n}$.
These sum rules are valid in any CFT, although they are particularly useful to bootstrap theories perturbatively around mean-field theory.
To compute the action of the functionals on a generic bulk block, note that acting with the functionals on \eqref{eq:PR-bulk} and \eqref{eq:PR-defect} gives
\begin{align}
 \a_{\ell,n}[f_{\Delta,\ell}] = -a_{\ell,n} \, , \qquad
 \b_{s,n}[f_{\Delta,\ell}] =  b_{s,n} \, , \qquad
 \gamma_{s,n}[f_{\Delta,\ell}] =  c_{s,n} \, .
\end{align}
Therefore, the expansion coefficients of the Polyakov-Regge block give the action of analytic functionals on bulk-channel conformal blocks.
A similar story holds for defect-channel Polyakov-Regge blocks $\wh P_{\Delta_1\Delta_2}^{\Dh,S}$.
These are defined using the ``defect-to-bulk'' dispersion relation, namely $\wh P_{\Delta_1\Delta_2}^{\Dh,S} = \int K \dDisc \wh f_{\Dh,S}$ with a kernel $K$ that can be found in \cite{Bianchi:2022ppi}.
Their expansion is of the form
\begin{align}
 \wh P_{\Delta_1\Delta_2}^{\Dh,S}(r,w)
&= \xi^{-\frac{\Delta_1+\Delta_2}{2}}
   \sum_{n,\ell=0}^\infty \wh a_{\ell,n}
   f_{\Delta_1+\Delta_2+\ell+2n,\ell}(r,w) \\
&= \wh f_{\Dh,S}
 + \sum_{n,\ell=0}^\infty \wh b_{s,n}
   \wh f_{\Delta_1+s+2n,s}(r,w)
 + \sum_{n,\ell=0}^\infty \wh c_{s,n}
   \wh f_{\Delta_2+s+2n,s}(r,w) \, ,
 \label{eq:PRh-exp}
\end{align}
which implies the action of analytic functionals
\begin{align}
 \a_{\ell,n}[\wh f_{\Dh,S}] = \wh a_{\ell,n} \, , \qquad
 \b_{s,n}[\wh f_{\Dh,S}] =  -\wh b_{s,n} \, , \qquad
 \gamma_{s,n}[\wh f_{\Dh,S}] =  -\wh c_{s,n} \, .
\end{align}
Besides the definition in terms of Polyakov-Regge blocks, the functionals $\a_{\ell,n}[f]$, $\b_{s,n}[f]$ and $\gamma_{s,n}[f]$ might also admit a representation as integrals of $\Disc f$ or $\dDisc f$ against certain kernels.
As in \cite{Mazac:2018biw,Mazac:2019shk,Giombi:2020xah}, one should obtain the integral form of the functionals by expanding in conformal blocks the dispersion relation kernel.
We do not perform this calculation here.

Having shown the utility of Polyakov-Regge blocks, we are faced with the question of how to compute them.
One method uses that Polyakov-Regge blocks are exchange Witten diagrams, with appropriate addition of contact terms to make them Regge superbounded \cite{Mazac:2018biw,Mazac:2019shk}.
For spin $\ell=0,2$, the relations should be of the form
\begin{align}
 P^{\Delta,0}_{\Delta_1\Delta_2}
 = k_1 E^{\Delta,0}_{\Delta_1\Delta_2} \, , \qquad
 P^{\Delta,2}_{\Delta_1\Delta_2}
 = k_1 k_2 \left( E^{\Delta,2}_{\Delta_1\Delta_2}
 + k_3 C_{\Delta_1\Delta_2} \right) \, .
 \label{eq:polyregge}
\end{align}
To extract the precise relation, we use the decomposition of exchange diagrams in terms of contact diagrams, see \eqref{eq:blkexch-conts} and appendix \ref{app:spintwo}.
We fix the overall normalization comparing to the block expansion \eqref{eq:PR-bulk},\footnote{This equation depends on the choice of normalization for the conformal blocks. Here we normalize blocks in the lightcone limit $|1-z| \ll |1-\zb| \ll 1$ as $f_{\Delta,\ell}(z,\zb) = (1-z)^{\frac{\Delta-\ell}{2}} (1-\zb)^{\frac{\Delta+\ell}{2}} + \ldots\,$.} and we fix the contact terms taking the Regge limit \eqref{eq:reggelim}.
In the end, we find
\begin{align}
 k_1
&= \frac{2^{\Delta+2} (-1)^{\frac{\Delta_1+\Delta_2-\Delta+2}{2}}
          \Gamma \left(\frac{\Delta +1}{2}\right)
          (1-\Delta_1)_{\frac{\Delta_1+\Delta_2-\Delta}{2}}
          (1-\Delta_2)_{\frac{\Delta_1+\Delta_2-\Delta}{2}}}
         {\pi^{\frac{p+1}{2}}
          \Gamma \! \left(\frac{\Delta -p}{2}\right)
          \left(\frac{\Delta_1+\Delta_2-\Delta-2}{2}\right)!
          \left(\frac{d-\Delta -\Delta_1-\Delta_2+2}{2}\right)
             _{\frac{\Delta_1+\Delta_2-\Delta -2}{2}}}\, , \\
 k_2
&= \frac{8 \left(\Delta ^2-1\right)}
        {(\Delta_1+\Delta_2-\Delta)
         (\Delta_1+\Delta_2+\Delta-d) (\Delta^2 -\Delta^2_{12})}
         \, , \\[0.4em]
 k_3
&= \frac{d+1}{2d}
  \!-\! \frac{\Delta_{12}^2 (\Delta_1+\Delta_2-d)^2}{2d\Delta  (\Delta -d)}
  + \frac{\left(\Delta_{12}^2-1\right) \left((\Delta_1+\Delta_2-d)^2-1\right)}{2d(\Delta -1) (\Delta-d+1)} \, .
\end{align}
These formulas hold for $\ell=0$, $\Delta_1+\Delta_2-\Delta \in 2\mathbb Z_{>0}$, and for $\ell=2$, $\Delta_1+\Delta_2-\Delta \in 2\mathbb Z_{\ge 0}$.
For general dimensions, we expect the formulas to be more complicated, and perhaps they can be obtained in Mellin space in analogy to \cite{Mazac:2018biw,Mazac:2019shk}.

Although the presentation in section \ref{sec:anfuncts} has been somewhat schematic, we hope it will encourage further exploration of these interesting issues.

\section{Witten diagrams in Mellin space}
\label{sec:mellin}

This section studies the Mellin representation of the two-point function of local operators in the presence of a defect, initially introduced in \cite{Goncalves:2018fwx}.
We show how the Witten diagrams of section \ref{sec:pos} translate to Mellin space, and in particular, we find closed formulas for exchange Mellin amplitudes with arbitrary scaling dimensions.

\subsection{Mellin space}

The Mellin amplitude is only valid for the connected correlator $F_c$, which we introduced in equation \eqref{eq:conn-F}.
Keeping this in mind, the Mellin amplitude $\Mm(\delta,\rho)$ is
\begin{align}
 F_c(\xi,\eta)
 = \int \frac{d\delta \, d\rho}{(2\pi i)^2} \,
   \frac{\Gamma (\delta ) \Gamma (\rho ) \Gamma \! \left(\frac{\Delta_1-\delta-\rho}{2} \right) \Gamma \! \left(\frac{\Delta_2-\delta -\rho}{2}\right)}{\xi^\delta \, (2\eta)^\rho} \,
   \Mm(\delta,\rho)
   \, .
 \label{eq:mellin-def}
\end{align}
This construction can be generalized to any correlator in defect CFT \cite{Goncalves:2018fwx}, but we do not need it here.
Although we are somewhat loose on what the integration contours are, suffice it to mention that they run parallel to the imaginary axis.
Actually, it is not obvious whether non-perturbative correlators admit a Mellin representation, and if they do, what is the correct integration contour.
This is not a problem to us, because we work in perturbation theory and apply the Mellin transform diagram by diagram, so it is possible to find a contour such that all manipulations are well defined.
However, it would be very interesting to analyze non-perturbative Mellin amplitudes along the lines of \cite{Penedones:2019tng,Bianchi:2021piu}.

Leaving these considerations aside, one motivation to work in Mellin space is that Mellin amplitudes are meromorphic in $\delta$ and $\rho$, at least in perturbation theory.
Furthermore, the location of poles corresponds to dimensions of operators exchanged in the OPEs.
More precisely, assume the OPE is of the form $\Om_1 \times \Om_2 \sim \Om + \ldots\,$, where operator $\Om$ has scaling dimension $\Delta$ and spin $\ell$.
Then the integrand of \eqref{eq:mellin-def} has poles at
\begin{align}
 \delta
 = \frac{\Delta_1 + \Delta_2 - \Delta + \ell - 2n}{2} \, , \qquad
 n = 0, 1, 2,\ldots \, .
 \label{eq:blkpoles}
\end{align}
Similarly, if the defect expansion is of the form $\Om_1 \times \Dm \sim \wh\Om\Dm + \ldots\,$, with $\wh\Om$ a local operator on the defect of dimension $\Dh$ and transverse-spin $s$, then the integrand in \eqref{eq:mellin-def} has poles at
\begin{align}
 \delta+\rho = \Dh-s + 2n \, , \qquad
 n = 0, 1, 2, \ldots \, .
 \label{eq:defpoles}
\end{align}
Let us stress that these are poles of the integrand in \eqref{eq:mellin-def}.
In holographic theories, the spectrum is approximately that of MFT
\begin{align}
 \text{Bulk:} \quad
 \Om_1 \Box^n \partial_{\mu_1} \ldots \partial_{\mu_\ell} \Om_2 \, , \qquad
 \text{Defect:} \quad
 \Box^n \partial_{i_1} \ldots \partial_{i_s} \Om_1 \, , \quad
 \Box^n \partial_{i_1} \ldots \partial_{i_s} \Om_2 \, .
\end{align}
The integrand in \eqref{eq:mellin-def} is such that poles by the MFT operators are taken care of by the gamma functions.
As a result, in holographic theories the Mellin amplitude $\Mm(\delta,\rho)$ has a simpler analytic structure than the full integrand, as we show in examples below.

\subsection{Contact diagram}

As a first application of Mellin space, let us obtain the Mellin transform of the contact diagram $C_{\Delta_1\Delta_2}$.
We start from \eqref{eq:cont-int} and use identity \eqref{eq:mell-barn-sum} twice, so the contact diagram has the form
\begin{align}
 C_{\Delta_1\Delta_2}(\xi,\eta)
 =
 \frac{\pi ^{p/2} \Gamma \left(\frac{\Delta_1+\Delta_2-p}{2}\right)}
      {2  \Gamma (\Delta_1) \Gamma (\Delta_2)}
&\int \frac{d\delta \, d\rho}{(2 \pi i)^2}
 \frac{\Gamma (\delta ) \Gamma (\rho )
       \Gamma \! \left(\frac{\Delta_1+\Delta_2}{2}-\delta-\rho \right)}
      { \xi^{\delta} (2\eta)^{\rho}} \notag \\
 & \times
 \int_0^\infty \frac{ds}{s} \, s^{\Delta_1-\delta-\rho}
 \frac{|x_1^i|^{\Delta_1-\delta-\rho } |x_2^i|^{\Delta_2-\delta-\rho } }
      {\left(s^2 |x_1^i|^2+|x_2^i|^2\right)^{\frac{\Delta_1+\Delta_2}{2}-\delta - \rho}} \, .
\end{align}
The $s$ integral is elementary, and after factoring out the prefactor \eqref{eq:mellin-def} we get a Mellin amplitude that is constant:
\begin{align}
 \Mm[C_{\Delta_1,\Delta_2}]
 = \frac{\pi^{\frac{p}{2}} \Gamma \! \left(\frac{\Delta_1+\Delta_2-p}{2} \right)}{4 \Gamma (\Delta_1) \Gamma (\Delta_2)} \, .
 \label{eq:mellin-cont}
\end{align}
This is completely analogous to four-point contact Mellin amplitudes, which are also constant.
Regarding the integration contour, observe that the manipulations that led to \eqref{eq:mellin-cont} are valid provided $\re\delta, \re\rho>0$ and $0 < \re\delta+\re\rho < \min(\Delta_1,\Delta_2)$.
As a result, we have to obey these constraints when choosing the integration contour in \eqref{eq:mellin-def}, or equivalently, the contour has to separate the ``left'' and ``right'' families of poles in the integrand.

The fact that contact amplitudes are constant is very desirable, and actually it motivates the choice of prefactors in the Mellin amplitude \eqref{eq:mellin-def}.
One reason is that large classes of exchange diagrams are sums of contact diagrams, and in this case, the Mellin amplitudes become rational functions of $\delta$ and $\rho$.
To combine contact amplitudes correctly, there are two observations to keep in mind.
The first observation is that contact diagrams are often multiplied in position space by powers of the cross-ratios $\xi^n$ and $\eta^m$, which in Mellin space are shifts of the variables $\delta \to \delta+n$ and $\rho \to \rho+m$.
However, because of our definition of the Mellin amplitude \eqref{eq:mellin-def}, the shift is accompanied by extra Pochhammer symbols:
\begin{align}
 \xi^n \eta^m  F(\xi,\eta)
&\;\leftrightarrow\;
 \frac{(\rho )_m (\delta )_n}{2^m} \left(\frac{\Delta_1-\delta-\rho}{2} \right)_{\frac{-m-n}{2}} \left(\frac{\Delta_2-\delta-\rho}{2} \right)_{\frac{-m-n}{2}}
 \mathcal{M}(\delta +n,\rho+m) \, .
 \label{eq:shiftMell}
\end{align}
The second observation is that the integrand in \eqref{eq:mellin-def} depends on $\Delta_1$ and $\Delta_2$, which gives an extra contribution when combining contact diagrams of different dimensions.
These two observations are best illustrated in practice.
For example, the Mellin transform of \eqref{eq:conts-ex} reads
\begin{align}
 \Mm\big[E_{2,2}^{4,2}\big]
 = \frac{\pi}{24} \left(
 1 + \frac{3 \rho -1}{\delta -1}
 \right) \, , \qquad
 \Mm\big[E_{3,2}^{5,2}\big]
 = \frac{\sqrt{\pi}}{16} \left(
   \frac{4}{5}
 + \frac{2 \rho -1}{\delta -1}
 \right) \, .
\end{align}
As anticipated, the exchange diagrams are remarkably simple rational functions of the Mellin variables.

\subsection{Bulk-exchange diagram}
\label{sec:mell-blk-ex}

We already mentioned in section \ref{sec:blk-exch-pos} that exchange Witten diagrams are hard to calculate due to the complicated formula for the bulk-to-bulk propagator \eqref{eq:blk-blk-G}.
Fortunately, there exists a ``split representation'' of the propagator as a double integral of bulk-to-boundary propagators
\begin{align}
 G_\Delta(z_1,z_2)
 = \int_{-i\infty}^{i\infty} \frac{d\nu}{2\pi i} \,
   \frac{2 \nu^2 \, \Cm_{d/2+\nu} \, \Cm_{d/2-\nu}}
        {\nu^2 - (\Delta-\frac d2)^2}
   \int_{\partial \AdS_{d+1}} d^d x \,
   K_{\frac d2+\nu}(z_1, x) K_{\frac d2 -\nu}(z_2, x) \, .
 \label{eq:splitrep}
\end{align}
Here $\Cm_\Delta$ is the normalization of the bulk-bulk propagator \eqref{eq:norm-G}, the $\nu$ integration goes along the contour $\re \nu = 0$, and the $x$ integral is over the boundary of $\AdS_{d+1}$, namely $\mathbb{R}^d$.
This formula is motivated and proven in the appendix of \cite{Penedones:2010ue}.
For our purposes, it suffices to note that thanks to this formula, exchange diagram factorize as products of lower-point diagrams, at the expense of adding extra integrations.

For example, the bulk-exchange diagram defined in \eqref{eq:def-blk-exch} is schematically
\begin{align}
 E_{\Delta_1\Delta_2}^{\Delta,0}
 \;\; = \;\;
 \begin{tikzpicture}[valign]
    \tikzstyle{every node}=[font=\small]
    \pgfmathsetmacro{\x}{-sqrt(2)/2}
    \pgfmathsetmacro{\y}{sqrt(2)/2}
    \pgfmathsetmacro{\z}{-0.5}
    \pgfmathsetmacro{\r}{0.4}
    \draw [thick] (0,0) circle [radius=1];
    \draw [thick, blue] (-\x,+\y) to[out=240,in=120] (-\x,-\y);
    \draw [dashed]      (+\x,+\y) -- (-\r,0);
    \draw [dashed]      (+\x,-\y) -- (-\r,0);
    \draw [dashed]      (-\r,  0) -- (-\z,0);
    \node at (0.1, 0.2) {$\Delta$};
    \node at (-0.95, 0.85) {$x_1$};
    \node at (-0.95,-0.95) {$x_2$};
 \end{tikzpicture}
 \;\; \sim \;
 \int d\nu \, d^d x_3 \;
 \begin{tikzpicture}[valign]
    \tikzstyle{every node}=[font=\small]
    \pgfmathsetmacro{\x}{-sqrt(2)/2}
    \pgfmathsetmacro{\y}{sqrt(2)/2}
    \pgfmathsetmacro{\z}{-0.5}
    \pgfmathsetmacro{\r}{0.2}
    \draw [thick] (0,0) circle [radius=1];
    \draw [dashed]      (+\x,+\y) -- (-\r,0);
    \draw [dashed]      (+\x,-\y) -- (-\r,0);
    \draw [dashed]      (-\r,  0) -- (1,0);
    \node at (0.35, 0.3) {$\frac d2+\nu$};
    \node at (-0.95, 0.85) {$x_1$};
    \node at (-0.95,-0.95) {$x_2$};
    \node at (1.3, -0.05) {$x_3$};
 \end{tikzpicture}
 \times
 \begin{tikzpicture}[valign]
    \tikzstyle{every node}=[font=\small]
    \pgfmathsetmacro{\x}{-sqrt(2)/2}
    \pgfmathsetmacro{\y}{sqrt(2)/2}
    \pgfmathsetmacro{\z}{-0.5}
    \pgfmathsetmacro{\r}{0.4}
    \draw [thick] (0,0) circle [radius=1];
    \draw [thick, blue] (-\x,+\y) to[out=240,in=120] (-\x,-\y);
    \draw [dashed]      (-1,  0) -- (-\z,0);
    \node at (-0.15, 0.3) {$\frac d2 -\nu$};
    \node at (-1.3, -0.05) {$x_3$};
 \end{tikzpicture}
 \; .
 \label{eq:blk-splitrep}
\end{align}
The first factor is a three-point function in the absence of branes
\begin{align}
 & \int_z K_{\Delta_1}(x_1, z) K_{\Delta_2}(x_2, z) K_{\Delta_3}(x_3, z)
 = \frac{\lambda_{\Delta_1\Delta_2\Delta_3}}{
        |x_{12}|^{\Delta_1+\Delta_2-\Delta_3}
        |x_{13}|^{\Delta_1+\Delta_3-\Delta_2}
        |x_{23}|^{\Delta_2+\Delta_3-\Delta_1}} \, .
\end{align}
Conformal symmetry fixes the form of the right-hand side, and the normalization follows from explicit calculation \cite{Freedman:1998tz,Dolan:2000ut}
\begin{align}
 \lambda_{\Delta_1\Delta_2\Delta_3}
 = \pi^{d/2}
 \frac{\Gamma \! \left(\frac{\Delta_1+\Delta_2-\Delta_3}{2} \right)
       \Gamma \! \left(\frac{\Delta_1-\Delta_2+\Delta_3}{2} \right)
       \Gamma \! \left(\frac{\Delta_2+\Delta_3-\Delta_1}{2}\right)
       \Gamma \! \left(\frac{\Delta_1+\Delta_2+\Delta_3-d}{2}\right)}
      {2 \Gamma (\Delta_1) \Gamma (\Delta_2) \Gamma (\Delta_3)} \, .
 \label{eq:threept}
\end{align}
For the bulk-exchange diagram, we need this formula with $\Delta_3 = \frac d2 + \nu$.

The second factor in \eqref{eq:blk-splitrep} is a one-point function of a scalar in the presence of the brane, which we calculated in \eqref{eq:onept}.
Summarizing, the split representation \eqref{eq:splitrep} allows to compute the $\AdS$ integrals, but we are left with one integral over $x \in \mathbb R^d$ and a Mellin-Barnes integral over $\nu$.
The $x$ integral reads
\begin{align}
 I
&=
 \int
 \frac{d^d x_3}
      {|x_{13}|^{\frac d2+\nu+\Delta_{12}}
       |x_{23}|^{\frac d2+\nu-\Delta_{12}}
       |x_{3}^i|^{\frac d2-\nu}} \, .
\end{align}
Now introduce three Schwinger parameters $s,t,u$, integrate over $x_3$, insert an identity $\int_\lambda \delta(\lambda-u)$, rescale $s,t,u$, and integrate over $\lambda$ and $u$ (see the paragraph before \eqref{eq:bd-interm}).
In the end, one finds the parametric integral
\begin{align}
  I
& = \frac{\pi ^{d/2} \Gamma\!\left(\frac{d+2\nu}{4}\right)}
         {\Gamma\!\left(\frac{d-2\nu}{4}\right)
          \Gamma\!\left(\frac{d+2\nu-2\Delta_{12}}{4}\right)
          \Gamma\!\left(\frac{d+2\Delta_{12}+2\nu}{4}\right)} \notag \\
& \quad \times
  \int_0^\infty \! \frac{ds\,dt}{s\,t}
  \frac{s^{\frac{d-2\nu}{4}}
        t^{\frac{d+2\Delta_{12}+2\nu}{4}}
        (t+1)^{\frac{d+2\nu-2p}{4}}
        (s+t+1)^{\frac{2\nu-d+2p}{4}}}{
        \left( s (t^2 |x_1^i|^2+|x_2^i|^2)
        + 2 \eta s t |x_1^i| |x_2^i|
        + \xi t (s+t+1) |x_1^i| |x_2^i|\right)^{\frac{d+2\nu}{4}}} \, ,
\end{align}
where we used the cross-ratios $\xi$ and $\eta$ in \eqref{eq:cross-ratios-xieta} to simplify the denominator.
Next, we apply identity \eqref{eq:mell-barn-sum} twice to split the denominator, and then integrate over $s$ and $t$
\begin{align}
I
&= \frac{\pi^{d/2}
         \Gamma\!\left(\frac{d+2\nu-2p}{4}\right)
         |x_1^i|^{-\frac{d}{4}-\frac{\Delta_{12}}{2}-\frac{\nu }{2}}
         |x_2^i|^{-\frac{d}{4}+\frac{\Delta_{12}}{2}-\frac{\nu }{2}} }{ 2
         \Gamma\!\left(\frac{d-2\nu}{4}\right)
         \Gamma\!\left(\frac{d+2\nu-2\Delta_{12}}{4}\right)
         \Gamma\!\left(\frac{d+2\Delta_{12}+2\nu}{4}\right) } \notag \\
 & \quad \times
   \int_{-i\infty}^{i\infty} \frac{d\bar\delta\,d\rho}{(2\pi i)^2} \,
   \frac{ \Gamma (\bar{\delta}) \Gamma (\rho ) \Gamma (\bar{\delta}-\nu )
          \Gamma\!\left(\frac{d+2\nu-2\Delta_{12}}{8}
                        -\frac{\bar{\delta}+\rho}{2}\right)
          \Gamma\!\left(\frac{d+2\nu+2\Delta_{12}}{8}
                        -\frac{\bar{\delta}+\rho}{2}\right)}
   {\xi^{\bar{\delta}} (2\eta)^{\rho}
    \Gamma\!\left(\bar{\delta}+\frac{d-2p-2\nu}{4}\right)} \, .
 \label{eq:posint}
\end{align}
We finally have all the ingredients for the bulk-exchange diagram \eqref{eq:blk-splitrep}.
The combination of the split representation \eqref{eq:splitrep}, the one-point function \eqref{eq:one-pt-diag}, the three-point function \eqref{eq:threept} and the position integral \eqref{eq:posint} has to be compared to the Mellin integral \eqref{eq:mellin-def}.
For the comparison to work, we shift the Mellin variable as $\bar \delta = \delta -\frac{\Delta_1+\Delta_2-\nu-d/2}{2}$.
All in all, the bulk-exchange amplitude $\Mm_{\Delta_1\Delta_2}^{\Delta,0} \equiv \Mm\big[E_{\Delta_1\Delta_2}^{\Delta,0}\big]$ is given by
\begin{align}
\Mm_{\Delta_1\Delta_2}^{\Delta,0}(\delta,\rho)
=
\frac{\pi^{p/2}}
     {\Gamma (\Delta_1) \Gamma (\Delta_2) \Gamma (\delta )
      \Gamma\!\left(\delta - \frac{\Delta_1+\Delta_2-d+p}{2}\right)}
\int_{-i\infty}^{i\infty} \frac{d\nu}{2\pi i}
\frac{ l_b(\nu ) l_b(-\nu)}{(\Delta -\frac{d}{2})^2-\nu ^2} \, ,
 \label{eq:blkexchmell}
\end{align}
where
\begin{align}
 l_b(\nu)
 =
 \frac{\Gamma\!\left(\frac{2 \nu + d-2 p}{4}\right)
       \Gamma\!\left(\frac{2\nu+2\Delta_1+2\Delta_2-d}{4}\right)
       \Gamma\!\left(\delta + \frac{2 \nu-2 \Delta_1-2 \Delta_2+d}{4}\right)}
      {4 \Gamma (\nu)} \, .
\end{align}
Note that the scalar Mellin amplitude only depends on $\delta$.
This is a consequence of our choice of prefactor in \eqref{eq:mellin-def}, because the prefactor captures the poles in $\delta+\rho$ corresponding to defect operators with MFT dimensions.
On the other hand, \eqref{eq:blkexchmell} captures new poles in $\delta$ due to the exchange of a bulk operator of dimension $\Delta$, recall \eqref{eq:blkpoles}.

Note also the similarity between our formula and the Mellin amplitude of a four-point scalar Witten diagram, see equation (38) in \cite{Penedones:2010ue}.
In fact, we can follow verbatim the analysis of reference \cite{Penedones:2010ue}, which observed that poles in $\delta$ arise when two poles in the integrand pinch the $\nu$ contour.
A simple application of the residue theorem then gives
\begin{align}
 \Mm_{\Delta_1\Delta_2}^{\Delta,0}(\delta,\rho)
 = \sum_{n=0}^\infty \frac{R_n}{\delta - \frac{\Delta_1+\Delta_2-\Delta-2 n}{2}} \, ,
 \label{eq:sumpolesblk}
\end{align}
where the residues are\footnote{To obtain this formula, we use that for a meromorphic function $g(z) = \sum_i \frac{f_i(z)}{z-a_i} = \sum_i \frac{f_i(a_i)}{z-a_i}$, provided $g(z)$ decays at infinity. We have checked numerically in examples that \eqref{eq:blkexchmell} and \eqref{eq:sumpolesblk} agree, which justifies using this identity.}
\begin{align}
 R_n
 = \frac{\pi^{p/2}}{16 n!}
 \frac{\Gamma\!\left(\frac{\Delta-p}{2}\right)
       \Gamma\!\left(\frac{\Delta+\Delta_1+\Delta_2-d}{2}\right)
       \left(1 + \frac{\Delta-d+p}{2}\right)_n
       \left(1 + \frac{\Delta -\Delta_1-\Delta_2}{2}\right)_n}
      {\Gamma (\Delta_1) \Gamma (\Delta_2)
       \Gamma\!\left(\Delta+n+1-\frac{d}{2}\right)} \, .
\end{align}
Here we see that when $\Delta_1+\Delta_2-\Delta \in 2 \mathbb Z_{>0}$, the infinite sum truncates to a finite sum.
This finite sum agrees with the sum over contact diagrams \eqref{eq:blkexch-conts} that we found in position space.
Although we shall not need it, the sum can be computed in terms of a generalized hypergeometric function
\begin{align}
 \Mm_{\Delta_1\Delta_2}^{\Delta,0}(\delta,\rho)
&= \frac{\pi ^{p/2}
         \Gamma\!\left(\frac{\Delta -p}{2}\right)
         \Gamma\!\left(\frac{\Delta_1+\Delta_2+\Delta-d}{2}\right)}
        {16 \Gamma (\Delta_1) \Gamma (\Delta_2)
         \Gamma\!\left(\frac{2 \Delta-d+2}{2}\right) \!
         \left(\delta - \frac{\Delta_1+\Delta_2-\Delta}{2}\right)} \notag \\
& \qquad \qquad \times \,
  _3F_2 \! \left( \!\! \begin{array}{*{20}{c}}
    {1+\frac{\Delta-d+p}{2},
     1+\frac{\Delta-\Delta_1-\Delta_2}{2},
    \delta +\frac{\Delta-\Delta_1-\Delta_2}{2}} \\
    {\Delta+1-\frac{d}{2},
     \delta+1+\frac{\Delta-\Delta_1-\Delta_2}{2}}
    \end{array}; 1\right) \, .
 \label{eq:mell-blk}
\end{align}

The $\ell > 0$ blocks for arbitrary dimensions can also be computed using the split representation, which was developed in \cite{Costa:2014kfa}.
However, that technology is somewhat involved and we choose not to explore it in the present work.
Actually, for applications to section \ref{sec:MWL}, all $\ell = 2$ exchanges truncate to a sum of contact diagrams, which can be easily Mellin transformed.

\subsection{Defect-exchange diagram}
\label{sec:mell-def-ex}

A similar calculation also applies to the defect-exchange diagram \eqref{eq:def-exch-def}.
The idea is to use the split representation for the brane-to-brane propagator, equation \eqref{eq:splitrep} with $d \to p$.
The exchange diagram factorizes as a product of bulk-brane diagrams
\begin{align}
 \wh E_{\Delta_1\Delta_2}^{\Dh,0}
 \;\; = \;
\begin{tikzpicture}[valign]
    \tikzstyle{every node}=[font=\small]
    \pgfmathsetmacro{\x}{sqrt(1)/2}
    \pgfmathsetmacro{\y}{sqrt(3)/2}
    \pgfmathsetmacro{\xx}{sqrt(2)/2}
    \pgfmathsetmacro{\yy}{sqrt(2)/2}
    \pgfmathsetmacro{\z}{0.5}
    \pgfmathsetmacro{\rx}{0.32}
    \pgfmathsetmacro{\ry}{0.45}
    \draw [thick] (0,0) circle [radius=1];
    \draw [dashed]      (-\xx,+\yy) -- (\rx,\ry);
    \draw [dashed]      (-\xx,-\yy) -- (\rx,-\ry);
    \draw [dashed,blue] (\rx,\ry) to[out=210,in=150] (\rx,-\ry);
    \draw [thick, blue] (\x,+\y) to[out=240,in=120] (\x,-\y);
    \node at (-0.95, 0.85) {$x_1$};
    \node at (-0.95,-0.95) {$x_2$};
    \node at (-0.2,0) { $\Dh$};
 \end{tikzpicture}
 \;\; \sim \;\;
 \int d\nu \, d^p \wh x
 \begin{tikzpicture}[valign]
    \tikzstyle{every node}=[font=\small]
    \pgfmathsetmacro{\x}{sqrt(1)/2}
    \pgfmathsetmacro{\y}{sqrt(3)/2}
    \pgfmathsetmacro{\xx}{sqrt(2)/2}
    \pgfmathsetmacro{\yy}{sqrt(2)/2}
    \pgfmathsetmacro{\z}{0.5}
    \pgfmathsetmacro{\rx}{0.32}
    \pgfmathsetmacro{\ry}{0.45}
    \draw [thick] (0,0) circle [radius=1];
    \draw [dashed]      (-\xx,+\yy) -- (\rx,\ry);
    \draw [dashed,blue] (\rx,\ry) to[out=230,in=150] (\x,-\y);
    \draw [thick, blue] (\x,+\y) to[out=240,in=120] (\x,-\y);
    \node at (-0.95, 0.85) {$x_1$};
    \node at (-0.4,-0.2) { {\footnotesize $\frac p2 + \nu$}};
    \node at (0.7,-1) {$\wh x$};
 \end{tikzpicture}
 \;\; \times \,
  \begin{tikzpicture}[valign]
    \tikzstyle{every node}=[font=\small]
    \pgfmathsetmacro{\x}{sqrt(1)/2}
    \pgfmathsetmacro{\y}{sqrt(3)/2}
    \pgfmathsetmacro{\xx}{sqrt(2)/2}
    \pgfmathsetmacro{\yy}{sqrt(2)/2}
    \pgfmathsetmacro{\z}{0.5}
    \pgfmathsetmacro{\rx}{0.32}
    \pgfmathsetmacro{\ry}{0.45}
    \draw [thick] (0,0) circle [radius=1];
    \draw [dashed]      (-\xx,-\yy) -- (\rx,-\ry);
    \draw [dashed,blue] (\rx,-\ry) to[out=-230,in=-150] (\x,\y);
    \draw [thick, blue] (\x,+\y) to[out=240,in=120] (\x,-\y);
    \node at (-0.95, -0.85) {$x_2$};
    \node at (-0.4,+0.2) { {\footnotesize $\frac p2 - \nu$}};
    \node at (0.7,+1) {$\wh x$};
 \end{tikzpicture} \;\; ,
\end{align}
at the expense of adding integrations over $\wh x$ and $\nu$.
The bulk-to-brane Witten diagrams are kinematically fixed, and their normalization is computed in section \ref{sec:bulk-def}.
As a result, we need to compute the integral
\begin{align}
 H = \int \frac{d^p \wh x_3 }
           {\big((x_1^i)^2 + (x_{1\hat3}^a)^2 \big)^{\frac p2+\nu}
            \big((x_2^i)^2 + (x_{2\hat3}^a)^2 \big)^{\frac p2-\nu}} \, .
 \label{eq:Hint}
\end{align}
The technique is by now standard.
Introduce two Schwinger parameters $s$ and $t$, integrate over $\wh x_3$, and integrate over $t$ with the help of an auxiliary parameter $\lambda$.
The result takes the form
\begin{align}
 H
 & = \frac{\pi^{\frac{p}{2}} \Gamma (\frac{p}{2})}
          {\Gamma (\frac{p}{2}-\nu ) \Gamma (\frac{p}{2}+\nu )}
 \int \frac{ds}{s}
 \frac{s^{\frac{p}{2}+\nu}}
      {\big(s^2 |x_1^i|^2+|x_2^i|^2
       + 2\eta s |x_1^i| |x_2^i|
       + \xi s |x_1^i| |x_2^i| \big)^{p/2}} \notag \\[0.3em]
& = \frac{\pi ^{\frac{p}{2}}}
         {|x_1^i|^{\frac{p}{2}+\nu} |x_2^i|^{\frac{p}{2}-\nu}}
    \int \frac{d\delta \, d\rho}{(2\pi i)^2}
    \frac{\Gamma(\delta) \Gamma (\rho)}{\xi^\delta (2\eta)^\rho}
    \frac{\Gamma\!\left(\frac{p/2-\nu-\delta -\rho}{2}\right)
          \Gamma\!\left(\frac{p/2+\nu-\delta -\rho}{2}\right)}
         {2 \Gamma (\frac{p}{2}-\nu ) \Gamma (\frac{p}{2}+\nu )} \, ,
 \label{eq:bdyint}
\end{align}
where we used \eqref{eq:mell-barn-sum} twice to factorize the denominator, and then computed an elementary integral over $s$.
Again, the combination of the split representation \eqref{eq:splitrep}, the bulk-brane diagram \eqref{eq:bulkbrane} and the boundary integral \eqref{eq:bdyint} is compared with the Mellin amplitude \eqref{eq:mellin-def}.
Using notation $\wh \Mm_{\Delta_1\Delta_2}^{\Dh,0} \equiv \Mm\big[\wh E_{\Delta_1\Delta_2}^{\Dh,0}\big]$, the defect-exchange amplitude reads
\begin{align}
 \wh \Mm_{\Delta_1\Delta_2}^{\Dh,0}(\delta,\rho)
 = \frac{\pi ^{\frac{p}{2}}}{\Gamma (\Delta_1) \Gamma (\Delta_2) \Gamma\!\left(\frac{\Delta_1-\delta -\rho}{2}\right) \Gamma\!\left(\frac{\Delta_2-\delta-\rho}{2} \right)}
 \int_{-i\infty}^{i\infty} \frac{d\nu}{2\pi i}
 \frac{l_d(\nu) l_d(-\nu)}{(\Dh-\frac{p}{2})^2-\nu ^2} \, ,
 \label{eq:melldefexch}
\end{align}
where
\begin{align}
 l_d(\nu)
 =
 \frac{\Gamma\!\left(\frac{2\nu + 2\Delta_1-p}{4}\right)
       \Gamma\!\left(\frac{2\nu + 2\Delta_2-p}{4}\right)
       \Gamma\!\left(\frac{2\nu + p - 2\delta - 2\rho}{4}\right)}
      {4 \Gamma(\nu)} \, .
\end{align}
In this case the amplitude depends only on $\delta+\rho$, and the $\delta$ dependence is accounted for by the prefactor in \eqref{eq:mellin-def}.
The interpretation is similar as before, namely the bulk-channel block expansion contains operators with MFT dimension $\Delta = \Delta_1+\Delta_2+\ell+2n$ which are completely captured by the prefactor.
Instead, in the defect-channel expansion there is a new operator with $\Dh$, whose poles are captured by \eqref{eq:melldefexch}.

Once again, we can obtain the poles in $\delta+\rho$ by picking the residues when two poles pinch the $\nu$ integration contour.
We find the following sum
\begin{align}
 \wh \Mm_{\Delta_1\Delta_2}^{\Dh,0}(\delta,\rho)
 =
 \sum_{n=0}^\infty \frac{\wh R_n}{\delta+\rho-\Dh-2n} \, ,
\end{align}
with residues
\begin{align}
 \wh R_n
 = - \frac{\pi ^{\frac{p}{2}}}{8 n!}
 \frac{\Gamma\!\left(\frac{\Dh+\Delta_1-p}{2}\right)
       \Gamma\!\left(\frac{\Dh+\Delta_2-p}{2}\right)
       \left(1+\frac{\Dh-\Delta_1}{2}\right)_n
       \left(1+\frac{\Dh-\Delta_2}{2}\right)_n}
      {\Gamma (\Delta_1) \Gamma (\Delta_2) \Gamma(\Dh+n+1-\frac{p}{2})} \, .
\end{align}
As expected, the sum truncates when $\Delta_1-\Dh, \Delta_2-\Dh \in \mathbb Z_{>0}$, analogously to equation \eqref{eq:Ehtrunc}.
More importantly, even when the sum does not truncate, our calculation goes through and we obtain
\begin{align}
 \wh \Mm_{\Delta_1\Delta_2}^{\Dh,0}
&=
 -\frac{\pi^{p/2}
        \Gamma\!\left(\frac{\Delta_1+\Dh-p}{2}\right)
        \Gamma\!\left(\frac{\Delta_2+\Dh-p}{2}\right)}
       {8 \Gamma (\Delta_1) \Gamma (\Delta_2) \Gamma(\Dh+1-\frac p2)
        (\delta+\rho-\Dh) }
  {}_3F_2 \! \left( \!\! \begin{array}{*{20}{c}}
  {1-\frac{\Delta_1-\Dh}{2}, 1-\frac{\Delta_2-\Dh}{2},
   \frac{\Dh-\delta-\rho}{2}} \\
  {\Dh+1-\frac{p}{2},1 + \frac{\Dh-\delta-\rho}{2}} \\
  \end{array}; 1 \right) \, .
  \label{eq:mell-def}
\end{align}

Although the discussion so far is restricted to scalar exchanges,  we showed in section \ref{sec:transpin} that generalization to fields with transverse spin $s>0$ is straightforward.
Indeed, a generic exchange diagrams factorizes as $\wh E_{\Delta_1,\Delta_2}^{\Dh,0} \sim \eta^s \wh E_{\Delta_1+s,\Delta_2+s}^{\Dh,0}$, where the scalar exchange has shifted arguments.
For example, the Mellin transform of the spin-one exchange \eqref{eq:tspinone} reads
\begin{align}
 \wh \Mm_{\Delta_1\Delta_2}^{\Dh,1}(\delta,\rho)
 = 2 \Delta_1 \Delta_2 \rho  \Mm_{\Delta_1+1,\Delta_2+1}^{\Dh,0}(\delta,\rho +1) \, ,
\end{align}
where the factor $\rho$ is explained by \eqref{eq:shiftMell}.
More generally, the spin-$s$ exchange diagram is of the form
\begin{align}
 \wh \Mm_{\Delta_1\Delta_2}^{\Dh,s}(\delta,\rho)
 \propto (\rho )_s \, \Mm_{\Delta_1+s,\Delta_2+s}^{\Dh,0}(\delta,\rho +s) \, .
\end{align}

\section{Half-BPS Wilson line in \texorpdfstring{$\Nm=4$}{N=4} SYM}
\label{sec:MWL}

In this section, we apply the position-space bootstrap to the half-BPS Wilson line in $\Nm=4$ SYM.
More precisely, we bootstrap the correlator of two chiral-primary operators and a half-BPS Wilson line at leading order in the supergravity approximation.
We start reviewing the setup from the CFT perspective, and then move on to the holographic description.
We derive the $\AdS$ effective actions for bulk and brane, and input this information to the position-space bootstrap method.
This allows us to determine the correlator in closed form.
The result agrees with a first-principles Witten diagram calculation, up to contact diagrams that we do not know how to fix in the first-principles method.
We conclude expressing our result in Mellin space, where it takes a remarkably simple form.

\subsection{Setup}
\label{sec:setupN4}

Before presenting the holographic calculation, let us review the system using field-theoretic language.
The same system was studied in \cite{Barrat:2021yvp,Barrat:2022psm}, where the interested reader can find further details.

We work with planar $\Nm=4$ SYM theory in the strong 't Hooft coupling limit $\lambda = g^2 N \to \infty$.
The local operators are chiral primaries in the $[0,k,0]$ representation of $SU(4)_R$, that in field theory language read\footnote{
Here $\ldots$ contains multi-trace operators that ensure that $S_k$ are operators dual to single-particle supergravity modes.
The most important property of this choice of operators is that extremal three-point functions vanish, namely the OPE coefficient $\lambda_{S_{k_1}S_{k_2}S_{k_1+k_2}} = 0$.
The difference between single-trace and $S_k$ operators plays only a tangential role in our story, and we refer to section 2.1 of \cite{Alday:2019nin} for a more detailed exposition.\label{foot:sugra}}
\begin{equation}
\label{eq:cpo}
 S_k(x,u) \propto \tr \big( u \cdot \phi(x) \big)^k + \ldots \, .
\end{equation}
As usual, we work with index-free notation, where $u$ is a null six-component vector $u^2 = 0$ that implements the tracelessness of $S_k$.
The chiral-primary operators are scalars with protected scaling dimension $\Delta = k$.

Besides local operators, we also consider a straight half-BPS Wilson line
\begin{equation}
\label{eq:wl}
 W(\theta)
 = \frac{1}{N} \, \tr \, \Pexp
 \int_{-\infty}^{\infty} d\tau\,
 \big( i \dot{x}^\mu A_\mu + |\dot{x}|\, \theta \cdot \phi \big) \, .
\end{equation}
This is a one-dimensional defect that lives in four-dimensional space, or in other words, the dual setup is $\AdS_5$ with a brane extending on $\AdS_2$. Therefore, from now on we set $d=4$ and $p = 1$.
In \eqref{eq:wl} there is a choice of $R$-symmetry direction which we parametrize with the unit six-component vector $\theta^2 = 1$.
For a straight contour, the expectation value of the Wilson operator is $\vev{W(\theta)} = 1$, so unlike section \ref{sec:dcft}, we do not need to normalize correlators by $\vev{W(\theta)}$.

Thanks to the index-free notation, we can write the two-point function of chiral-primary operators as
\begin{align}
\label{eq:two-pt}
 \vev{ S_{p_1}(x_1, u_1) S_{p_2}(x_2, u_2) W(\theta)}
 = \frac{(u_1 \cdot \theta)^{p_1} (u_2 \cdot \theta)^{p_2}}
        {|x_1^i|^{p_1} |x_2^i|^{p_2}} \,
   \Fm(\xi,\eta,\sigma) \, .
\end{align}
The function $\Fm(\xi,\eta,\sigma)$ captures the non-trivial information of this correlator.
The first two variables $\xi$, $\eta$ are the spacetime cross-ratios \eqref{eq:cross-ratios-xieta}, while $\sigma$ is an $R$-symmetry cross-ratio.
The dependence on $\sigma$ is polynomial, namely
\begin{equation}
 \Fm(\xi,\eta,\sigma)
 = \sum_{n=0}^{\pmin} \sigma^n \, \Fm_n(\xi,\eta)
 \quad\; \text{with} \quad\;
 \sigma
 = \frac{u_1 \cdot u_2}{(u_1 \cdot \theta)(u_2 \cdot \theta)} \,.
 \label{eq:invariant}
\end{equation}
Here and below we define $\pmin = \min(p_1,p_2)$.
We will be mostly concerned with the connected correlator $\Fm_c$, which is defined analogously to \eqref{eq:conn-two}.
After extracting the overall prefactor in \eqref{eq:two-pt}, the connected correlator is
\begin{align}
 \Fm(\xi,\eta,\sigma)
 = \Big(\frac{\sigma}{\xi}\Big)^{\frac{p_1+p_2}{2}}
 + a_{p_1} a_{p_2}
 + \Fm_c(\xi,\eta,\sigma) \, ,
 \label{eq:conncorrN4}
\end{align}
where $a_p$ is the one-point function $\vev{ S_p \, W } = a_p \big(\frac{u_1 \cdot \theta}{|x^i|}\big)^{p}$.
For reference, observe that the one-point function is known explicitly
\begin{align}
 a_p = \frac{\sqrt{p \lambda } I_p(\sqrt{\lambda })}{2^{\frac{p}{2}+1} N I_1(\sqrt{\lambda })} + O\big(\tfrac{1}{N^2}\big) \, .
\end{align}
Also note that our definition of connected correlator is different than the one used e.g. in \cite{Giombi:2012ep}.

The Wilson line is a half-BPS object that breaks supersymmetry according to
\begin{align}
 PSU(2,2|4) \;\supset\;
 OSp(4^*|4) \;\supset\;
 SL(2,\mathbb R) \oplus SO(3)_{\text{trans.}} \oplus SO(5)_R \, .
 \label{eq:osp44}
\end{align}
The $SL(2, \mathbb R)$ is the group of global conformal transformation along the line, $SO(3)_{\text{trans.}}$ are rotation in the transverse spacetime directions, while $SO(5)_R$ is the leftover $R$-symmetry.
The preserved $OSp(4^*|4)$ symmetry has important consequences for the correlation function under study, namely the superconformal Ward identity
\begin{equation}
\left. \left( \partial_z + \frac{1}{2} \partial_\alpha \right)
\Fm(z,\zb,\alpha) \right|_{z = \alpha} = 0\,
\label{eq:WI}
\end{equation}
must be obeyed \cite{Liendo:2016ymz}.
This has to be supplemented by another equation with $z \leftrightarrow \zb$.
The Ward identity is expressed in terms of variables $z,\zb,\a$ defined as
\begin{align}
 \xi  = \frac{(1-z)(1-\zb)}{\sqrt{z \zb}} \, , \qquad
 \eta = \frac{z + \zb}{2\sqrt{z \zb}} \, , \qquad
 \sigma = -\frac{(1-\a)^2}{2\a} \, .
 \label{eq:xieta2zzb}
\end{align}
The superconformal Ward identity is key for the position-space bootstrap of section \ref{sec:pos-space-boot}.

\subsection{Effective action at strong coupling}
\label{sec:effect-act}

With this preliminary information in mind, we can now start the holographic calculation.
The chiral-primary operator $S_k$ is dual to a supergravity field $s_k$, which arises from the KK reduction of IIB supergravity on $\AdS_5 \times S^5$.
As in the simple example of section \ref{sec:prelim}, the first step is to obtain the effective action for $s_k$.
We start in section \ref{sec:act5} reviewing the $\AdS_5$ effective action, which is already known in the literature \cite{Kim:1985ez,Lee:1998bxa,Arutyunov:1998hf}.
We then derive the $\AdS_2$ effective action.
Besides rederiving vertices that were previously known \cite{Berenstein:1998ij,Giombi:2017cqn}, we also obtain new bulk-brane vertices that, to the best of our knowledge, have not been previously computed.

\subsubsection{Effective action in \texorpdfstring{$\AdS_5$}{AdS5}}
\label{sec:act5}

We use the ten-dimensional metric $G_{MN} = g_{MN} + h_{MN}$, where $g_{MN}$ is the background $AdS_5 \times S_5$ metric, and $h_{MN}$ are fluctuations around it.
We use Poincare coordinates $z^\mu$ on $\AdS_5$, and stereographic coordinates $y^A$ on the $S^5$, so the background metric reads
\begin{align}
 \label{eq:met-AdS5-S5}
 ds^2_{\AdS_5 \times S^5}
 = \frac{dz^\mu dz^\mu}{(z^0)^2}
 + \frac{dy^A dy^A}{(1 + \frac14 y^B y^B)^2} \, , \qquad
 \mu,\nu = 0, 1, \ldots, 4 \, , \quad
 A,B = 5, \ldots, 9 \, .
\end{align}
Let us consider the fluctuations $h_{\mu\nu}$ of the $\AdS_5$ metric, which are expanded in $S^5$ spherical harmonics to perform the KK reduction
\begin{align}
 h_{\mu\nu}(z,y)
 = \sum_{k,I} (h^I_k)_{\mu\nu}(z) Y_k^I(y) \, .
 \label{eq:hmunu}
\end{align}
The spherical harmonic $Y_k^I$ for $k = 0,1,2,\ldots$ transforms in the rank-$k$ symmetric traceless representation of $SO(6)$, with the index $I=1,\ldots,d_k$ labeling the elements in the representation.
Appendix \ref{sec:spherical} gives an explicit definition of spherical harmonics, and provides simple rules to map them to the index-free notation in equations \eqref{eq:cpo}-\eqref{eq:invariant}.

Besides fluctuations of the $\AdS_5$ metric in \eqref{eq:hmunu}, there are fluctuations for the metric on the sphere $h_{AB}$, and for the Ramond-Ramond four form $C_4 = \bar C_4 + \delta C_4$, which are also expanded in spherical harmonics.
Upon inserting the spherical harmonic expansions in the IIB supergravity action, one obtains mixing in the kinetic terms.
The mixing problem was first solved in \cite{Kim:1985ez}, where the reader can find a table of all fields, and their properties after unmixing.
Of all the supergravity fields, only the ones with even spin and in symmetric-traceless irreps of $SO(6)_R$ couple to the fundamental string.
This leaves us with the $s$, $\varphi_{\mu\nu}$ and $t$ fields, whose properties are summarized in table \ref{tab:stp}.
The standard AdS/CFT dictionary shows that $s_k$ is the supergravity field dual to the CFT operator $S_k$, while $\varphi_{\mu\nu}$ and $t$ are superconformal descendants.

For future reference, let us mention that the fields $s$, $\varphi_{\mu\nu}$ and $t$ are related to $\AdS_5$ metric fluctuations as \cite{Arutyunov:1998hf}
\begin{align}
 (h^I_k)_{\mu\nu}
&= (\varphi^I_k)_{\mu\nu}
 + \frac{4}{k+1}
   \left( \nabla_{\mu}\nabla_{\nu}
 - \frac{1}{2} k (k-1) g_{\mu\nu} \right) s_k^I \notag \\
&\quad
 + \frac{4}{k+3}
   \left( \nabla_{\mu}\nabla_{\nu}
 - \frac{1}{2} (k+4) (k+5) g_{\mu\nu} \right) t_k^I
 \, .
 \label{eq:hfluct}
\end{align}
There exist similar formulas for the fluctuations $h_{AB}$, $\delta C$, but they will not play a role in our calculation.
\begin{table}
 \centering
 \begin{tabular}{c||c|c|c|c}
  Field  & $m^2$ & $\Delta$ & $\;\;\ell\;\;$ & $SU(4)$-irrep \\ \hline \hline
  $s$ & $k(k-4)$ & $k$ & 0 & $[0,k,0]$\\ \hline
  $\varphi_{\mu\nu}$ & $(k+2)(k-2)$ & $k+2$ & 2 & $[0,k-2,0]$ \\ \hline
  $t$ & $k(k+4)$ & $k+4$ & 0 & $[0,k-4,0]$ \\ \hline
 \end{tabular}
 \caption{Kaluza-Klein modes that couple to the string worldsheet}
 \label{tab:stp}
\end{table}
Inserting formula \eqref{eq:hfluct} and its analogs for $h_{AB}$, $\delta C$ in the IIB supergravity action, one extracts kinetic terms, cubic couplings, and so on:
\begin{align}
 S_{\text{IIB}}
&= \frac{4N^2}{(2\pi)^5} \int \! \frac{d^5 z}{(z_0)^5} \, \Big(
      L^{(2)}_{\text{IIB}}
    + L^{(3)}_{\text{IIB}}
    + \ldots \Big) \, .
 \label{eq:actIIB}
\end{align}
Recall that we are interested in the scattering of $s_k$ with a fundamental string, which is dual to the CFT correlator $\vev{ S_{p_1} S_{p_2} W}$.
A simple power-counting argument shows that only cubic couplings contribute at leading order in the supergravity approximation.
Since we shall not attempt to compute corrections beyond leading order, the quadratic and cubic terms in \eqref{eq:actIIB} are all we need.\footnote{This is a fortunate state of affairs, since quartic and higher couplings are remarkably complicated \cite{Arutyunov:1999fb}.}
The kinetic terms for $s$, $\varphi$ and $t$ are \cite{Kim:1985ez,Lee:1998bxa,Arutyunov:1998hf}
\begin{align}
 L^{(2)}_{\text{IIB}}
&= \sum_{k,I} \Big(
      \zeta^s_k \, L^{(0)}_2(s^I_k)
    + \zeta^t_k \, L^{(0)}_2(t^I_k)
    + \zeta^\varphi_k \, L^{(2)}_2(\varphi^I_k)
    + \ldots \Big) \, ,
\end{align}
where $\zeta^X_k$ are normalizations presented in appendix \ref{sec:cubic-coupl}.
Besides the overall normalization, the kinetic terms are standard
\begin{align}
 L^{(0)}_2(s)
&= \frac{1}{2} \nabla_\mu s \nabla^\mu s
 + \frac{1}{2} m^2 s^2 \, , \\
 L^{(2)}_2(\varphi)
&= \frac14 \nabla_\mu \varphi_{\nu\rho} \nabla^\mu \varphi^{\nu\rho}
 - \frac12 \nabla_\mu \varphi^{\mu\nu} \nabla^\rho \varphi_{\nu\rho}
 + \frac12 \nabla_\mu \varphi \nabla_\rho \varphi^{\mu\rho}
   \notag \\
& \quad
 - \frac14 \nabla_\mu \varphi \nabla^\mu \varphi
 - \frac14 (2-m^2) \varphi_{\mu\nu} \varphi^{\mu\nu}
 - \frac14 (2+m^2) \varphi^2 \, ,
\end{align}
where the trace of the spin-two field is $\varphi = g^{\mu\nu} \varphi_{\mu\nu}$, and the masses appear in table \ref{tab:stp}.
In a similar way, one can write down the cubic couplings. For our purposes, we only need vertices that contain two $s$ fields, and one $X=s,\varphi,t$ field:
\begin{align}
 L_{\text{IIB}}^{(3)}
&= - \sum_{k,p,q} \Big(
    S^{IJK}_{k p q} s^I_{k} \, s^J_{p} \, s^K_{q}
  + T^{IJK}_{k p q} s^I_{k} \, s^J_{p} \, t^K_{q}
  + G^{IJK}_{k p q} T_{\mu\nu}(s_{k}^I,s_{p}^J) (\varphi_{q}^{K})^{\mu\nu}
  + \ldots
  \Big) \, .
\end{align}
The precise form of the vertices was obtained in \cite{Arutyunov:1999en} and is reviewed in appendix \ref{sec:cubic-coupl}.

\subsubsection{Effective action in \texorpdfstring{$\AdS_2$}{AdS2}}

Let us now derive the effective action of fluctuations on top of the fundamental string F1, which lives in the background $\AdS_5 \times S^5$ just described.
Because in this background the NS-NS forms are turned-off, the string action reads
\begin{align}
 S_{\text{F1}}
&= \frac{\sqrt{\lambda}}{2\pi} \int d^2 \wh z \,
   \sqrt{\det \big( G_{MN}(X) \partial_\a X^M \partial_\b X^N \big) } \, ,
 \label{eq:NG}
\end{align}
where we also ignore fermions.
The worldsheet coordinates are $\wh z = (\wh z^{\,0}, \wh z^{\,1})$, and $X^M(\wh z)$ describes the embedding of the string in ten dimensions.
The string intersects the boundary of $\AdS_5$ in a contour $\gamma$, and sits at a particular point in the $S^5$.
From the CFT perspective, $\gamma$ is the location of the Wilson loop, while the point in $S^5$ is related to the polarization $\theta$ in \eqref{eq:wl}.
The expectation value of the Wilson loop is computed by the string path integral, which in the limit $\lambda \gg 1$ is dominated by the saddle point where the string has minimal area.\footnote{
See section 2 of \cite{Giombi:2022pas} for a recent pedagogical review of a similar setup.}
Here we consider small fluctuations around the minimal-area configuration.

From now on, we restrict our attention to the case when $\gamma$ is a straight line, a configuration that preserves the $OSp(4^*|4)$ group, recall equation \eqref{eq:osp44}.
More precisely, the string intersects the boundary of $\AdS_5$ along the $x^1$ direction, and is located at $x^{i=2,3,4} = 0$.
Furthermore, we choose the string to be located at point $y^A = 0$ on $S^5$.
It is convenient to work in static gauge, in which case the embedding reads
\begin{align}
 X^0
 = \wh z^{\,0} \, , \quad
 X^1
 = \wh z^{\,1} \, ,  \quad
 X^{i=2,3,4}
 = X^{A=5,\ldots,9}
 = 0 \, .
 \label{eq:static}
\end{align}
This embedding represent the static configuration of the string. However, we are interested in allowing small fluctuations around this configuration. In particular, we allow the orthogonal directions $X^{i}$ and $X^{A}$ to fluctuate, which can be implemented in a change of the boundary conditions \eqref{eq:static}.
Motivated by \cite{Giombi:2017cqn},\footnote{To be precise, the authors of \cite{Giombi:2017cqn} split the coordinates as $z^\mu = (z^\a, z^i)$ for $\a = 0,1$ and $i = 2,3,4$. They used the metric
\begin{align}
 ds^2_{\AdS_5 \times S^5}
 = \left(\frac{1 + \frac14 (z^i)^2}{1 - \frac14 (z^i)^2} \right)^2 \,
   \frac{dz^\a dz^\a}{(z^0)^2}
 + \frac{dz^i dz^i}{(1 - \frac14 (z^i)^2)^2}
 + \frac{dy^A dy^A}{(1 + \frac14 y^B y^B)^2} \, ,
\end{align}
and chose static gauge with $X^\a = \wh z^\a$, $X^i = x^i$ and $X^A = y^A$. This is equivalent to our choice \eqref{eq:embedd-string} because we work in Poincare coordinates, see \eqref{eq:met-AdS5-S5}.} we require the string to be embedded as
\begin{align}
\label{eq:embedd-string}
 X^0
 = \wh z^{\,0} \, \frac{1 - \frac14(x^i)^2}{1 + \frac14(x^i)^2} \, , \quad
 X^1
 = \wh z^{\,1} \, ,  \quad
 X^i
 = \frac{x^i \wh z^{\,0}}{1 + \frac14(x^i)^2} \, , \quad
 X^A
 = y^A \, .
\end{align}
For $x^i = y^A = 0$ this naturally reduces to the static configuration \eqref{eq:static}.
However, we instead treat $x^i(\wh z)$ and $y^A(\wh z)$ as fields that live on the worldsheet, and expand \eqref{eq:NG} around the background configuration to find their action.
The fields $x^i$ and $y^A$ have standard kinetic terms, as well as self-interactions and interactions with bulk fields.
The general structure of the action is
\begin{align}
 S_{\text{F1}}
&= \frac{\sqrt{\lambda}}{2\pi} \int \frac{d^2 \wh z}{(\wh z^{\,0})^2} \,
   \Big(
     1
     + L_{\text{F1}}^{(2,0)}
     + L_{\text{F1}}^{(0,1)}
     + L_{\text{F1}}^{(1,1)}
     + L_{\text{F1}}^{(0,2)}
     + \ldots
   \Big) \, ,
 \label{eq:actF1}
\end{align}
where $L_{\text{F1}}^{(m,n)}$ corresponds to terms with $m$ worldsheet fields and $n$ bulk fields.
For the kinetic part, we find
\begin{align}
 L_{\text{F1}}^{(2,0)}
 = \frac12 g_2^{\a\b} \partial_\a x^i \, \partial_\b x^i
 + x^i x^i
 + \frac12 g_2^{\a\b} \partial_\a y^A \, \partial_\b y^A \, , \qquad
 g_2^{\a\b} = \delta^{\a\b} (\wh z^{\,0})^2 \, .
\end{align}
This result is in perfect agreement with \cite{Giombi:2017cqn}.
Similarly to the bulk case, these fields are part of a short multiplet of $OSp(4^*|4)$, with quantum numbers reported in table \ref{tab:xy}.
In defect CFT terminology, the field $y^A$ is dual to the tilt operator, which appears in the Ward identity for the broken $R$-symmetry $SO(6)_R \to SO(5)_R$ \cite{Cuomo:2021cnb,Padayasi:2021sik}.
Similarly, the field $x^i$ is dual to the displacement operator, which appears in the Ward identity for the the broken translation symmetry \cite{Billo:2016cpy}.

\begin{table}
 \centering
 \begin{tabular}{c||c|c|c|c}
  ~Field~ & $\;\;m^2\;\;$ & $\;\;\Delta\;\;$ & $\,SO(3)_{\text{trans.}}\,$ & $\,\;SO(5)_R\;\,$ \\ \hline \hline
  $y^A$ & 0 & 1 & singlet & vector \\ \hline
  $x^i$ & 2 & 2 & vector & singlet
 \end{tabular}
 \caption{Bosonic open-string modes on the fundamental string worldsheet.}
 \label{tab:xy}
\end{table}

One can proceed to extract the rest of couplings in a similar manner.
For example, reference \cite{Giombi:2017cqn} computed all quartic vertices of worldsheet fields, in our notation $L_{\text{F1}}^{(4,0)}$.
These terms do not contribute to the two-point function of chiral-primary operators at leading order, so we do not present the formulas here.
The first vertex of interest couples a bulk field directly to the string worldsheet
\begin{align}
 L_{\text{F1}}^{(0,1)}
&= \sum_{k,I}
   \left(
      \frac12 (\varphi_k^I)_\a^\a
    - \frac{2 k (k-1)}{k+1} s_k^I
    - \frac{2 (k+4) (k+5)}{k+3} t_k^I
     + \ldots
   \right) Y^I_k \, .
 \label{eq:terms01}
\end{align}
These vertices allow bulk fields to be ``absorbed'' by the string, recall \eqref{eq:example-one-pt}.
The vertex for $s_k^I$ has appeared multiple times in the literature, see e.g. \cite{Berenstein:1998ij,Giombi:2006de,Giombi:2009ds}, but to the best of our knowledge, the vertices for $\varphi$ and $t$ are new.
In \eqref{eq:terms01} we neglect total derivatives $\partial_\a V^\a$, since they drop from Witten diagrams.
It is also implicit in the notation that the spherical harmonic is evaluated at zero $Y_k^I = Y_k^I(y=0)$, and the bulk fields live on the worldsheet, for example $s_k^I = s_k^I(\wh z^{\,0}, \wh z^{\,1},0,0,0)$.

Finally, we need vertices where the bulk field $s$ excites a worldsheet fluctuation.
These vertices have not appeared before in the literature, and a calculation exactly as above gives
\begin{align}
 L_{\text{F1}}^{(1,1)}
 = - \sum_{k,I} \frac{2k(k-1)}{k+1} \Big(
      \, y^A \partial_A
   + \wh z_0 \, x^i \partial_i \,
   + \ldots
 \Big) s^I_k Y^I_k \, .
 \label{eq:blkdef}
\end{align}
To obtain this formula we used multiple integration by parts, and dropped total derivatives and terms proportional to the equations of motion (EOM).
When computing the correlator $\vev{S_{p_1} S_{p_2} W}$, the EOM terms will contribute contact diagrams like the one in \eqref{eq:def-cont-diag}.
However, in section \ref{eq:explcalc} we neglect all contact diagrams, and only fix them later with the bootstrap approach of section \ref{sec:pos-space-boot}.
A more careful first-principle calculation, that aims at computing the correlator including contact terms, should keep track of the EOM terms in \eqref{eq:blkdef}.

\subsection{Explicit calculation (up to contact terms)}
\label{eq:explcalc}

\begin{figure}
 \begin{align*}
   \vev{ S_{p_1} S_{p_2} W }_{\text{c}}
   \; = \;
   \sum_{\Om = \{s_k, \varphi_k, t_k \} } \;\;
   \begin{tikzpicture}[baseline={([yshift=-.55ex]current bounding box.center)},scale=1.2]
    \pgfmathsetmacro{\x}{-sqrt(2)/2}
    \pgfmathsetmacro{\y}{sqrt(2)/2}
    \pgfmathsetmacro{\z}{-0.5}
    \pgfmathsetmacro{\r}{0.4}
    \draw [thick] (0,0) circle [radius=1];
    \draw [thick, blue] (-\x,+\y) to[out=-120,in=120] (-\x,-\y);
    \draw [dashed]      (+\x,+\y) -- (-\r,0);
    \draw [dashed]      (+\x,-\y) -- (-\r,0);
    \draw [dashed]      (-\r,  0) -- (-\z,0);
    \node at (-0, 0.2) {{\footnotesize $\Om$}};
    \node at (-0.95, 0.85) {$s_{p_1}$};
    \node at (-0.95,-0.85) {$s_{p_2}$};
    \end{tikzpicture}
    \;\; +
    \begin{tikzpicture}[baseline={([yshift=-.55ex]current bounding box.center)},scale=1.2]
    \pgfmathsetmacro{\x}{-sqrt(2)/2}
    \pgfmathsetmacro{\y}{sqrt(2)/2}
    \pgfmathsetmacro{\z}{-0.5}
    \pgfmathsetmacro{\rx}{-0.6}
    \pgfmathsetmacro{\ry}{+0.45}
    \draw [thick] (0,0) circle [radius=1];
    \draw [dashed]      (+\x,+\y) -- (-\rx,\ry);
    \draw [dashed]      (+\x,-\y) -- (-\rx,-\ry);
    \draw [dashed,blue] (-\rx,\ry) to[out=210,in=-210] (-\rx,-\ry);
    \draw [thick, blue] (-\x,+\y) to[out=240,in=-240] (-\x,-\y);
    \node at (-0.95, 0.85) {$s_{p_1}$};
    \node at (-0.95,-0.85) {$s_{p_2}$};
    \node at (0,0) {\footnotesize $x,y$};
    \end{tikzpicture}
    \;\; +
    \begin{tikzpicture}[baseline={([yshift=-.55ex]current bounding box.center)},scale=1.2]
    \pgfmathsetmacro{\x}{-sqrt(2)/2}
    \pgfmathsetmacro{\y}{sqrt(2)/2}
    \pgfmathsetmacro{\z}{-0.5}
    \draw [thick] (0,0) circle [radius=1];
    \draw [thick, blue] (-\x,+\y) to[out=240,in=-240] (-\x,-\y);
    \draw [dashed]      (+\x,+\y) -- (-\z,0);
    \draw [dashed]      (+\x,-\y) -- (-\z,0);
    \node at (-0.95, 0.85) {$s_{p_1}$};
    \node at (-0.95,-0.85) {$s_{p_2}$};
    \end{tikzpicture}
    \;\; + \;\; \ldots
  \end{align*}
  \caption{Diagrams that contribute to the connected correlator $\vev{ S_{p_1} S_{p_2} W }_{\text{c}}$ at leading order in the supergravity limit. The quantum numbers of exchanged fields appear in tables \ref{tab:stp} and \ref{tab:xy}.}
  \label{fig:SSW}
\end{figure}
Having derived the $\AdS_5$ and $\AdS_2$ effective actions, we are ready for a first attempt to calculate $\vev{S_{p_1} S_{p_2} W}_c$.
Recall that we are interested in the connected correlator, so we neglect disconnected diagrams as in \eqref{eq:example-two-pt-conn}.
Figure \ref{fig:SSW} shows the connected diagrams in the leading supergravity approximation, which consist of exchange diagrams with $s$, $\varphi_{\mu\nu}$ and $t$ in $\AdS_5$, exchange diagrams with $x$, $y$ in $\AdS_2$, and contact diagrams.
The quantum numbers of the exchanged fields are summarized in tables \ref{tab:stp} and \ref{tab:xy}, while contact interactions are ignored in this section for reasons to be explained later.

To calculate $\vev{ S_{p_1} S_{p_2} W }$, one should view the on-shell supergravity action with certain boundary conditions as the generating functional of CFT correlators \cite{Gubser:1998bc,Witten:1998qj}, and then take functional derivatives.
In practice, however, this amounts to computing the correlator with standard position-space Feynman rules, using propagators \eqref{eq:blk-blk-G}-\eqref{eq:K-blk-bdy} and the effective actions of section \ref{sec:effect-act}.
For the bulk-to-boundary propagator, we choose the normalization \cite{Berenstein:1998ij}
\begin{align}
 \Pi_p(x,z)
 = \Nm_p K_p(x,z) \, , \qquad
 \Nm_p
 = 2^{p/2-2} \, \frac{p+1}{N \sqrt{p}} \, ,
\end{align}
which ensures that in the absence of the defect, the two-point function $\vev{ S_p S_p}$ is unit normalized.
Besides this, there are several numerical factors that need to be accounted for.
For example, because of \eqref{eq:actIIB}, bulk-to-bulk propagators come with a factor $\frac{(2\pi)^5}{4N^2 \zeta}$, where the normalizations $\zeta$ are shown in \eqref{eq:zeta-def}.
Similarly, brane-to-brane propagators come with a factor $\frac{2\pi}{\sqrt{\lambda}}$, see \eqref{eq:actF1}.
Regarding bulk three-point vertices, they contain an overall factor $\frac{4N^2}{(2\pi)^5}$.
There are three bulk vertices $X = S, \, T, \, G$, which come with extra factors $6$, $2$ and $2$ respectively, that count the number of different Wick contractions.
Each three-point vertex is of the form
\begin{align}
 X_{p_1p_2k}^{IJK} Y_k^I(0)
 \; \to \;
 X_{p_1p_2k} h_{p_1p_2}^k(\sigma) \, ,
 \label{eq:Rsym-contr}
\end{align}
where $X_{p_1p_2k}$ are found in \eqref{eq:threept-coup}.
In equation \eqref{eq:Rsym-contr} we have mapped the spherical harmonic indices $I,J,K$ to the index-free notation of section \ref{sec:setupN4}.
The derivation of this map is presented in appendix \ref{sec:spherical}, but here it suffices to note that the result is exactly the $R$-symmetry block
\begin{align}
\label{eq:Rsymblock}
 h^{k}_{p_1 p_2}(\sigma)
 = \sigma ^{\frac{p_1+p_2-k}{2}}
  \, _2F_1 \! \left(-\frac{p_{12}+k}{2},\frac{p_{12}-k}{2};-k-1; \frac{\sigma}{2}\right) \, .
\end{align}
For integer values of $p_1$, $p_2$ and $k$ the $R$-symmetry block is a polynomial in $\sigma$.
Finally, the bulk-boundary vertices are proportional to $\frac{\sqrt{\lambda}}{2\pi}$, with the precise values in \eqref{eq:terms01}.
As before, appendix \ref{sec:spherical} translates the $R$-symmetry contractions to index-free notation
\begin{align}
 Y_{p_1}^I(0) Y_{p_2}^J(0)
 \; \to \;
 1  \, , \quad
 \partial_A Y_{p_1}^I(0) \partial_A Y_{p_2}^J(0)
 \; \to \;
 p_1 p_2 (\sigma-1)  \, .
\end{align}
All in all, we have the following contributions to the connected correlator
\begin{align}
 \Fm_c(\xi,\eta,\sigma)
&\;\sim\; \Nm_{p_1} \Nm_{p_2} \frac{\sqrt{\lambda}}{2\pi} \Bigg[
     \sum_{k}
     \frac{6 S_{p_1p_2k}}{\zeta^s_k}
     \frac{2k(k-1)}{k+1} E_{p_1p_2}^{k,0} h_{p_1p_2}^{k} \notag \\
& \quad
   + \sum_{k}
     \frac{2 T_{p_1p_2k}}{\zeta^t_k}
     \frac{2(k+4)(k+5)}{k+3} E_{p_1p_2}^{k+8,0} h^{k}_{p_1p_2}
   + \sum_{k}
     \frac{2G_{p_1p_2k}}{\zeta^\varphi_k} \frac12
     E_{p_1p_2}^{k+4,2} h^{k}_{p_1p_2} \notag \\
&  \quad
   + \frac{4 p_1 p_2(p_1-1)(p_2-1)}{(p_1+1)(p_2+1)} \left(
     p_1 p_2 (\sigma-1) \wh E_{p_1p_2}^{1,0} + \wh E_{p_1p_2}^{2,1}
 \right)\Bigg] \, ,
 \label{eq:tentative-corr}
\end{align}
where the sum runs over $k$'s allowed by $SU(4)_R$ symmetry.
The couplings $S$, $T$, $G$ and $\zeta$ are found in appendix \ref{sec:cubic-coupl}, while $E_{\Delta_1\Delta_2}^{\Delta,\ell}$ and $\wh E_{\Delta_1\Delta_2}^{\Dh,s}$ are bulk-exchange and brane-exchange Witten diagrams respectively, which we computed in section \ref{sec:pos}.

Combining all these ingredients, one can check that \eqref{eq:tentative-corr} cannot be the correct result, because it does not obey the superconformal Ward identity \eqref{eq:WI}.
The reason \eqref{eq:tentative-corr} is incorrect is that we have completely ignored contact diagrams.
Instead, the correct result takes the form
\begin{align}
 \Fm_c(\xi,\eta,\sigma)
 \,=\, \eqref{eq:tentative-corr}
 \,+\; C_{p_1p_2}(\xi,\eta) \sum_{n=0}^{\pmin} c_n \, \sigma^n \, .
 \label{eq:tentative-corr-conts}
\end{align}
Unfortunately, starting from the IIB and fundamental string actions, we have been unable to determine coefficients $c_n$ that generate correlators consistent with the Ward identities.
The naive way to obtain $c_n$ is to expand the string action up to terms like $\sim \int_{\AdS_2} s_{p_1} s_{p_2}$, and keep track of the EOM terms neglected in \eqref{eq:blkdef}.
Because of their $R$-symmetry structure, these terms can contribute to $c_{0}$ and $c_{1}$, but as we show below, all $c_n$ are non-vanishing in the supersymmetric correlator.
This suggests there are other sources of contact terms, and raises the question of how to do a first-principles calculation that correctly accounts for them.
One possibility is that in the derivation of the IIB effective action, one generates the missing contact terms when integrating by parts or in doing certain field redefinitions.
We hope future work will clarify this question, while in the rest of the paper we determine $c_n$ with a bootstrap method.

\subsection{Position space bootstrap}
\label{sec:pos-space-boot}

Although we got far using traditional methods, in the end we failed to find a correlator consistent with the supersymmetry of the system.
To solve this problem, in this section we determine the correlator with a bootstrap method.
The main idea was put forth in \cite{Rastelli:2017udc}: write an ansatz with all Witten diagrams allowed by the effective action, and fix the relative coefficients with the superconformal Ward identity.

Let us spell out the method in detail.
First, recall there are bulk-exchange diagrams with the fields $s$, $\varphi$ and $t$ in table \ref{tab:stp}.
These fields belong to a superconformal multiplet, where the superprimary $s$ transforms in the representation $[0,k,0]$ of $SU(4)_R$.
The value of $k$ is restricted by $SU(4)_R$ symmetry to be $|p_{12}| \le k \le p_1 + p_2$ and to run in steps of two.
Furthermore, when the superprimary satisfies $k = p_1 + p_2$ or $k = |p_{12}|$, the three-point vertices $S$, $T$, $G$ in \eqref{eq:threept-coup} vanish, so we take $|p_{12}| < k < p_1 + p_2$.
As a result, each multiplet $(s,\varphi,t)$ in table \ref{tab:stp} contributes for $k \in \Sm$, where
\begin{align}
 \Sm =
 \Big\{ |p_{12}|+2, |p_{12}|+4, \ldots, p_1 + p_2 - 4, p_1 + p_2 - 2 \Big\} \, .
 \label{eq:krange}
\end{align}
This concludes the discussion of bulk-exchange diagrams.

Regarding defect-exchange diagrams, the fields $x$, $y$ in table \ref{tab:xy} always contribute, because they are always allowed by $R$-symmetry.
Finally, there can be contact diagrams, so the full ansatz for the connected correlator is\footnote{One issue we glossed over is what contact diagrams to include in the ansatz. In fact, \eqref{eq:pos-ansatz} ignores contact diagrams with derivatives, which would come from $\int_{\AdS_2} \nabla^n s_1 \nabla^m s_2$.
One justification is that spin-two diagrams $E^{\Delta,2}_{\Delta_1\Delta_2}$ and contact diagrams $C_{\Delta_1\Delta_2}$ have the same Regge behavior.
One the other hand, derivative contact diagrams grow faster in the Regge limit, and because we do not expect the Regge limit to be controlled only by contact diagrams, they should be absent.
We thank Xinan Zhou for this argument.
}
\begin{align}
 \Fm_c(\xi,\eta,\sigma)
&= \sum_{k\in\Sm} \Big(
     A^s_k       \, E_{p_1p_2}^{k,0}(\xi,\eta)  \, h_{p_1p_2}^k(\sigma)
   + A^\varphi_k \, E_{p_1p_2}^{k+2,2}(\xi,\eta)\, h_{p_1p_2}^{k-2}(\sigma)
   \notag \\[-0.5em]
& \qquad \qquad
   + A^t_k       \, E_{p_1p_2}^{k+4,0}(\xi,\eta)\, h_{p_1p_2}^{k-4}(\sigma)
 \Big)
 +  C_{p_1p_2}(\xi,\eta) \sum_{n=0}^{\pmin} A_n^{ss} \, \sigma^n \notag \\
&\quad
 + A^y \, \wh E_{p_1p_2}^{1,0}(\xi,\eta) (\sigma-1)
 + A^x \, \wh E_{p_1p_2}^{2,1}(\xi,\eta) \, ,
 \label{eq:pos-ansatz}
\end{align}
with the sum ranges in \eqref{eq:krange} and $A$'s numerical coefficients to be determined.
The bulk-exchange diagrams $E_{\Delta_1\Delta_2}^{\Delta,\ell}$ can be evaluated as sums of contact diagrams, see \eqref{eq:blkexch-conts} and appendix \ref{app:spintwo}.
Furthermore, contact diagrams are known in closed form \eqref{eq:cont-form}.
Finally, the defect-exchange diagrams $\wh E_{\Delta_1\Delta_2}^{\Dh,s}$ can be computed from the differential equation \eqref{eq:diffeq-Eh}.

For a given choice of $p_1$ and $p_2$, we build the ansatz as just explained and impose the superconformal Ward identity \eqref{eq:WI} order by order in an expansion around $r=0$.
The resulting equations can be solved, and they fully determine the correlator \eqref{eq:pos-ansatz} up to an overall normalization, which of course cannot be fixed by the Ward identity only.
In order to fix the overall normalization, note that in the strong 't Hooft coupling limit we have \cite{Barrat:2021yvp}
\begin{align}
 \lim_{\lambda\to\infty} \lim_{N\to\infty}
 \lambda_{S_{p_1} S_{p_2} S_{k}} a_{S_k}
 = \frac{\sqrt{\lambda}}{N^2}
   \frac{\sqrt{p_1 p_2} \, k}{2^{k/2+1}}
 + \ldots \, ,
 \label{eq:CFT-data}
\end{align}
where $a_\Om \propto \vev{\Om W}$ and $\lambda_{\Om_1\Om_2\Om_3} \propto  \vev{\Om_1\Om_2\Om_3}$.
The Euclidean OPE limit $\xi \to 0$, $\eta \to 1$ is dominated by the lowest-dimension chiral primary operator $S_{k_{\text{min}}}$, and we find
\begin{align}
 \lim_{\xi \to 0} \lim_{\eta \to 1} \Fm_c(\xi,\eta,\sigma)
 = \frac{\sqrt{\lambda}}{N^2}
   \frac{\sqrt{p_1 p_2} \, k_{\text{min}}}{2^{k_{\text{min}}/2+1}} \,
   \xi^{\frac{k_{\text{min}}-p_1-p_2}{2}} \,
   h_{p_1p_2}^{k_{\text{min}}}(\sigma) + \ldots \, .
\end{align}
Comparing this to ansatz \eqref{eq:pos-ansatz} fixes the overall normalization of $\Fm_c$.

We bootstrapped many correlators with $p_1,p_2 \lesssim 30$, and in every case all coefficients were fixed.
All cases with $p_1,p_2 \le 4$ were computed previously in \cite{Barrat:2021yvp,Barrat:2022psm}, and we found perfect agreement.\footnote{The careful reader will note that our correlator for $(p_1,p_2) = (2,4)$ differs with the one in \cite{Barrat:2022psm}. The difference is only the contribution $\lambda_{242} a_2 f_{2,0}^{-2}$ in the bulk OPE. The reason is that \cite{Barrat:2022psm} considered the two-point function of single-trace operators $\Om_p$, for which $\Om_2 \times \Om_4 \sim \Om_2$. However, we work in the basis natural for supergravity, such that $\lambda_{S_{2} S_{4} S_{2}} = 0$, recall footnote \ref{foot:sugra}.}
At this point, it is interesting to compare our method to the analytic bootstrap in \cite{Barrat:2021yvp,Barrat:2022psm}.
The analytic bootstrap calculation reconstructed the correlator $\Fm_c$ using a dispersion relation or inversion formula.
These formulas suffer from low-spin ambiguities, which in practice means that only part of the correlator is reconstructed.
In the present language, the dispersion relation reconstructs only the sum over bulk-exchange diagrams, because both defect-exchange and contact diagrams have vanishing discontinuity.
In \cite{Barrat:2021yvp,Barrat:2022psm}, the missing part of the correlator was determined making a very general ansatz and fixing it from superconformal Ward identities.
Remarkably, this ad-hoc procedure managed to reconstruct precisely the defect-exchange diagrams for the $x^i$ and $y^A$ fluctuations.
Needless to say, the present method is superior to the one employed in \cite{Barrat:2021yvp,Barrat:2022psm}.
On one hand, here we have a more clear physical picture of all the terms that form the correlator.
On the other, the present method is easier and faster, thanks to the efficient methods to compute Witten diagrams.

Because of the efficiency of our method, it is possible to generate many correlation functions, and by looking at all the examples, we could guess a closed formula
\begin{align}
 \Fm_c(\xi,\eta,\sigma)
&= \frac{\sqrt{\lambda}}{N^2}
   \sum_{k\in\Sm} \frac{\sqrt{p_1 p_2} \, k}{2^{\frac{k}{2}+1}} \Bigg(
   P^{k,0}_{p_1p_2} h^k_{p_1p_2}
 + \frac{(p^2_{12}-k^2) \big(p^2_{12}-(k+2)^2\big)}
        {128 k (k+1)^2 (k+3)}
        P^{k+2,2}_{p_1p_2} h^{k-2}_{p_1p_2} \notag \\
& \qquad \qquad \quad
 + \frac{(p_{12}^2-(k+2)^2) (p_{12}^2-(k-2)^2)
         (p_{12}^2-k^2)^2}
        {16384 (k-2) (k-1)^2 k^2 (k+1) (k+2) (k+3)}
   P^{k+4,0}_{p_1p_2} h^{k-4}_{p_1 p_2}
 \Bigg) \notag \\
& \quad
  + \frac{\sqrt{\lambda}}{N^2}
   \frac{(p_1 p_2)^{\frac12} (p_1-1) (p_2-1)}{2^{3-\frac{p_1+p_2}{2}} \pi } \bigg(
    p_1 p_2 (\sigma-1) \wh E_{p_1p_2}^{1,0}
    + \wh E_{p_1,p_2}^{2,1}
    + (1-2\sigma) C_{p_1p_2}
 \bigg) \, .
 \label{eq:pos-res}
\end{align}
The sum over $k$ is described in \eqref{eq:krange}, and $P_{\Delta_1\Delta_2}^{\Delta\ell}$ are Polyakov-Regge blocks.
Recall that, up to contact terms, Polyakov-Regge blocks are bulk-exchange Witten diagrams with a different normalization, see \eqref{eq:polyregge}.
The bootstrap result \eqref{eq:pos-res} agrees with the explicit calculation \eqref{eq:tentative-corr}, up to the contact terms that we could not determine in the first-principles calculation.
This provides a strong consistency check for our results.
Let us stress that to obtain \eqref{eq:pos-res}, the only input from section \ref{sec:effect-act} is the field content in the effective actions, but not the precise value of vertices.

The position-space correlator \eqref{eq:pos-res} has a simple interpretation in terms of conformal blocks.
Because Polyakov-Regge blocks are normalized as $P_{\Delta_1\Delta_2}^{\Delta\ell} = \xi^{-\frac{\Delta_1+\Delta_2}{2}} f_{\Delta,\ell} + \ldots$, the coefficient of $P_{p_1p_2}^{k,0} h^k_{p_1p_2}$ in \eqref{eq:pos-res} is the OPE coefficient $\lambda_{S_{p_1} S_{p_2} S_k} a_{S_{k}}$, in perfect agreement with the known value \eqref{eq:CFT-data}.
Also note that $\varphi_{\mu\nu}$ and $t$ are superdescendants of $s$, so their OPE coefficients are fixed in terms of the superprimary.
This is also explicit in the correlator, because the relative coefficients of the first two lines of \eqref{eq:pos-res} precisely form a superblock (compare with equation (5.12) of \cite{Barrat:2022psm}).
Finally, the last line of \eqref{eq:pos-res} contains bulk-defect OPE coefficients of chiral-primaries $S_k$ with tilt $\wh t$ and displacement $\wh D^i$ operators.
To present the OPE coefficients, we use index-free notation $t(0,\wh v) = t^A(0) \wh v_A$, with the polarization vector satisfying $\wh v^{\,2} = \wh v \cdot \theta = 0$, and we find\footnote{We use the tilt and displacement operators with unit-normalized two-point functions. As a result, the Ward identities that define these operators have extra normalization factors
\begin{align}
 \partial_\mu T^{\mu i} = C_D^{1/2} \, \wh D^i \, \delta_W \, , \qquad
 \partial_\mu J^{\mu}_A = C_t^{1/2} \, \wh t_A \, \delta_W \, .
\end{align}
}
\begin{align}
 \vev{ S_p(x,u) \, \wh t(0,\wh v) \, W(\theta)}
&= \frac{(u \cdot \theta)^{p-1} (u\cdot \wh v)}{x^2 |x^i|^{p-1}}
   \left(
   \frac{\lambda^{1/4}}{N} \frac{p^{3/2}}{2^{\frac{p+1}{2}}}
 + \ldots \right) \, , \\
 \vev{ S_p(x,u) \, \wh D^i(0) \, W(\theta)}
&= \frac{x^i (u \cdot \theta)^p}{x^4 |x^j|^{p}}
   \left(
   \frac{\lambda^{1/4}}{N} \frac{p^{3/2}}{2^{\frac{p}{2}} \sqrt{3}}
 + \ldots
   \right) \, .
\end{align}
To obtain this formula, we used the normalization of defect-exchange Witten diagrams in \eqref{eq:normEh}.
The correlator $\vev{ S_p \, \wh t \, W }$ is captured by the topological sector of \cite{Giombi:2018qox,Giombi:2018hsx}, and we find perfect agreement.
Summarizing, the correlator \eqref{eq:pos-res} is a sum over protected operators, each of them contributing an exchange Witten diagram, supplemented by extra contact diagrams fixed by superconformal Ward identities.
Besides these highly non-trivial sanity checks, in appendix \ref{sec:topsect} we successfully compare to the localization calculation of \cite{Giombi:2012ep}, which captures only the topological part of the correlator.

\subsection{Mellin space}

There are at least three good reasons to map the above results to Mellin space.
First, it would be good to have a bootstrap method directly in Mellin space, since this could be necessary in setups where the position-space bootstrap does not work.
For example, bulk-exchange diagrams do not truncate in theories with an $\AdS_4$ bulk, making the position space bootstrap unfeasible.
Second, the Mellin amplitude $\Mm$ should be closely related to a flat-space scattering amplitudes of supergravitons off a string or brane, see \cite{Pufu:2023vwo} for a recent discussion.
By making this link more precise, one could input information of flat-space amplitudes to bootstrap subleading corrections in $\frac{1}{\sqrt{\lambda}}$, see similar examples \cite{Goncalves:2014ffa,Binder:2019jwn,Chester:2020dja}.
Third, thanks to the properties of Mellin amplitudes, it is plausible the result might be simpler in Mellin space, a hypothesis that we confirm below.

Keeping this motivation in mind, we Mellin transform the position-space correlator \eqref{eq:pos-res}.
The first two lines in \eqref{eq:pos-res} contain Polyakov-Regge blocks, which are combinations of bulk-exchange and contact diagrams, see \eqref{eq:polyregge}.
The Mellin transforms of bulk-exchange and contact diagrams are presented in \eqref{eq:mell-blk} and \eqref{eq:mellin-cont} respectively, and they always reduce to rational functions of the Mellin variables $\delta$ and $\rho$.
The last line in \eqref{eq:pos-res} contains defect-exchange diagrams, that in Mellin space do not always reduce to rational functions, recall \eqref{eq:mell-def}.
For example, for $p_1 = p_2 = 2$, the Mellin amplitude reads
\begin{align}
 \Mm(\delta,\rho,\sigma)
 = \frac{\sqrt{\lambda}}{8N^2} \Bigg( &
   \frac{(\rho \sigma - \sigma + 4) \sigma }{\delta -1}
 - \frac{2\rho + 4 \sigma - 4}{\delta +\rho}
 + \frac{\rho }{\delta +\rho +2}
 + (\sigma -1)^2  \notag \\
&+ \frac{2^{\delta +\rho + 2} \, \Gamma (-\delta-\rho)}
        {\Gamma\!\left(\frac{2-\delta-\rho }{2}\right)^2}
   \left(\frac{\delta +2}{\delta +\rho +2}-\sigma \right)
 \Bigg) \, .
 \label{eq:mell22}
\end{align}
In this case, the defect-exchange amplitudes $\wh\Mm_{2,2}^{1,0}$ and $\wh\Mm_{2,2}^{2,1}$ simplify in terms of gamma functions, but we have not checked if this holds for arbitrary $p_1,p_2>2$.

In a similar way, we can Mellin transform the correlator for many values $p_1$, $p_2$, and guess a closed formula for $\Mm$.
We observe that the Mellin transform of the first two lines in \eqref{eq:pos-res} simplifies drastically, but there are no important simplifications in the last line:
\begin{align}
 \Mm(\delta,\rho,\sigma)
&= \sum_{n=1}^{\pmin} \sum_{i=1}^n
   \frac{ b_{n,i} \, \rho + c_{n,i}}{\delta-i} \, \sigma^n
 + e \, p_1 p_2 (\sigma-1) \wh \Mm_{p_1p_2}^{1,0}
 + e \wh \Mm_{p_1p_2}^{2,1}
 + a \, (\sigma-1)^2 \, .
 \label{eq:mell-res}
\end{align}
The numerical constants $a$, $b_{n,i}$, $c_{n,i}$ and $e$ were guessed in closed form, and take a relatively simple form:
\begin{align}
 a
 & =
   - \frac{\sqrt{\lambda}}{N^2} \,
     \frac{2^{\frac{p_1+p_2}{2}} \sqrt{p_1 p_2}
           \left(-\frac{1}{2}\right)_{\frac{p_1+p_2}{2}}}
          {16 (p_1-2)! (p_2-2)!} \, , \notag \\
 b_{n,i}
 & = a \,
     \frac{(2-n)_{i-1} (2-p_1)_{n-2} (2-p_2)_{n-2}}
          {2^{n-2} (i-1)! (n-2)!
           \left(\frac{3-p_1-p_2}{2}\right)_{n-2}} \,, \notag \\
 c_{n,i}
 & = \frac{b_{n,i}}{4} \left(\frac{p_{12}^2-1 - 8 (n-p_1) (n-p_2) (2 n-p_1-p_2)/(i-n)}{p_1+p_2-2 n+1}+4 i+6 n-5p_1-5p_2+1\right) \, , \notag \\
 e
 & = \frac{\sqrt{\lambda}}{N^2} \,
     \frac{2^{\frac{p_1+p_2}{2}}}{8 \pi} \,
     (p_1 p_2)^{1/2} (p_1-1) (p_2-1) \, .
\end{align}
The formula for $c_{n,i}$ is singular for $n=i$, but the value of $c_{n,n}$ in \eqref{eq:mell-res} is finite and given by
\begin{align}
 c_{n,n}
 & = a \, \frac{\left(\frac{p_1+p_2}{2}-n\right)
           (2-p_1)_{n-1} (2-p_2)_{n-1}}
          {(-2)^{n-3} (n-1)! \left(\frac{3-p_1-p_2}{2}\right)_{n-1}}
 \, .
\end{align}

The Mellin amplitude \eqref{eq:mell-res} has a simple interpretation in terms of the OPE.
The finite sum captures contributions of single-trace operators in the OPE $S_{p_1} \times S_{p_2} \sim S_k + \ldots$, compare with the general formula \eqref{eq:blkpoles}.
Similarly, the last two terms correspond to the exchange of tilt and displacement in the defect expansion $S_p \times W \sim \wh t \, W + \wh D^i W + \ldots$.
On the other hand, we do not have a clear interpretation for the term $a(\sigma-1)^2$, which should correspond to a contact interaction, because it is constant.

The result \eqref{eq:mell-res} deserves several comments.
First, it would be interesting to find the prescription for taking the flat-space limit and comparing to flat-space string amplitudes.
Second, the sum in \eqref{eq:mell-res} is remarkably simple, and is reminiscent of the elegant formula found in \cite{Rastelli:2016nze,Rastelli:2017udc} for the four-point function $\vev{S_{p_1} S_{p_2} S_{p_3} S_{p_4}}$.
In that case, the simplicity is related to the hidden ten-dimensional conformal symmetry \cite{Caron-Huot:2018kta}, and it would be fascinating if a similar structure can be found here.
Finally, it is unclear to us whether \eqref{eq:mell-res} could have been bootstrapped directly in Mellin space.
In any case, if such bootstrap approach is possible, it certainly requires superconformal Ward identities in Mellin space, a subject that we now discuss.

\subsubsection{Ward identity}

Superconformal Ward identities were first implemented in Mellin space in \cite{Zhou:2017zaw}, see also \cite{Alday:2020dtb,Alday:2021odx}.
We adapt these methods to the present setup in appendix \ref{sec:wi-mell}.
The result is that given the decomposition \eqref{eq:invariant} of the correlation function, the $R$-symmetry channels satisfy
\begin{align}
 \sum_{j=0}^{\pmin} \left[
 \zeta_{\pm}^{(j)} \big(
     4 \xi \partial_\xi
   - \xi \partial_\eta
   + 4 j \big)
 +\zeta_\pm^{(j+1)} (\xi \partial_\xi + \eta \partial_\eta +j)
 \right] \frac{\Fm_j(\xi,\eta)}{(-2)^j}
 = 0 \, ,
 \label{eq:WImellin}
\end{align}
where $\zeta_{+}^{(i)}$ are polynomials of the cross-ratios
\begin{subequations}
\label{eq:zetaPM}
 \begin{align}
  \zeta_{+}^{(i)}
& = \sum _{j=0}^{\left\lfloor i/2\right\rfloor } \binom{i}{2 j}
  \left(\eta ^2-1\right)^j \left((2 \eta +\xi )^2-4\right)^j
  (2 \eta^2 +\xi \eta-2)^{i-2 j} \, , \\
 \zeta_{-}^{(i)}
& = \sum _{j=0}^{\left\lfloor i/2\right\rfloor } \binom{i}{2 j+1}
  \left(\eta ^2-1\right)^j \left((2 \eta +\xi )^2-4\right)^j
  (2 \eta^2 +\xi \eta-2)^{i-2 j-1} \, .
\end{align}
\end{subequations}
This form of the superconformal Ward identity has a direct translation in Mellin space, because derivatives map to
\begin{align}
 \xi \partial_\xi
&\;\leftrightarrow\;
 -\delta   \, , \qquad
 \eta \partial_\eta
 \;\leftrightarrow\;
 -\rho  \, ,
\end{align}
while products by $\xi^m \eta^n$ act as shifts, see \eqref{eq:shiftMell}.
As a result, the superconformal Ward identity maps into equations that involve shifts of the Mellin amplitudes and products by rational functions of the Mellin variables.
For example, when $p_1=p_2=2$ the Ward identity for $\zeta_-^{(j)}$ reads
\begin{align}
0 = & \,(\delta+\rho ) \Mm_0(\delta ,\rho )
 + 2 \rho  \Mm_1(\delta ,\rho )
 + 4 \rho  \Mm_2(\delta ,\rho )
 - \frac{2 \rho  (\rho +1) (\delta +\rho +1)}
        {(\delta +\rho )^2}
        \Mm_1(\delta ,\rho +2) \notag \\
&+ \delta \Mm_1(\delta +1,\rho -1)
 + 2 \delta  \Mm_2(\delta +1,\rho -1)
 - \frac{2 \delta  \rho  (\delta +\rho +1)}
        {(\delta +\rho )^2}
        \Mm_1(\delta +1,\rho +1) \notag \\
&- \frac{4 \rho  (\rho +1) (\delta +2 \rho +2)}
        {(\delta +\rho )^2}
        \Mm_2(\delta ,\rho +2)
 - \frac{2 \delta  \rho  (2 \delta +5 \rho +3)}
        {(\delta +\rho )^2}
        \Mm_2(\delta +1,\rho +1) \notag \\
&- \frac{\delta  (\delta +1) (\delta +3 \rho )}
        {(\delta +\rho )^2}
        \Mm_2(\delta +2,\rho )
 + \frac{4 \rho  (\rho +1) (\rho +2) (\rho +3)}
        {(\delta +\rho )^2 (\delta +\rho +2)}
        \Mm_2(\delta ,\rho +4) \notag \\
&+ \frac{8 \delta  \rho  (\rho +1) (\rho +2)}
        {(\delta +\rho )^2 (\delta +\rho +2)}
        \Mm_2(\delta +1,\rho +3)
 + \frac{4 \delta  (\delta +1) \rho  (\rho +1)}
        {(\delta +\rho )^2 (\delta +\rho +2)}
        \Mm_2(\delta +2,\rho +2) \, .
\end{align}
It is not hard to check that the Mellin amplitude \eqref{eq:mell22} indeed satisfies the Ward identity.
Note that $\Mm_0$ can be solved directly as a function of $\Mm_1$ and $\Mm_2$.
This seems to be a general feature, although we leave a more detailed analysis of the Ward identities for future work.

\section{Conclusions}
\label{sec:conclusion}

In this work we proposed a bootstrap method for holographic defect correlators, inspired by the position-space bootstrap of \cite{Rastelli:2016nze,Rastelli:2017udc}.
Our focus was on two-point functions of local operators in the presence of the holographic defect, a setup that is relatively unexplored in the literature.
The most important ingredient for the bootstrap is knowledge of certain Witten diagrams, that were partially studied in \cite{RastZhouMell,Mazac:2018biw,Kaviraj:2018tfd,Goncalves:2018fwx}.
The present paper extends these results, by analyzing systematically all Witten diagrams that should contribute in the leading supergravity approximation to setups of arbitrary dimension and codimension.
In this respect, the spin-two bulk-exchange diagram is the most challenging diagram, and its computation in appendix \ref{app:spintwo} constitutes an important new technical result.

To illustrate the position-space bootstrap, we revisited the half-BPS Wilson line in $\Nm=4$ SYM.
References \cite{Barrat:2021yvp,Barrat:2022psm} initiated the study of bulk two-point functions with the Wilson line using analytic bootstrap.
Our calculation clarifies the results in \cite{Barrat:2021yvp,Barrat:2022psm}, because we show that their ``low-spin ambiguities'' correspond to defect-exchange and contact Witten diagrams.
Furthermore, because the present method is more streamlined, we have succeeded in obtaining closed formulas for correlator of chiral-primary operators of arbitrary length.
These formulas involve finite sums with known coefficients, see the position space result \eqref{eq:pos-res} and its Mellin counterpart \eqref{eq:mell-res}.
The position-space formula agrees with the explicit calculation \eqref{eq:tentative-corr}, up to contact terms that we were unable to fix without the bootstrap.
Although this provides a strong sanity check for our method, it still begs the question of how to obtain the correlators from first principles.
We hope this will be clarified in future work.

Our results open a number of new research directions.
One possibility is to extend the calculation in $\Nm=4$ SYM with the half-BPS Wilson line to subleading orders in the large-$N$ or large-$\lambda$ expansion.
Presumably this requires the use of integrated correlators and a better understanding of the flat-space limit, as \cite{Pufu:2023vwo} proposes inspired by the success of similar work, e.g. \cite{Binder:2019jwn,Chester:2020dja}.
Another option is to study other holographic defects at leading order in the supergravity limit.
Some examples in $\Nm=4$ SYM are Wilson lines in symmetric or antisymmetric representations \cite{Drukker:2005kx,Gomis:2006sb,Giombi:2006de,Yamaguchi:2006tq}, 't Hooft lines \cite{Kapustin:2005py,Chen:2006iu}, surface operators \cite{Gukov:2006jk,Buchbinder:2007ar,Harvey:2008zz}, or interfaces \cite{Karch:2000gx,Karch:2000ct,DeWolfe:2001pq}.
The present methods are not restricted to four dimensions, and they should also apply to the $3d$ ABJM or $6d$ $(2,0)$ theories.
In this respect, recently \cite{Meneghelli:2022gps} studied the M2-brane surface defect of $6d$ $(2,0)$ with the analytic bootstrap of \cite{Barrat:2021yvp,Barrat:2022psm}, and it seems this setup is also amenable to the Witten diagram bootstrap \cite{WIP20}.

Finally, we have outlined in section \ref{sec:anfuncts} how some of our results are interconnected with other analytic bootstrap methods, such as analytic functionals.
However, these links could be extended and made more precise.
For example, it would be good to understand under what circumstances the Mellin representation is valid for non-perturbative defect correlators, in the spirit of \cite{Penedones:2019tng,Bianchi:2021piu}, and also to derive dispersion relations directly in Mellin space.
Alternatively, one could explore the crossing-symmetric Mellin bootstrap of \cite{Gopakumar:2016wkt,Gopakumar:2016cpb,Dey:2016mcs}, or the factorization properties of Mellin amplitudes in analogy with \cite{Paulos:2011ie,Fitzpatrick:2011ia,Goncalves:2014rfa}.
Finally, one might attempt to bootstrap higher-point correlators, see \cite{Chen:2023oax} for recent progress in BCFT.

\section*{Acknowledgements}

I am particularly grateful to Julien Barrat and Pedro Liendo for collaborations that inspired this work.
I am also really thankful to Yifan Wang and Xinan Zhou for many useful discussions, Simone Giombi for correspondence and comments on the draft, and Junding Chen and Xinan Zhou for collaboration on related work.
I thank Pietro Ferrero, Vasco Gon\c{c}alves, Apratim Kaviraj, Leonardo Rastelli, Victor Rodriguez and Junchen Rong for stimulating discussions.
Preliminary versions of this work were presented in Porto, SCGP, NYU and Torino, and I would like to thank these groups for their interesting questions and comments.
This work was supported by a grant from the Simons Foundation (915279, IHES) and (733758, Simons Bootstrap Collaboration).

\appendix

\section{Spin-two bulk exchange}
\label{app:spintwo}

In this appendix we compute the spin-two bulk-exchange diagram with the method of \cite{DHoker:1999mqo}.
The original paper considered only graviton exchange, and later this was generalized to massive spin-two exchanges \cite{Arutyunov:2002fh,Rastelli:2017udc} and unequal external scalars \cite{Berdichevsky:2007xd}.
By a slightly modification of their method, we compute the Witten diagram much more explicitly than reference \cite{Berdichevsky:2007xd}.
This result is relevant for setups with and without defects, and we hope it will find use in future calculations.

Our initial goal is to compute
\begin{align}
 A_{\mu\nu}(w,x_1,x_2)
 = \int \frac{d^{d+1}{z}}{z_0^{d+1}}
   G_{\mu\nu\mu'\nu'}(w,z) T^{\mu'\nu'}(z,x_1,x_2) \, ,
 \label{eq:Aintegral}
\end{align}
where $G_{\mu\nu\mu'\nu'}(w,z)$ is the bulk-to-bulk propagator for a massive spin-two particle.
The form of the tensor $T_{\mu\nu}$ is dictated by the coupling of the tensor and the scalars \eqref{eq:ssT-coup}.
More specifically
\begin{align}
 T_{\mu\nu}(z,x_1,x_2)
 = \frac12 \nabla_{(\mu} K_{\Delta_1} \nabla_{\nu)} K_{\Delta_2}
 - \frac{g_{\mu\nu}}{2} \Big[
  \nabla_\rho K_{\Delta_1} \nabla^\rho K_{\Delta_2}
  + \frac12(m_1^2 + m_2^2 - f) K_{\Delta_1} K_{\Delta_2}
 \Big] \, ,
\end{align}
where the mass of the external scalars is $m_i^2 = \Delta_i(\Delta_i-d)$, the mass of the exchanged tensor is $f = \Delta(\Delta-d)$, and we normalize symmetrization as $a_{(\mu} b_{\nu)} = a_{\mu} b_\nu + a_\nu b_\mu$.
For the sake of clarity, we suppressed arguments of bulk-to-boundary propagators.

The result for $A_{\mu\nu}$ allows to compute exchange diagrams for four-point functions, see for example \cite{Rastelli:2017udc}.
In this paper, we are interested instead in a process where the spin-two particle is absorbed by a brane in the bulk.
Because the absorption vertex is $\int_{\AdS_{p+1}} g^{ij} \varphi_{ij}$, the diagram is
\begin{align}
 \int \frac{d^{p+1} w}{w_0^{p+1}} \, g^{ij} A_{ij}(w,x_1,x_2)
 \equiv \frac{E^{\Delta,2}_{\Delta_1,\Delta_2}(\xi,\eta)}
        {|x^i_1|^{\Delta_1} |x^i_2|^{\Delta_2}} \, .
 \label{eq:spin2diag}
\end{align}
Recall that indices $i,j=p+1,\ldots,d$ are orthogonal to the brane.

The computation of $A_{\mu\nu}$ proceeds in several steps.
First act with the equations of motion on the integral \eqref{eq:Aintegral}, so the bulk-to-bulk propagator reduces to a delta function and the integral trivializes.
Next make an ansatz for the result of integral \eqref{eq:Aintegral}, and compute the action of the equations of motion on this ansatz.
Comparing the two sides gives differential equations for the ansatz that can be solved in closed form.
Finally, compute the defect integral \eqref{eq:spin2diag} using the value of $A_{\mu\nu}$ previously determined.
We carry out the key steps of the calculation, but more detail can be found in the original references.

\paragraph{EOM on the RHS.}
Instead of working with $A_{\mu\nu}$, it is convenient to exploit conformal invariance to simplify its form.
Using a translation $w \to w-x_1$ and $x_2 \to x_{21}$, and a conformal inversion $x'_\mu = x_\mu/x^2$ and $z'_\mu = z_\mu/z^2$, one finds
\begin{align}
 A_{\mu\nu}(w,x_1,x_2)
 = \frac{1}{(x_{12}^2)^{\Delta_2}}
   \frac{J_{\mu\lambda}(w) J_{\nu\rho}(w) I^{\lambda\rho}(w'-x_{12}')}
        {w^4} \, , \quad
 J_{\mu\nu}(w)
 = \delta_{\mu\nu} - \frac{2w_\mu w_\nu}{w^2} \, .
\end{align}
The function $I_{\mu\nu}(w)$ is essentially the integral $A_{\mu\nu}$
\begin{align}
 I_{\mu\nu}(w)
 = \int \frac{d^{d+1}{z}}{z_0^{d+1}}
   G_{\mu\nu\mu'\nu'}(w,z) \bar T^{\mu'\nu'}(z) \, ,
 \label{eq:Iintegral}
\end{align}
but the ``stress-tensor'' no longer depends on the position of external operators:
\begin{align}
 \bar T_{\mu\nu}
 = \frac12 \nabla_{(\mu} z_0^{\Delta_1} \nabla_{\nu)}
 \left( \frac{z_0}{z^2} \right)^{\Delta_2}
 \!\!- \frac{g_{\mu\nu}}{2} \Big[
  \nabla_\rho z_0^{\Delta_1} \nabla^\rho
  \left( \frac{z_0}{z^2} \right)^{\Delta_2}
  \!+ \frac12(m_1^2 + m_2^2 - f) z_0^{\Delta_1}
    \left( \frac{z_0}{z^2} \right)^{\Delta_2}
 \Big] \, .
 \label{eq:Tbar}
\end{align}
Now act with the equations of motion on $I_{\mu\nu}(w)$.
The operator $W_{\mu\nu}^{\ph{\mu\nu}\lambda\rho}$ implements the equation of motion for the spin-two propagator, and in particular
\begin{align}
 W_{\mu\nu}^{\ph{\mu\nu}\lambda\rho}[G_{\lambda\rho\mu'\nu'}]
 = \left(
   g_{\mu\mu'} g_{\nu\nu'}
 + g_{\mu\nu'} g_{\nu\mu'}
 - \frac{2}{d-1} g_{\mu\nu} g_{\mu'\nu'}
 \right) \delta(w,z) \, .
\end{align}
Comparing with \eqref{eq:Tbar} and evaluating the derivatives one finds
\begin{align}
 W_{\mu\nu}^{\ph{\mu\nu}\lambda\rho}[I_{\lambda\rho}(w)]
 = w_0^{\Delta_{12}} t^{\Delta_2} \left(
  \frac{m_1^2+m_2^2-f}{d-1} \, g_{\mu\nu}
 - 2 \Delta_1 \Delta_2 \frac{P_{(\mu} w_{\nu)}}{w^2}
 + 2 \Delta_1 \Delta_2 P_{\mu} P_{\nu} \right) \, .
 \label{eq:WactI}
\end{align}
Here and below we use the notation $P_\mu = \delta_{0\mu}/w_0$.

\paragraph{EOM on the LHS.}
In order to proceed, we make an ansatz for the integral $I_{\mu\nu}(w)$
\begin{align}
 I_{\mu\nu}(w)
 = w_0^{\Delta_{12}} \big( g_{\mu\nu} h(t) + P_\mu P_\nu \phi(t) \big)
 + \nabla_\mu \nabla_\nu \big[ w_0^{\Delta_{12}} X(t) \big]
 + \nabla_{(\mu} \big[ P_{\nu)} w_0^{\Delta_{12}} Y(t) \big] \, ,
 \label{eq:Iansatz}
\end{align}
where $t = w_0^2 / w^2$.
This ansatz is somewhat different than in \cite{Berdichevsky:2007xd}, and it leads to cleaner equations that can be solved explicitly.
Now act with the differential operator $W_{\mu\nu}^{\ph{\mu\nu}\lambda\rho}$ on the ansatz using
\begin{align}
 W_{\mu\nu}^{\ph{\mu\nu}\lambda\rho}[\phi_{\lambda\rho}]
 =
 - \nabla_\rho \nabla^\rho \phi_{\mu\nu}
 + \nabla_{(\nu|}  \nabla^\rho \phi_{\rho|\mu)}
 - \nabla_\mu  \nabla^\nu  \phi^\rho_\rho
 - (2-f) \phi_{\mu\nu}
 + \frac{2d-2+f}{d-1} g_{\mu\nu} \phi^\rho_\rho \, ,
\end{align}
and then compare the result with \eqref{eq:WactI}.
This calculation is tedious, but fortunately it can be implemented in \mathematica.
It is not hard to solve the resulting differential equations, and we find
\begin{align}
 & h(t)
 = \frac{f X(t)-\phi (t)}{d-1} \, , \\
 f & Y(t)
 = 2 (t-1) t \phi '(t)
 + (d-\Delta_{12}-2) \phi (t)
 + \Delta_{1} t^{\Delta_2}
 + a \, , \\
 \frac{d f (d+f-1)}{d-1} & X(t)
 = 2 (t-1) (2 t-1) t^2 \phi ''(t)
 + t \big((4 t-3) (d+2 t-2)-2 \Delta_{12} (t-1) \big) \phi '(t) \notag \\
&+ \left(
    \frac{(d+1) f}{2 (d-1)}
   -\frac{\Delta_{12} (3 d-\Delta_{12}-4)}{2}
   +(d-2) (d-1)
 \right) \phi (t)
 +  a (2 d-\Delta_{12}-1) \notag \\
&+ \frac{t^{\Delta_2}}{2} \big(
    \Delta_2 (\Delta_2-d)
  - \Delta_{1} (\Delta_{1}+2 \Delta_2-3 d+2)
  + 4 \Delta_{1} \Delta_2 t - f \big) \, .
\end{align}
Here $a$ is an integration constant that will drop from the final result.
The function $\phi(t)$ is still the solution to the differential equation
\begin{align}
  4 & (t-1) t^2 \phi ''(t)
 + t \big(2d +12 t-12 + 4 \Delta_{12} (t-1) \big) \phi'(t) \notag \\
&+ \big(\Delta_{12}(d-4-\Delta_{12})+2 d+f-4 \big) \phi (t)
 + 2 a (\Delta_{12}+1)
 + 2 \Delta_{1} (\Delta_{1}+1) t^{\Delta_{2}}
 = 0 \, .
\end{align}
Provided $\Delta_1 + \Delta_2 - \Delta \in 2 \mathbb{Z}_{\ge 0}$, the solution is a finite power series in $t$:
\begin{align}
  \phi(t)
& = \frac{2 a (\Delta_{12}+1)}
           {(\Delta -\Delta_{12}-2) (d-\Delta -\Delta_{12}-2)}
  - \!\!\! \sum_{n=1}^{\frac{\Delta_1+\Delta_2-\Delta+2}{2}}
    \frac{\left(\frac{\Delta-\Delta_1-\Delta_2}{2} \right)_{n-1}
          \left(\frac{d-\Delta -\Delta_1-\Delta_2}{2}\right)_{n-1}}
         {2 (1-\Delta_1)_{n-2} (1-\Delta_2)_n} \, t^{\Delta_2-n} \, .
\end{align}

\paragraph{Restoring $A_{\mu\nu}$.}
Now we have an explicit formula for $\phi(t)$, $h(t)$, $X(t)$ and $Y(t)$.
To obtain a formula for $I_{\mu\nu}$, we should simply unwrap the ansatz \eqref{eq:Iansatz} using
\begin{align}
 \nabla_\mu \nabla_\nu & \big[ w_0^{\Delta_{12}} X(t) \big]
 = w_0^{\Delta_{12}} \Big[ \!
 - g_{\mu\nu} \left(2 t \partial_t + \Delta_{12} \right)
 - \frac{P_{(\mu} w_{\nu)}}{w^2}
   \left(4 t^2 \partial_t^2 + (2 \Delta_{12}+6) t \partial_t \right)
   \notag \\
&+ \frac{w_\mu w_\nu}{w^4}
   \left(4 t^2 \partial_t^2 +8  t \partial_t\right)
 + P_\mu P_\nu \left(4 t^2 \partial_t^2+ (4 \Delta_{12}+6) t \partial_t+\Delta_{12}(\Delta_{12}+1)\right)
 \Big] X(t) \, , \\
 \nabla_{(\mu} \big[ & P_{\nu)} w_0^{\Delta_{12}} Y(t) \big]
 = w_0^{\Delta_{12}} \Big[
    2 (\Delta_{12}+1) P_\mu P_\nu
  - 2 g_{\mu\nu}
  + 4 P_\mu P_\nu \, t \partial_t
  - \frac{2 P_{(\mu} w_{\nu)}}{w^2} t \partial_t
 \Big] Y(t) \, .
\end{align}
Finally, to obtain $A_{\mu\nu}$ from $I_{\mu\nu}$ we need to undo the coordinate transformation.
The transformation is simple for the cross-ratio $t$ and $w_0$
\begin{align}
 t' \; \to \; x_{12}^2 K_1(x_1,w) K_1(x_2,w) \, , \qquad
 w_0' \; \to \; K_1(x_1,w) \, ,
\end{align}
where $K_\Delta(x,w)$ is the bulk-to-boundary propagator \eqref{eq:K-blk-bdy}.
Furthermore, the contractions with $J_{\mu\nu}(w)$ were worked out in \cite{Rastelli:2017udc}:
\begin{align}
&P'_\nu \frac{J_{\mu\nu}}{w^2}
 \; \to \; R_\mu \equiv P_\mu - 2 \frac{(w-x_1)_\mu}{(w-x_1)^2} \, , \\
&\frac{J_{\mu\nu}}{w^2} \frac{(w'-x')_\mu}{(w'-x')^2}
 \; \to \; Q_\mu \equiv
 - \frac{(w-x_1)_\mu}{(w-x_1)^2}
 + \frac{(w-x_2)_\mu}{(w-x_2)^2} \, , \\
&\frac{J_{\mu\rho}}{w^2} g'_{\rho\lambda} \frac{J_{\lambda\nu}}{w^2}
 \; \to \; g_{\mu\nu} \, .
\end{align}
For applications to four-point functions, it is now straightforward to express spin-two exchange Witten diagram as sums of $D$-functions, as explained in more detail in \cite{Rastelli:2017udc}.

\paragraph{Computing the defect integral.}
Finally, to obtain the exchange diagram $E^{\Delta,2}_{\Delta_1\Delta_2}$ we need to contract $g^{ij} A_{ij}$, where $i,j=p+1,\ldots,d$ are directions orthogonal to the brane.
Since we integrate $w$ along the brane, the orthogonal components of $w$ vanish $w^i =0$, and we find
\begin{align}
 g_i^{\ph ii}
& \, = \, d-p \, , \qquad
 R_i R^i
 \, = \, 4 |x_1^i|^2 K_2(x_1,w) \, , \label{eq:line1} \\
 Q_i Q^i
& \, = \, |x_1^i|^2 K_2(x_1,w)
     + |x_2^i|^2 K_2(x_2,w)
     - 2 x^i_1 x^i_2 K_1(x_1,w) K_1(x_2,w) \, , \\
 Q_i R^i
&\, = \, 2 |x_1^i|^2 \, K_2(x_1,w)
   - 2 x_1^i x_2^i \, K_1(x_1,w) K_1(x_2,w) \, . \label{eq:line3}
\end{align}
Finally, the $w$-integral is given by the contact diagram formula \eqref{eq:def-cont-diag}, and the terms with $x_{1,2}^i$ in \eqref{eq:line1}-\eqref{eq:line3} lead to powers of the cross-ratios $\xi$ and $\eta$, see \eqref{eq:cross-ratios-xieta}.
This process gives a closed-form expression for $E^{\Delta,2}_{\Delta_1\Delta_2}$ as a finite sum of contact diagrams, but because the expression is not particularly illuminating, we give it in an ancillary notebook.

\section{Details on the \texorpdfstring{$\AdS_5$}{AdS5} effective action}
\label{sec:cubic-coupl}

This appendix gives formulas omitted in section \ref{sec:act5}.
The normalizations of kinetic terms read
\begin{align}
\label{eq:zeta-def}
 \zeta^s_k
 = \frac{32 \pi^3 k(k-1)}{2^{k-1} (k+1)^2} \, , \quad
 \zeta^t_k
 = \frac{32 \pi^3 (k+4)(k+5)}{2^{k-1} (k+1)(k+3)} \, , \quad
 \zeta^\varphi_k
 = \frac{\pi^3}{2^{k-1} (k+1)(k+2)} \, .
\end{align}
For the three-point couplings, it is convenient to introduce
\begin{align}
\label{eq:sigma-alpha}
 \Sigma = k_1 + k_2 + k_3 \, , \quad
 \a_1 = \frac{k_2 + k_3 - k_1}{2} \, , \quad
 \a_2 = \frac{k_1 + k_3 - k_2}{2} \, , \quad
 \a_3 = \frac{k_1 + k_2 - k_3}{2} \, .
\end{align}
All three-point couplings have the same $R$-symmetry structure, which we factor out
\begin{align}
 X^{IJK}_{k_1 k_2 k_3}
 = X_{k_1 k_2 k_3}
 \langle \Cm^I_{k_1} \Cm^J_{k_2} \Cm^K_{k_3} \rangle
 \qquad \text{for} \quad
 X = S, T, G \, .
 \label{eq:Rsymstruct}
\end{align}
The factored term is defined in equation \eqref{eq:CCC-contr} below, while the couplings read
\begin{subequations}
\label{eq:threept-coup}
 \begin{align}
 S_{k_1 k_2 k_3}
&= \frac{2^7 \Sigma((\Sigma/2)^2-1) ((\Sigma/2)^2-4) \a_1\a_2\a_3}
        {3(k_1+1)(k_2+1)(k_3+1)} \,
   a_{k_1k_2k_3} \, , \\[0.2em]
 T_{k_1 k_2 k_3}
&= \frac{2^7 (\Sigma+4) (\a_1+2)(\a_2+2)
         \a_3(\a_3-1)(\a_3-2)(\a_3-3)(\a_3-4)}
        {(k_1+1)(k_2+1)(k_3+3)} \,
   a_{k_1k_2k_3} \, , \\[0.2em]
 G_{k_1 k_2 k_3}
&= \frac{4 (\Sigma+2)(\Sigma+4)\a_3(\a_3-1)}
        {(k_1+1)(k_2+1)} \,
   a_{k_1k_2k_3} \, , \\[0.2em]
 a_{k_1k_2k_3}
&= \frac{\pi^3}{(\Sigma/2+2)! 2^{(\Sigma-2)/2}}
   \frac{k_1! k_2! k_3!}{\a_1! \a_2! \a_3!} \, .
\end{align}
\end{subequations}
The massive spin-two field $\varphi_{\mu\nu}$ couples to the tensor
\begin{align}
 T_{\mu\nu}(s_1,s_2)
 = \frac12 \nabla_{(\mu} s_1 \nabla_{\nu)} s_2
 - \frac12 g_{\mu\nu} \Big[
  \nabla_\rho s_1 \nabla^\rho s_2
  + \frac12(m_1^2 + m_2^2 - f) s_1 s_2
 \Big] \, .
 \label{eq:ssT-coup}
\end{align}

\section{Spherical harmonics}
\label{sec:spherical}

This appendix discusses $S^5$ spherical harmonics.
The ultimate goal is to map the spherical harmonics $Y^I$ often found in the supergravity literature, to the index-free notation of section \ref{sec:setupN4}, that is more convenient in field-theory calculations.

\subsection{Basic definition}

Spherical harmonics provide a basis of scalar functions in $S^5$, that can be organized in irreducible representations of $SO(6)$. For each $k = 0,1,2,\ldots$, we introduce a basis of tensors $\Cm_{m_1 \ldots m_k}^I$ that are symmetric and traceless in the indices $m_i$, where each index runs over $m_i = 1,\ldots,6$.
The index $I=1,\ldots,d_k$ labels the $d_k = \frac{1}{12} (k+1) (k+2)^2 (k+3)$ elements in the irrep.
We choose the basis $\Cm$ such that
\begin{align}
 \Cm^{I}_{m_1 \ldots m_k} \Cm^{J}_{m_1 \ldots m_k}
 = \delta^{IJ} \, , \qquad
 \Cm^{I}_{m_1 \ldots m_k} \Cm^{I}_{n_1 \ldots n_k}
 = \Pi^{m_1\ldots m_k}_{n_1 \ldots n_k} \, ,
 \label{eq:C-complete}
\end{align}
where $\Pi$ is normalized as $\Pi^2 = \Pi$, and it projects tensors to their symmetric-traceless part.
Then, we define the spherical harmonics as
\begin{align}
 Y^I_k(y)
 = \Cm^{I}_{m_1 \ldots m_k} \theta^{m_1} \ldots \theta^{m_k}
 \qquad \text{for} \qquad
 \theta^m = \left(
   \frac{y^A}{1 + \frac14 y^2},
   \frac{1-\frac14 y^2}{1 + \frac14 y^2}
 \right) \, .
\end{align}
Here $\theta^m$ is a six-dimensional unit vector $\theta^2 = 1$ that labels a point in $S^5$, while $y^A$ for $A = 1,\ldots,5$ are the stereographic coordinates of the sphere used in section \ref{sec:effect-act}.
In particular, we took the string solution in \eqref{eq:static} to be located at $y^A = 0$, which corresponds to $\theta = (0, \ldots, 0, 1)$.

\subsection{Relation to index-free notation}

In section \ref{sec:MWL} we study correlators of chiral-primary operators $S_k$ in rank-$k$ symmetric-traceless representations of $SO(6)$.
In $\Nm=4$ SYM these operators are constructed from $k$ fundamental scalars. One possible description of $S_k$ consists on contracting with the tensors $\Cm^I$
\begin{align}
 S_k^I(x) \,\propto\, \Cm^I_{m_1 \ldots m_k}
 \tr \Big[ \phi^{m_1}(x) \ldots \phi^{m_k}(x) \Big] \, .
\end{align}
The constant of proportionality is fixed requiring unit-normalized two-point function.
On the other hand, chiral-primary operators can also be described using index-free notation
\begin{align}
 S_k(x,u) \,\propto\, u_{m_1} \ldots u_{m_k}
 \tr \Big[ \phi^{m_1}(x) \ldots \phi^{m_k}(x) \Big] \, ,
\end{align}
where $u_m$ is a null six-component vector $u^2 = 0$.
It follows from the completeness relation \eqref{eq:C-complete} that the two notations are related as
\begin{align}
 S_k(x,u) = u^{m_1} \ldots u^{m_k} \Cm^I_{m_1\ldots m_k} S^I_k(x) \, .
 \label{eq:toidxfree}
\end{align}
Finally, given an operator in index-free notation, we can free the vector indices \cite{Costa:2011mg}
\begin{align}
\label{eq:todorov}
 S_{m_1\ldots m_k}(x)
 = \frac{1}{k! (k+1)!} D_{m_1} \ldots D_{m_k} S_k(x,u) \, ,
\end{align}
where we use the Todorov operator
\begin{align}
 D_m
 = \left( 2 + u \cdot \frac{\partial}{\partial u} \right)
 \frac{\partial}{\partial u^m}
 - \frac{u_m}{2} \frac{\partial^2}{\partial u \cdot \partial u} \, .
\end{align}

\subsection{Index contractions}

Using this preliminary information, we are ready to map the $R$-symmetry structures that use spherical harmonics to index free notation.

\paragraph{Bulk-exchange diagrams} In the holographic calculation, one encounters bulk-bulk-bulk vertices \eqref{eq:Rsymstruct} that are proportional to
\begin{align}
 \label{eq:CCC-contr}
 \langle \Cm^I_{k_1} \Cm^J_{k_2} \Cm^K_{k_3} \rangle
 = \Cm^I_{m_1\ldots m_{\a_3}n_1\ldots n_{\a_2}}
   \Cm^J_{m_1\ldots m_{\a_3}l_1\ldots l_{\a_1}}
   \Cm^K_{n_1\ldots n_{\a_2}l_1\ldots l_{\a_1}} \, ,
\end{align}
where $\a_i$ were defined in \eqref{eq:sigma-alpha}.
As a result, the correlator of interest contains terms
\begin{align}
 \vev{S_{k_1}^I S_{k_2}^J W} \supset
 \langle \Cm^{I}_{p_1} \Cm^{J}_{p_2} \Cm^{K}_k \rangle Y_k^{K}(0) \, .
 \label{eq:tenstruct}
\end{align}
When we map this formula to index-free notation using \eqref{eq:toidxfree}, we encounter contractions of the form $C^I C^I = \Pi$, with the projector $\Pi$ introduced in \eqref{eq:C-complete}.
This projector can be written in terms of the Todorov operator \cite{Costa:2011mg} in terms of an auxiliary null polarization vector $u$:
\begin{align}
 \Pi_{m_1\ldots m_k}^{n_1\ldots n_k}
 = \frac{1}{k! (k+1)!} D_{m_1} \ldots D_{m_k} u^{n_1} \ldots u^{n_1} \, .
\end{align}
Using this information, one sees that the tensor structure in \eqref{eq:tenstruct} is equivalent to the following expression
\begin{align}
 \langle \Cm^{I}_{p_1} \Cm^{J}_{p_2} \Cm^{K}_k \rangle Y_k^{K}(0)
 \;\; &\to \;\;
 \frac{(u_1 \cdot u_2)^{\frac{p_1+p_2-k}{2}}}{k! (k+1)!} \,
 (u_1 \cdot D)^{\frac{k+p_{12}}{2}}
 (u_2 \cdot D)^{\frac{k-p_{12}}{2}}
 (u \cdot \theta)^k \notag \\
 & =\;\; (u_1 \cdot \theta)^{p_1} (u_2 \cdot \theta)^{p_2} \times
   h_{p_1p_2}^k(\sigma) \, .
 \label{eq:bulkrsym}
\end{align}
Here $\sigma$ is the $R$-symmetry cross-ratio in \eqref{eq:invariant}, the $R$-symmetry block is defined in \eqref{eq:Rsymblock}, and we use $\theta = \theta(y=0)$.
The right-hand side of \eqref{eq:bulkrsym} was previously encountered in equation (A.6) of \cite{Barrat:2021yvp}, where it was observed that it equals an $R$-symmetry block.

\paragraph{Defect-exchange diagrams}
The calculation for the defect exchanges, proceeds in exactly the same way.
The exchange of a $x^i$ mode comes with tensor structure
\begin{align}
 Y_{p_1}^I(y) Y_{p_2}^J(y) \;\;\to\;\;
 (u_1 \cdot \theta)^{p_1} (u_2 \cdot \theta)^{p_2}
 \times 1 \, .
\end{align}
For the exchange of $y^A$, the tensor structure is instead
\begin{align}
 \frac{\partial Y_{p_1}^I}{\partial y^A}
 \frac{\partial Y_{p_2}^J}{\partial y^A}
 \Bigg|_{y=0}
 \;\;\to\;\;
 \frac{\partial (u_1 \cdot \theta)^{p_1}}{\partial y^A}
 \frac{\partial (u_2 \cdot \theta)^{p_2}}{\partial y^A}
 \Bigg|_{y=0}
 =
 (u_1 \cdot \theta)^{p_1} (u_2 \cdot \theta)^{p_2} \times
 p_1 p_2 ( \sigma - 1) \, .
\end{align}
As expected, these results are proportional to $R$-symmetry block for defect operator that are scalars and vectors of $SO(5)_R$.
This follows simply from the formula for the blocks $\wh h_{\wh K}(\sigma)$ given in \cite{Barrat:2021yvp,Barrat:2022psm}, evaluated at $\wh K = 0,1$.

\section{Comparison to topological sector}
\label{sec:topsect}

\begin{figure}
\centering
\def\svgwidth{.3\linewidth}
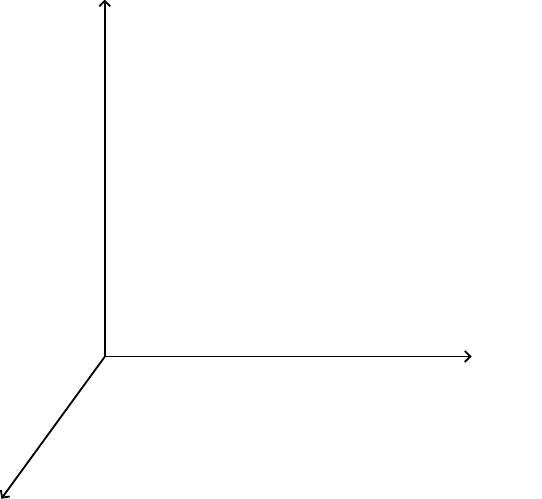
$\qquad\qquad$
\def\svgwidth{.3\linewidth}
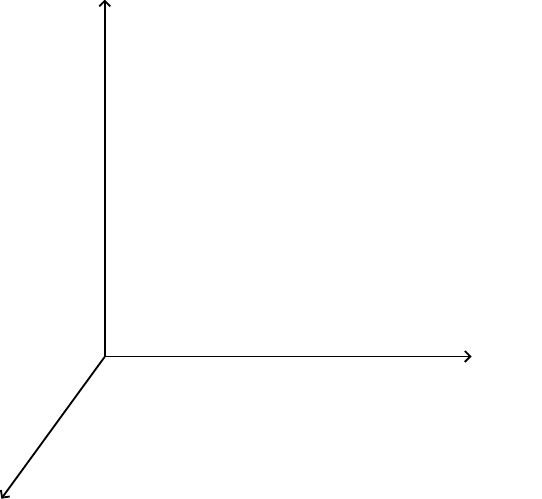
 \caption{Comparison of the closed-chain (left) and open-chain (right) topology.}
 \label{fig:openclose}
\end{figure}

In this appendix, we compare our prediction for the two-point function in the supegravity limit \eqref{eq:pos-res}, with results from supersymmetric localization \cite{Giombi:2009ds,Giombi:2012ep}.

The main observation is that correlators of chiral-primary operators and Wilson loops restricted to $S^2 \subset \mathbb{R}^4$ become topological when applying an appropriate twist.
This topological subsector is the reduction of the full theory to the cohomology of a certain nilpotent supercharge \cite{topologicalNotes}.
The topological sector admits an OPE, which is essentially the truncation of the usual OPE to chiral-primary operators.
The truncated OPE leads to so-called microbootstrap equations, that impose non-trivial constraints on the CFT data of chiral-primary operators in the presence of a defect \cite{Barrat:2023}.
The story is completely analogous to other protected sectors in the literature, see for example \cite{Drukker:2009sf,Beem:2013sza,Chester:2014mea}.
Here, one has the added feature that the topological sector is described by a two-dimensional topological field theory, and correlation functions can be evaluated in terms of Gaussian matrix models \cite{Giombi:2009ds,Giombi:2012ep,Giombi:2018qox,Giombi:2018hsx}.

To describe the topological sector, we put the Wilson line at $x_2 =x_3=x_4 =0$, so that it extends in the ${\rm t} = x_1$ direction, and also choose $\theta = (0, \ldots, 0, 1)$.
Furthermore, chiral-primary operators are restricted to the plane $({\rm t},{\rm x})$, where ${\rm x} = x_2$.
We call the topological operators $\Sm_p$, which are chiral-primary operators with $R$-symmetry polarization correlated with the spacetime position
\begin{align}
 \Sm_p({\rm t},{\rm x})
 \equiv S_p(x({\rm t},{\rm x}); u({\rm t},{\rm x})) \, .
\end{align}
The form of $u({\rm t},{\rm x})$ can be determined by studying the cohomology of the nilpotent supercharge, and it reads \cite{topologicalNotes}
\begin{align}
 x({\rm t},{\rm x})
&= ({\rm t},{\rm x},0,0) \, , \\
 u({\rm t},{\rm x})
&= \big( 1-{\rm t}^2-{\rm x}^2, i (1 + {\rm t}^2+{\rm x}^2), 0, 0, -2{\rm t}, -2{\rm x} \big) \, .
\end{align}
For the correlator we are interested in $\vev{ \Sm_{p_1}({\rm t}_1,{\rm x}_1) \Sm_{p_2}({\rm t}_2,{\rm x}_2) W(\theta) }$, we can use conformal transformations to set ${\rm t}_1 = {\rm t}_2 = 0$ and ${\rm x}_2 = 1$. Then the only free parameter is ${\rm x}_1 \equiv z$, and the correlator reads
\begin{align}
 \Fm_{\text{top}}(\vartheta)
 = \Fm_c\left(
  \xi    = \frac{(1-z)^2}{|z|}, \,
  \eta   = \frac{z}{|z|}, \,
  \sigma = - \frac{(1-z)^2}{2z}
 \right) \, .
 \label{eq:top-sect}
\end{align}
The statement that the correlator becomes topological is that it only depends on
\begin{align}
 \vartheta
 = \sgn z
 =
 \begin{cases}
  -1 \quad \text{for open chain} \\
  +1 \quad \text{for closed chain}
 \end{cases} \, .
\end{align}
In other words, the topological correlator only depends on whether the two local operators are on the same side (closed chain) or opposite sides (open chain) of the Wilson loop.
The open/closed chain terminology, which we borrow from \cite{Giombi:2012ep}, is explained in figure \ref{fig:openclose}.
At this point, it is possible to check that the correlator \eqref{eq:pos-res} indeed reduces to a constant in the kinematics \eqref{eq:top-sect}, giving different results depending on $\vartheta = \sgn z$.

We can perform an even stronger check, because the topological correlator was computed in the planar limit at arbitrary 't Hooft coupling in \cite{Giombi:2012ep}.
To compare to their result, we need to note two important differences.
First, their definition of connected correlator differs from ours by the factors of $a_{p_1} a_{p_2}$ in \eqref{eq:conncorrN4}.
Second, they consider the two-point function of single-trace operators $\Om_p = \tr (u \cdot \phi)^{p_1}$ in $U(N)$ gauge theory, while we consider fields $S_p$ dual to supergravity modes.
The difference is that for supergravity fields extremal OPE coefficients vanish, in particular $\lambda_{S_{p_1} S_{p_2} S_{|p_{12}|}} = 0$, whereas for single-trace operators $\lambda_{\Om_{p_1} \Om_{p_2} \Om_{|p_{12}|}} \ne 0$.
These two differences with the analysis of Giombi and Pestun (GP), can be easily compensated as $\Fm_{\text{top}}
 \sim \Fm_{\text{GP}}
 - a_{p_1} a_{p_2}
 - \lambda_{p_1p_2|p_{12}|} a_{|p_{12}|}$.
More precisely, the prediction of \cite{Giombi:2012ep} becomes in our conventions
\begin{align}
 \Fm_{\text{top}}&(\vartheta)
 = \frac{\sqrt{p_1 p_2}}{2^{\frac{p_1+p_2+2}{2}}I_1(\sqrt{\lambda })}
    \frac{\sqrt{\lambda}}{N^2} \Bigg[
  - \frac{\sqrt{\lambda } I_{p_1}(\sqrt{\lambda}) I_{p_2}(\sqrt{\lambda})}{2 I_1(\sqrt{\lambda})}
  - (-\vartheta )^{\frac{p_1+p_2-|p_{12}|}{2}}
    |p_{12}| I_{|p_{12}|}(\sqrt{\lambda}) \notag \\
& + \frac{\sqrt{\lambda}}{2} I_{1+\veps}(\sqrt{\lambda})
  + \sum _{k=1}^{\pmin} (-\vartheta)^k (p_1+p_2-2 k)
    I_{p_1+p_2-2 k}(\sqrt{\lambda})
  - \!\!\! \sum _{k=1}^{\frac{p_1+p_2-\veps -2}{2}} \!\!\!
    (2 k+\veps ) I_{2 k+\veps }(\sqrt{\lambda})
 \Bigg],
 \label{eq:pred-topcorr}
\end{align}
where we introduced $\veps = p_1 + p_2 \; (\textrm{mod}\, 2)$.
The first term in the first line corrects for the extra $a_{p_1} a_{p_2}$, the second term in the first line corrects for the extra $\lambda_{p_1p_2|p_{12}|} a_{|p_{12}|}$.
The last line corresponds to equation (4.42) in \cite{Giombi:2012ep}, where we rewrote the infinite sums as finite sums, see also their equation (4.46).
The result \eqref{eq:pred-topcorr} should be valid at any $\lambda$, and at leading order as $\lambda \to \infty$, one checks that it is in perfect agreement with our prediction \eqref{eq:pos-res}.

\section{Ward identity in Mellin space}
\label{sec:wi-mell}

In this appendix, we rewrite the superconformal Ward identity in Mellin space, adapting the method of \cite{Zhou:2017zaw}.
Recall that the superconformal Ward identity has a simple expression in terms of $z,\zb,\a$ defined in \eqref{eq:xieta2zzb}:
\begin{equation}
 W(z,\zb)
 = \left. \left( \partial_z + \frac{1}{2} \partial_\alpha \right)
   \Fm(z,\zb,\a) \right|_{z = \alpha} = 0\, .
 \label{eq:WIap}
\end{equation}
Furthermore, the correlator $\Fm$ decomposes as a sum of $R$-symmetry channels
\begin{align}
 \Fm(z,\zb,\a)
 = \sum_{j=0}^{\pmin}
   \frac{(1-\alpha )^{2j}}{\alpha^j }
   \frac{\Fm_j(z,\zb)}{(-2)^j} \, .
\end{align}
It is clear that \eqref{eq:WIap} relates the different channels $\Fm_j$ through  linear differential equations.
In particular, we want the result to be polynomial in $\xi$ and $\eta$, because polynomials have simple action on the Mellin representation \eqref{eq:mellin-def}.
A straightforward application of the chain rule gives
\begin{align}
 W(z,\zb)
 = \sum_{j=0}^{\pmin}
 \frac{(1-z)^{2j}}{z^j} \left(
   \frac{\partial_\eta}{2\sqrt{z \zb}}
 + \frac{\zb-1}{\sqrt{z \zb}} \partial_\xi
 - \frac{\xi \partial_\xi + \eta \partial_\eta}{2 z}
 + \frac{j (z+1)}{2 (z-1) z}
 \right) \frac{\Fm_j(\xi ,\eta )}{(-2)^j}
 = 0 \, .
 \label{eq:WI1}
\end{align}
The crucial observation is that $\xi$ and $\eta$ are invariant under $z \leftrightarrow \zb$ and $z,\zb \leftrightarrow \frac1z, \frac1\zb$, but the Ward identity \eqref{eq:WI1} is not.
Thus, to rewrite the Ward identity as a polynomial in $\xi$ and $\eta$, we have to take linear combinations of \eqref{eq:WI1} that respect these symmetries.
First we implement the $z \to \frac1z$ invariance by considering
\begin{align}
 W(z,&\zb) + W\left(\tfrac1z, \tfrac1\zb \right) 
 = \sum_{j=0}^{\pmin}
 \frac{(1-z)^{2j}}{2z^j} \left(
 \frac{(1-z)^2}{z} (\eta  \partial_\eta  + \xi \partial_\xi + j)
 - \xi \partial_\eta
 + 4 \xi \partial_\xi
 + 4 j
 \right) \frac{\Fm_j(\xi ,\eta )}{(-2)^j}
 \, ,
 \label{eq:WI2}
\end{align}
and then we implement $z \leftrightarrow \zb$ invariance taking the two linear combinations
\begin{align}
 W(z,\zb) + W(\tfrac1z, \tfrac1\zb)
 \pm \left(W(\zb,z) + W(\tfrac1\zb, \tfrac1z) \right)
 = 0 \, .
 \label{eq:WIPM}
\end{align}
The two equations in \eqref{eq:WIPM} are equivalent to formula \eqref{eq:WImellin} in the main text.
This identification follows from the definition of $\zeta_{\pm}^{(j)}$ in terms of $z,\zb$:
\begin{align}
 \zeta_{+}^{(j)}
 = \frac{(1-z)^{2j}}{2 z^j}
 + \frac{(1-\zb)^{2j}}{2 \zb^j} \, , \quad
 \zeta_{-}^{(j)}
 = \frac{z \zb}{(\zb-z) (1-z \zb)} \left(
   \frac{(1-z)^{2j}}{z^j} - \frac{(1-\zb)^{2j}}{\zb^j}
 \right) \, .
 \label{eq:defzeta}
\end{align}
Although not manifestly so, these objects are polynomial in $\xi$ and $\eta$.
In fact, the prefactor of $\zeta_-^{(j)}$ ensures it is indeed a polynomial.
Finally, a combination of \eqref{eq:xieta2zzb}, \eqref{eq:defzeta} and the binomial theorem gives formula \eqref{eq:zetaPM} for $\zeta_\pm^{(j)}$.

\providecommand{\href}[2]{#2}\begingroup\raggedright\endgroup

\end{document}

%% file: closed-chain.pdf_tex
\begingroup%
  \makeatletter%
  \providecommand\color[2][]{%
    \errmessage{(Inkscape) Color is used for the text in Inkscape, but the package 'color.sty' is not loaded}%
    \renewcommand\color[2][]{}%
  }%
  \providecommand\transparent[1]{%
    \errmessage{(Inkscape) Transparency is used (non-zero) for the text in Inkscape, but the package 'transparent.sty' is not loaded}%
    \renewcommand\transparent[1]{}%
  }%
  \providecommand\rotatebox[2]{#2}%
  \newcommand*\fsize{\dimexpr\f@size pt\relax}%
  \newcommand*\lineheight[1]{\fontsize{\fsize}{#1\fsize}\selectfont}%
  \ifx\svgwidth\undefined%
    \setlength{\unitlength}{155.88356078bp}%
    \ifx\svgscale\undefined%
      \relax%
    \else%
      \setlength{\unitlength}{\unitlength * \real{\svgscale}}%
    \fi%
  \else%
    \setlength{\unitlength}{\svgwidth}%
  \fi%
  \global\let\svgwidth\undefined%
  \global\let\svgscale\undefined%
  \makeatother%
  \begin{picture}(1,0.92141914)%
    \lineheight{1}%
    \setlength\tabcolsep{0pt}%
    \put(0,0){\includegraphics[width=\unitlength,page=1]{closed-chain.pdf}}%
    \put(0.21886957,0.85574612){\color[rgb]{0,0,0}\makebox(0,0)[lt]{\lineheight{1.25}\smash{\begin{tabular}[t]{l}$x_1=\tau$\end{tabular}}}}%
    \put(0.05744641,-0.00027755){\color[rgb]{0,0,0}\makebox(0,0)[lt]{\lineheight{1.25}\smash{\begin{tabular}[t]{l}$x_3,x_4$\end{tabular}}}}%
    \put(0.60547147,0.77821944){\color[rgb]{0,0,0}\makebox(0,0)[lt]{\lineheight{1.25}\smash{\begin{tabular}[t]{l}$W$\end{tabular}}}}%
    \put(0.30179795,0.65281484){\color[rgb]{0,0,0}\makebox(0,0)[lt]{\lineheight{1.25}\smash{\begin{tabular}[t]{l}$S_{p_1}$\end{tabular}}}}%
    \put(0.25168506,0.47610206){\color[rgb]{0,0,0}\makebox(0,0)[lt]{\lineheight{1.25}\smash{\begin{tabular}[t]{l}$S_{p_2}$\end{tabular}}}}%
    \put(0.75819953,0.30886258){\color[rgb]{0,0,0}\makebox(0,0)[lt]{\lineheight{1.25}\smash{\begin{tabular}[t]{l}$x_2=\sigma$\end{tabular}}}}%
    \put(0,0){\includegraphics[width=\unitlength,page=2]{closed-chain.pdf}}%
  \end{picture}%
\endgroup%

%% file: open-chain.pdf_tex
\begingroup%
  \makeatletter%
  \providecommand\color[2][]{%
    \errmessage{(Inkscape) Color is used for the text in Inkscape, but the package 'color.sty' is not loaded}%
    \renewcommand\color[2][]{}%
  }%
  \providecommand\transparent[1]{%
    \errmessage{(Inkscape) Transparency is used (non-zero) for the text in Inkscape, but the package 'transparent.sty' is not loaded}%
    \renewcommand\transparent[1]{}%
  }%
  \providecommand\rotatebox[2]{#2}%
  \newcommand*\fsize{\dimexpr\f@size pt\relax}%
  \newcommand*\lineheight[1]{\fontsize{\fsize}{#1\fsize}\selectfont}%
  \ifx\svgwidth\undefined%
    \setlength{\unitlength}{155.88356078bp}%
    \ifx\svgscale\undefined%
      \relax%
    \else%
      \setlength{\unitlength}{\unitlength * \real{\svgscale}}%
    \fi%
  \else%
    \setlength{\unitlength}{\svgwidth}%
  \fi%
  \global\let\svgwidth\undefined%
  \global\let\svgscale\undefined%
  \makeatother%
  \begin{picture}(1,0.92141914)%
    \lineheight{1}%
    \setlength\tabcolsep{0pt}%
    \put(0,0){\includegraphics[width=\unitlength,page=1]{open-chain.pdf}}%
    \put(0.21886957,0.85574612){\color[rgb]{0,0,0}\makebox(0,0)[lt]{\lineheight{1.25}\smash{\begin{tabular}[t]{l}$x_1=\tau$\end{tabular}}}}%
    \put(0.05744641,-0.00027755){\color[rgb]{0,0,0}\makebox(0,0)[lt]{\lineheight{1.25}\smash{\begin{tabular}[t]{l}$x_3,x_4$\end{tabular}}}}%
    \put(0.60547147,0.77821944){\color[rgb]{0,0,0}\makebox(0,0)[lt]{\lineheight{1.25}\smash{\begin{tabular}[t]{l}$W$\end{tabular}}}}%
    \put(0.30179795,0.65281484){\color[rgb]{0,0,0}\makebox(0,0)[lt]{\lineheight{1.25}\smash{\begin{tabular}[t]{l}$S_{p_1}$\end{tabular}}}}%
    \put(0.62696553,0.47610206){\color[rgb]{0,0,0}\makebox(0,0)[lt]{\lineheight{1.25}\smash{\begin{tabular}[t]{l}$S_{p_2}$\end{tabular}}}}%
    \put(0.75819953,0.30886258){\color[rgb]{0,0,0}\makebox(0,0)[lt]{\lineheight{1.25}\smash{\begin{tabular}[t]{l}$x_2=\sigma$\end{tabular}}}}%
    \put(0,0){\includegraphics[width=\unitlength,page=2]{open-chain.pdf}}%
  \end{picture}%
\endgroup%